\documentclass[11 pt,onecolumn]{IEEEtran}

\IEEEoverridecommandlockouts     

\usepackage{graphicx,algorithm,algorithmic,subfigure,latexsym,amsmath,amsfonts,amssymb,cite,mathrsfs}
\usepackage{epsfig,color}
\usepackage{psfig,psfrag,wrapfig}
\usepackage{pstricks}
\usepackage{epstopdf}

\newtheorem{lemma}{Lemma}[section]
\newtheorem{theorem}{Theorem}[section]

\newtheorem{definition}{Definition}[section]

\def\rrr#1\\{\par
\medskip\hbox{\vbox{\parindent=2em\hsize=6.12in
\hangindent=4em\hangafter=1#1}}}
\allowdisplaybreaks

\begin{document}



%

\title{\LARGE Semistability-Based Convergence Analysis for Paracontracting Multiagent Coordination Optimization}

\author{Qing Hui and Haopeng Zhang\\
\small{Control Science and Engineering Laboratory\\
Department of Mechanical Engineering \\
Texas Tech University \\
Lubbock, TX 79409-1021\\
{\tt\small (qing.hui@ttu.edu; haopeng.zhang@ttu.edu)}\\
Technical Report CSEL-08-13, August 2013}
\thanks{This work was supported by the Defense Threat Reduction Agency, Basic Research Award \#HDTRA1-10-1-0090 and Fundamental Research Award \#HDTRA1-13-1-0048, to Texas Tech University.}
}

\maketitle
\IEEEpeerreviewmaketitle
\pagestyle{empty}
\thispagestyle{empty}
\baselineskip 22pt

\begin{abstract}
This sequential technical report extends some of the previous results we posted at arXiv:1306.0225.
\end{abstract}

\section{Introduction}   

Recently we have proposed a new class of swarm optimization algorithms called the Multiagent Coordination Optimization
(MCO) \cite{ZH:CEC:2013,ZH:CASE:2013,HZ:TR:2013} algorithm inspired by swarm intelligence and consensus protocols for multiagent coordination in \cite{HHB:TAC:2008,HHB:TAC:2009,Hui:IJC:2010,HH:AUT:2008}. This new algorithm is a new optimization technique based not only on swarm
intelligence \cite{BDT:1999} which simulates the bio-inspired
behavior, but also on cooperative control of autonomous agents. The MCO algorithm
starts with a set of random solutions for agents which can
communicate with each other. The agents then move through the
solution space based on the evaluation of their cost functional and
neighbor-to-neighbor rules like multiagent consensus protocols
\cite{HHB:TAC:2008,HHB:TAC:2009,HH:AUT:2008,Hui:IJC:2010,Hui:TAC:2011,Hui:AUT:2011}.
 Detailed convergence analysis for MCO has been conducted in the companion report \cite{HZ:TR:2013}. In this sequential report, we first propose a \textit{paracontraction} \cite{EKN:LAA:1990} based MCO algorithm and then implement the paracontracting MCO algorithm in a parallel computing way by introducing MATLAB\textsuperscript{\textregistered} built-in function $\mathtt{parfor}$ into the paracontracting MCO algorithm. Then we rigorously analyze the global convergence of the paracontracting MCO algorithm by means of \textit{semistability theory} \cite{HHB:TAC:2008,Hui:TAC:2013}. This sequential report can be viewed as an addendum to the companion report \cite{HZ:TR:2013}.

\section{Mathematical Preliminaries}\label{mp}

\subsection{Graphs}

Let $\mathbb{R}$ denote the set of real numbers and $\mathbb{R}^{n\times n}$ denote the set of $n\times n$ real matrices.
In this sequential report, we use algebraic graph-related notation to describe our paracontracting MCO algorithm. More specifically, let $\mathcal{G}(t)= (\mathcal{V},
\mathcal{E}(t), \mathcal{A}(t))$ denote a \textit{node-fixed dynamic directed graph} (or \textit{node-fixed dynamic digraph}) with the set
of vertices $\mathcal{V}= \{v_1,v_2,\ldots,v_{n}\}$ and
$\mathcal{E}(t)\subseteq \mathcal{V}\times \mathcal{V}$
represent the set of edges, where $t\in\overline{\mathbb{Z}}_{+}=\{0,1,2,\ldots\}$. The time-varying matrix $\mathcal{A}(t)\in\mathbb{R}^{n\times n}$ with
nonnegative adjacency elements $a_{i,j}(t)$ serves as the weighted
adjacency matrix. The node index of $\mathcal{G}(t)$ is denoted as a
finite index set $\mathcal{N}=\{1,2,\ldots,n\}$. An edge of $\mathcal{G}(t)$ is
denoted by $e_{i,j}(t)=(v_i,v_j)$ and the adjacency elements associated
with the edges are positive. We assume $e_{i,j}(t)\in
\mathcal{E}(t)\Leftrightarrow a_{i,j}(t)=1$ and $a_{i,i}(t)=0$ for all $i\in
\mathcal{N}$. The set of neighbors of the node $v_i$ is denoted by
$\mathcal{N}^{i}(t)=\{v_j \in \mathcal {V}:(v_i,v_j)\in \mathcal {E}(t),
j=1,2, \ldots, |\mathcal{N}|, j\not = i\}$, where $|\mathcal{N}|$ denotes the cardinality of $\mathcal{N}$. The degree matrix of a node-fixed dynamic digraph
$\mathcal{G}(t)$ is defined as
$\Delta(t) =[\delta_{i,j}(t)]_{i,j=1,2,\ldots,|\mathcal{N}|}$, where\
\begin{eqnarray*}
\delta_{i,j}(t)=\left\{
      \begin{array}{ll}
        \sum_{j=1}^{|\mathcal{N}|} a_{i,j}(t), & \hbox{if $i=j$,} \\
        0, & \hbox{if $i\neq j$.}
      \end{array}
    \right.
\end{eqnarray*}
The \textit{Laplacian matrix} of the node-fixed dynamic digraph $\mathcal{G}(t)$ is defined by
$L(t)=\Delta(t) - \mathcal {A}(t)$.
If $L(t)=L^{\mathrm{T}}(t)$, then $\mathcal{G}(t)$ is called a \textit{node-fixed dynamic undirected graph} (or simply \textit{node-fixed dynamic graph}). If there is a path from any node to any other node in a node-fixed dynamic digraph,
then we call the dynamic digraph \textit{strongly connected}. Analogously, if there is a path from any node to any other node in a node-fixed dynamic graph,
then we call the dynamic graph \textit{connected}. From now on we use short notations $L_{t},\mathcal{G}_{t},\mathcal{N}^{i}_{t}$ to denote $L(t),\mathcal{G}(t),\mathcal{N}^{i}(t)$, respectively. 

\subsection{Paracontraction}

Paracontraction is a nonexpansive property for a class of linear operators which can be used to guarantee convergence of linear iterations \cite{EKN:LAA:1990}. The following definition due to \cite{EKN:LAA:1990} gives the notion of paracontracting matrices.

\begin{definition}[\hspace{-0.01em}\cite{EKN:LAA:1990}]
Let $\mathbb{R}^{n}$ denote the set of $n$-dimensional real column vectors and $W\in\mathbb{R}^{n\times n}$. $W$ is called \textit{paracontracting} if for any $x\in\mathbb{R}^{n}$, $Wx\neq x$ is equivalent to $\|Wx\|<\|x\|$, where $\|\cdot\|$ denotes the 2-norm in $\mathbb{R}^{n}$.
\end{definition}

Recall from \cite{Bernstein:2009,HCH:2009,Hui:TAC:2013} that a matrix $A\in\mathbb{R}^{n\times n}$ is called \textit{discrete-time semistable} if ${\rm{spec}}(A)\subseteq\{s\in\mathbb{C}:|s|<1\}\cup\{1\}$, and if $1\in{\rm{spec}}(A)$, then $1$ is semisimple, where ${\mathrm{spec}}(A)$ denotes the spectrum of $A$. Hence, $A$ is discrete-time semistable if and only if $\lim_{k\to\infty}A^{k}$ exists.
$A\in\mathbb{R}^{n\times n}$ is called \textit{nontrivially discrete-time semistable} \cite{Hui:TAC:2013} if $A$ is discrete-time semistable and $A\neq I_{n}$, where $I_{n}\in\mathbb{R}^{n\times n}$ denotes the $n\times n$ identity matrix. The following result shows a close relationship between paracontracting matrices and discrete-time semistable matrices under certain circumstances. To state this result, let $\ker(A)$ denote the kernel of $A$.

\begin{lemma}[\hspace{-0.01em}\cite{HZ:TR:2013}]\label{lemma_para}
Let $W\in\mathbb{R}^{n\times n}$. Then $W$ is nontrivially discrete-time semistable, $\|W\|\leq 1$, and $\ker((W-I_{n})^{\mathrm{T}}(W-I_{n})+W^{\mathrm{T}}-I_{n}+W-I_{n})=\ker((W-I_{n})^{\mathrm{T}}(W-I_{n})+(W-I_{n})^{2})$  if and only if $W$ is paracontracting. 
\end{lemma} 

Let $\mathbb{R}^{m\times n}$ denote the set of $m\times n$ real matrices. The following definition is due to \cite{HZ:TR:2013}.

\begin{definition}[\hspace{-0.01em}\cite{HZ:TR:2013}]
Let $A_{k}\in\mathbb{R}^{n\times n}$, $k=0,1,2,\ldots$, and $C\in\mathbb{R}^{m\times n}$. The set of pairs $\{(A_{k},C)\}_{k\in\overline{\mathbb{Z}}_{+}}$ is called \textit{discrete-time approximate semiobservable with respect to some matrix $A\in\mathbb{R}^{n\times n}$} if
\begin{eqnarray}
\bigcap_{k=0}^{\infty}\ker(C(I_{n}-A_{k}))=\ker(I_{n}-A).
\end{eqnarray} 
\end{definition}

Finally, using the above definition and Theorem 1 of \cite{EKN:LAA:1990}, one can show the following key results which are needed for the main convergence result in this technical report. The detailed proofs can be found in \cite{HZ:TR:2013}.

\begin{lemma}[\hspace{-0.01em}\cite{HZ:TR:2013}]\label{lemma_DTSS}
Let $J$ be a (possibly infinite) countable index set and $P_{k}\in\mathbb{R}^{n\times n}$, $k\in J$, be discrete-time semistable, $\|P_{k}\|\leq1$, and $\ker(P_{k}^{\mathrm{T}}P_{k}-I_{n})=\ker((P_{k}-I_{n})^{\mathrm{T}}(P_{k}-I_{n})+(P_{k}-I_{n})^{2})$. Consider the sequence $\{x_{i}\}_{i=0}^{\infty}$ defined by the iterative process $x_{i+1}=Q_{i}x_{i}$, $i=0,1,2,\ldots$, where $Q_{i}\in\{P_{k}:\forall k\in J\}$. 
\begin{itemize}
\item[$i$)] If $|J|<\infty$, then $\lim_{i\to\infty}x_{i}$ exists. If in addition, $P_{k}\in\mathbb{R}^{n\times n}$ is nontrivially discrete-time semistable for every $k\in J$, then $\lim_{i\to\infty}x_{i}$ is in $\bigcap_{k\in\mathcal{I}}\ker(I_{n}-P_{k})$, where $\mathcal{I}$ is the set of all indexes $k$ for which $P_{k}$ appears infinitely often in $\{Q_{i}\}_{i=0}^{\infty}$.
\item[$ii$)] If there exists $s\in J$ such that $P_{s}$ is nontrivially discrete-time semistable, $\{(Q_{k},I_{n})\}_{k\in\overline{\mathbb{Z}}_{+}}$ is discrete-time approximate semiobservable with respect to some nontrivially discrete-time semistable matrix $Q_{r}$,  $r\in\overline{\mathbb{Z}}_{+}$, and for every positive integer $N$, there always exists $j\geq N$ such that $Q_{j}=Q_{r}$, then $\lim_{i\to\infty}x_{i}$ exists and the limit is in $\ker(I_{n}-Q_{r})$.
\end{itemize}
\end{lemma}

\section{Paracontracting Multiagent Coordination Optimization}\label{pmco}

\subsection{Paracontracting MCO with Node-Fixed Dynamic Graph Topology}

\begin{wrapfigure}{r}{.40\linewidth}
\centering
\epsfig{file=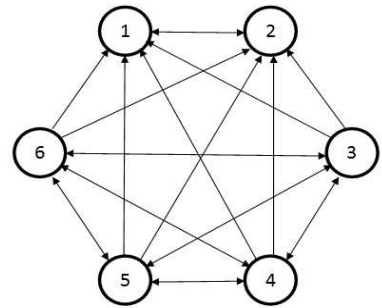,width=0.8\linewidth,height=0.6\linewidth}
\caption{Grouped directed structure.}\label{structure}
\end{wrapfigure}

The MCO algorithm with static graph topology, proposed in \cite{ZH:CEC:2013} to solve a given optimization problem $\min_{\textbf{x}\in\mathbb{R}^{n}}f(\textbf{x})$, can be described in a vector form as follows:
\begin{eqnarray}
\textbf{v}_{k}(t+1)&=&\textbf{v}_{k}(t)+\eta\sum_{j\in\mathcal{N}^{k}}(\textbf{v}_{j}(t)-\textbf{v}_{k}(t))+\mu\sum_{j\in\mathcal{N}^{k}}(\textbf{x}_{j}(t)-\textbf{x}_{k}(t))+\kappa(\textbf{p}(t)-\textbf{x}_{i}(t)),\label{PSO_1}\\
\textbf{x}_{k}(t+1)&=&\textbf{x}_{k}(t)+\textbf{v}_{k}(t+1),\label{PSO_2}\\
\textbf{p}(t+1)&=&\left\{\begin{array}{ll}
\textbf{p}(t)+\kappa(\textbf{x}_{\min}(t)-\textbf{p}(t)), & {\mathrm{if}}\,\,\textbf{p}(t)\not\in\mathcal{Z},\\
\textbf{x}_{\min}(t), & {\mathrm{if}}\,\,\textbf{p}(t)\in\mathcal{Z},\\
\end{array}\right.\label{PSO_3}
\end{eqnarray}
where $k=1,\ldots,q$, $t\in\overline{\mathbb{Z}}_{+}$, $\textbf{v}_{k}(t)\in\mathbb{R}^{n}$ and $\textbf{x}_{k}(t)\in\mathbb{R}^{n}$ are the velocity
and position of particle $k$ at iteration $t$, respectively,
$\textbf{p}(t)\in\mathbb{R}^{n}$ is the position
of the global best value that the swarm of the particles can achieve
so far,
$\eta$, $\mu$, and $\kappa$ are three scalar random coefficients which are usually
selected in uniform distribution in the range $[0,1]$, $\mathcal{Z}=\{\textbf{y}\in\mathbb{R}^{n}:f(\textbf{x}_{\min})<f(\textbf{y})\}$, and $\textbf{x}_{\min}=\arg\min_{1\leq k\leq q}f(\textbf{x}_{k})$. Later in \cite{HZ:TR:2013} we have extended (\ref{PSO_1}) to the dynamic graph case where $\mathcal{N}^{k}$ becomes $\mathcal{N}^{k}(t)=\mathcal{N}_{t}^{k}$. In this sequential report, we further extend (\ref{PSO_1}) to the form with dynamic graph topology sequence $\{\mathcal{G}_{t}\}_{t=0}^{\infty}$ given by
\begin{eqnarray}
\textbf{v}_{k}(t+1)&=&P(t)\textbf{v}_{k}(t)+\eta P(t)\sum_{j\in\mathcal{N}^{k}_{t}}(\textbf{v}_{j}(t)-\textbf{v}_{k}(t))+\mu P(t)\sum_{j\in\mathcal{N}^{k}_{t}}(\textbf{x}_{j}(t)-\textbf{x}_{k}(t))+\kappa P(t)(\textbf{p}(t)-\textbf{x}_{i}(t)),\nonumber\\
\label{PMCO_1}
\end{eqnarray} where $P(t)\in\mathbb{R}^{n\times n}$ is a paracontracting matrix, and $\mathcal{N}^{k}(t)=\mathcal{N}_{t}^{k}$ represents the node-fixed dynamic or time-varying graph topology. Here we use a specific dynamic neighborhood structure called Grouped Directed Structure (GDS) \cite{LH:ACC:2013} to generate a neighboring set sequence $\{\mathcal{N}_{t}^{k}\}_{t=0}^{\infty}$. The reason of using GDS for the neighboring set sequence $\{\mathcal{N}_{t}^{k}\}_{t=0}^{\infty}$ is to prevent all the
particles in paracontracting MCO from being trapped to local optima other than the global
optimum. In this structure, we divide all particles into different groups at every time instant. In each
group, particles have the strongly-connected graphical structure. The information
exchange between the two groups is directed. For example, in Figure
\ref{structure}, 
we divide the 6 particles into two groups, one contains particles 1,2 called ``all-information" group and the
other includes particles 3--6 called ``half information"
group. In each group, the graphical structure is strongly-connected. Particles
1,2 can know the information of all the other particles and
particles 3--6 cannot know the information of particles 1,2. With
this technique, if the information from the particle 1 or 2 is not
desirable then we can limit the information inside of the group of
particles 1,2. Meanwhile, if the information from the particle in
``all-information" group is desirable then it is highly possible to
lead the particles in ``all-information" group to global optima.

The function of introducing $P(t)$ in (\ref{PMCO_1}) is to use contraction mapping to guarantee the convergence of MCO. A natural question arising from (\ref{PSO_2})--(\ref{PMCO_1}) is the following: Can we always guarantee the convergence of (\ref{PSO_2})--(\ref{PMCO_1}) for a given optimization problem $\min_{\textbf{x}\in\mathbb{R}^{n}}f(\textbf{x})$? Here convergence means that all the limits $\lim_{t\to\infty}\textbf{x}_{k}(t)$, $\lim_{t\to\infty}\textbf{v}_{k}(t)$, and $\lim_{t\to\infty}\textbf{p}(t)$ exist for every $k=1,\ldots,q$. This sequential report tries to answer this question by giving some sufficient conditions to guarantee the convergence of (\ref{PSO_2})--(\ref{PMCO_1}). The basic idea borrowing from \cite{SHH:CDC:2011} is to convert the iterative algorithm into a discrete-time switched linear system and then discuss its semistability property. 

\subsection{Parallel Implementation}

Similar to \cite{HZ:TR:2013}, in this section a parallel implementation of the paracontracting MCO algorithm is introduced, which is described as Algorithm \ref{MCO} in the MATLAB language format. The command $\mathtt{matlabpool}$ opens or closes a pool of MATLAB sessions for parallel computation, and enables the parallel language features within the MATLAB  language (e.g., $\mathtt{parfor}$) by starting a parallel job which connects this MATLAB client with a number of labs.

The command $\mathtt{parfor}$ executes code loop in parallel. Part of the $\mathtt{parfor}$ body is executed on the MATLAB client (where the $\mathtt{parfor}$ is issued) and part is executed in parallel on MATLAB workers. The necessary data on which $\mathtt{parfor}$ operates is sent from the client to workers, where most of the computation happens, and the results are sent back to the client and pieced together. In Algorithm \ref{MCO}, the command $\mathtt{parfor}$ is used for loop of the update formula of all particles. Since the update formula needs the neighbors' information, so two temporary variables $C$ and $D$ are introduced for storing the global information of  position and velocity, respectively, $P_{k}$ is a (time-dependent) paracontracting matrix, and $L_{k}$ is the (time-dependent) Laplacian matrix for the communication topology $\mathcal{G}_{k}$ for MCO.

\begin{algorithm}[ht]
\caption{Parallel Paracontracting MCO Algorithm}\label{MCO}
\begin{center}
\small\tt\begin{algorithmic}
\FOR{each agent $i=1,\ldots,q$}
  \STATE Initialize the agent's position with a uniformly distributed random vector: $x_{i}\sim U(\underline{x},\overline{x})\in
\mathbf{R}^{n\times 1}$, where $\underline{x}$ and $\overline{x}$ are the lower and upper boundaries of the search space;
  \STATE Initialize the agent's velocity: $v_{i}\sim U(\underline{v},\overline{v})$, where $\underline{v}$ and $\overline{v} \in
\mathbf{R}^{n\times 1}$ are the lower and upper boundaries of the search speed;
  \STATE Update the agent's best known position to its initial position: $p_{i}\leftarrow x_{i}$;
  \STATE If $f(p_{i})<f(p)$ update the multiagent network's best known position: $p\leftarrow p_i$.
\ENDFOR
  \REPEAT
   \STATE $k \leftarrow k+1$;\\
       \FOR {each agent $i=1,\ldots,q$}
       \STATE $C=[x_1,x_2,\cdots, x_q]^{\rm{T}}$, $D=[v_1,v_2,\cdots, v_q]^{\rm{T}}$;\\
       \STATE $\mathtt{parfor}$ {each agent $i=1,\ldots,q$}
      \STATE Choose random parameters: $\eta\sim U(0,1)$, $\mu\sim U(0,1)$, $\kappa\sim U(0,1)$;
      \STATE Update the agent's velocity: $v_{i}\leftarrow P_{k}v_{i}+\eta P_{k}(L_{k}(i,:)D)^{\rm{T}}+\mu P_{k}(L_{k}(i,:)C)^{\rm{T}}+\kappa P_{k}(p-x_{i})$;
      \STATE  Update the agent's position: $x_{i}\leftarrow x_{i}+v_{i}$;\\
          $\mathtt{endparfor}$\\
    \FOR{$f(x_{i})<f(p_{i})$}
    \STATE Update the agent's best known position: $p_{i}\leftarrow x_{i}$;
    \STATE Update the multiagent network's best known position: $p\leftarrow p+\kappa(p_{i}-p)$;
    \STATE If $f(p_{i})<f(p)$ update the multiagent network's best known position: $p\leftarrow p_{i}$;\\
     \ENDFOR
     \ENDFOR
   \UNTIL{$k$ is large enough or the value of $f$ has small change}
  \RETURN{$p$}
\end{algorithmic}
\end{center}
\end{algorithm}

\section{Convergence analysis}\label{scr}
In this section, we present some theoretic results on global convergence of the iterative process in Algorithm~\ref{MCO}. We follow the steps and key ideas in \cite{HZ:TR:2013}. In particular, we view the randomized paracontracting MCO algorithm as a discrete-time switched linear system and then use semistability theory to rigorously show its global convergence. 
To proceed with presentation, we need the following definition.

\begin{definition}\label{def_odot}
Let $x\in\mathbb{R}^{n}$ be a column vector and $S,K\subseteq\mathbb{R}^{m}$ be subspaces. Define $x\otimes S=\{x\otimes y:y\in S\}$, $x\odot S=\{[x_{1}y_{1}^{\mathrm{T}},\ldots,x_{n}y_{n}^{\mathrm{T}}]^{\mathrm{T}}:[x_{1},\ldots,x_{n}]^{\mathrm{T}}=x,x_{i}\in\mathbb{R},y_{i}\in S,i=1,\ldots,n\}$, and $S+K=\{x+y:x\in S,y\in K\}$.
\end{definition}

The following property about the operation ``$\odot$'' is immediate.

\begin{lemma}\label{lemma_odot}
Let $x=[x_{1},\ldots,x_{n}]^{\mathrm{T}}\in\mathbb{R}^{n}$ and $S$ be a subspace. Then $x\odot S=\sum_{i=1}^{n}x_{i}\textbf{e}_{i}\otimes S$, where $[\textbf{e}_{1},\ldots,\textbf{e}_{n}]=I_{n}$.
\end{lemma}

\begin{IEEEproof}
By definition, $x\odot S=\{[x_{1}y_{1}^{\mathrm{T}},\ldots,x_{n}y_{n}^{\mathrm{T}}]^{\mathrm{T}}:y_{i}\in S,i=1,\ldots,n\}$. On the other hand, $\sum_{i=1}^{n}x_{i}\textbf{e}_{i}\otimes S=\{\sum_{i=1}^{n}x_{i}\textbf{e}_{i}\otimes y_{i}:y_{i}\in S,i=1,\ldots,n\}$. Since $\sum_{i=1}^{n}x_{i}\textbf{e}_{i}\otimes y_{i}=[x_{1}y_{1}^{\mathrm{T}},\ldots,x_{n}y_{n}^{\mathrm{T}}]^{\mathrm{T}}$, it follows that $x\odot S=\sum_{i=1}^{n}x_{i}\textbf{e}_{i}\otimes S$.
\end{IEEEproof}

Next, using the new operations defined in Definition~\ref{def_odot}, we have the following results.

\begin{lemma}\label{lemma_EW}
Let $n,q$ be positive integers and $q\geq 2$. For every $j=1,\ldots,q$, let $E_{n\times nq}^{[j]}\in\mathbb{R}^{n\times nq}$ denote a block-matrix whose $j$th block-column is $I_{n}$ and the rest block-elements are all zero matrices, i.e., $E_{n\times nq}^{[j]}=[\textbf{0}_{n\times n},\ldots,\textbf{0}_{n\times n},I_{n},\textbf{0}_{n\times n},\ldots,\textbf{0}_{n\times n}]$, $j=1,\ldots,q$, where $\textbf{0}_{m\times n}$ denotes the $m\times n$ zero matrix. Define $W^{[j]}=(\textbf{1}_{q\times 1}\otimes P)E_{n\times nq}^{[j]}$ for every $j=1,\ldots,q$, where $\otimes$ denotes the Kronecker product, $P\in\mathbb{R}^{n\times n}$ is a paracontracting matrix, and $\textbf{1}_{m\times n}$ denotes the $m\times n$ matrix whose entries are all ones. Then the following statements hold:
\begin{itemize}
\item[$i$)] For every $j=1,\ldots,q$, ${\mathrm{rank}}(I_{q}\otimes P-W^{[j]})=(q-1){\mathrm{rank}}(P)$, where ${\mathrm{rank}}(A)$ denotes the rank of $A$. 
\item[$ii$)] For any $\textbf{w}=[w_{1},\ldots,w_{q}]^{\mathrm{T}}\in\mathbb{R}^{q}$, $W^{[j]}(\textbf{w}\otimes\textbf{e}_{i})=w_{j}(I_{q}\otimes P)(\textbf{1}_{q\times 1}\otimes\textbf{e}_{i})$ for every $j=1,\ldots,q$ and every $i=1,\ldots,n$. In particular, $W^{[j]}(\textbf{1}_{q\times 1}\otimes\textbf{e}_{i})=(I_{q}\otimes P)(\textbf{1}_{q\times 1}\otimes\textbf{e}_{i})$ and $\ker(W^{[j]}-I_{q}\otimes P)=\textbf{1}_{q\times 1}\otimes{\mathrm{span}}\{\textbf{e}_{1},\ldots,\textbf{e}_{n}\}+(\textbf{1}_{q\times 1}-\textbf{g}_{j})\odot{\mathrm{span}}\{\textbf{j}_{1},\ldots,\textbf{j}_{n-{\mathrm{rank}}(P)}\}$ for every $j=1,\ldots,q$ and every $i=1,\ldots,n$, where $[\textbf{g}_{1},\ldots,\textbf{g}_{q}]=I_{q}$, ${\mathrm{span}}\,S$ denotes the span of a subspace $S$, and ${\mathrm{span}}\{\textbf{j}_{1},\ldots,\textbf{j}_{n-{\mathrm{rank}}(P)}\}=\ker(P)$.
\item[$iii$)] For any $\textbf{w}=[w_{1},\ldots,w_{q}]^{\mathrm{T}}\in\mathbb{R}^{q}$, $E_{n\times nq}^{[j]}(\textbf{w}\otimes\textbf{e}_{i})=w_{j}\textbf{e}_{i}$ for every $j=1,\ldots,q$ and every $i=1,\ldots,n$. In particular, $E_{n\times nq}^{[j]}(\textbf{1}_{q\times 1}\otimes\textbf{e}_{i})=\textbf{e}_{i}$ for every $j=1,\ldots,q$ and every $i=1,\ldots,n$. Next, for any $A\in\mathbb{R}^{n\times n}$, $E_{n\times nq}^{[j]}(\textbf{1}_{q\times 1}\otimes A)=A$, $E_{n\times nq}^{[j]}(\textbf{g}_{s}\otimes\textbf{j}_{r})=\textbf{j}_{r}$ if $s=j$, and $E_{n\times nq}^{[j]}(\textbf{g}_{s}\otimes\textbf{j}_{r})=\textbf{0}_{n\times 1}$ if $s\neq j$ for every $j=1,\ldots,q$, every $s=1,\ldots,q$, and every $r=1,\ldots,n-{\mathrm{rank}}(P)$. Finally, $W^{[j]}(\textbf{g}_{s}\otimes\textbf{j}_{r})=\textbf{0}_{nq\times 1}$ for every $j=1,\ldots,q$, every $s=1,\ldots,q$, and every $r=1,\ldots,n-{\mathrm{rank}}(P)$.
\end{itemize}
\end{lemma}

\begin{IEEEproof}
$i$) First note that by Fact 7.4.3 of \cite[p.~445]{Bernstein:2009}, $W^{[j]}=(\textbf{1}_{q\times 1}\otimes P)E_{n\times nq}^{[j]}=\textbf{1}_{q\times 1}\otimes PE_{n\times nq}^{[j]}$ for every $j=1,\ldots,q$. Now it follows from Fact 7.4.20 of \cite[p.~446]{Bernstein:2009} that 
\begin{eqnarray}\label{Wj2}
&&\hspace{-2.3em}W^{[j]}=\textbf{1}_{q\times 1}\otimes PE_{n\times nq}^{[j]}=(\textbf{1}_{q\times 1}\otimes[\textbf{0}_{n\times n},\ldots,\textbf{0}_{n\times n},P,\textbf{0}_{n\times n},\ldots,\textbf{0}_{n\times n}])\nonumber\\
&&\hspace{0.9em}=[\textbf{1}_{q\times 1}\otimes\textbf{0}_{n\times n},\ldots,\textbf{1}_{q\times 1}\otimes\textbf{0}_{n\times n},\textbf{1}_{q\times 1}\otimes P,\textbf{1}_{q\times 1}\otimes\textbf{0}_{n\times n},\ldots,\textbf{1}_{q\times 1}\otimes\textbf{0}_{n\times n}]\nonumber\\
&&\hspace{0.8em}=\small\left[\begin{array}{ccccccc}
\textbf{0}_{n\times n} & \ldots & \textbf{0}_{n\times n} & P & \textbf{0}_{n\times n} & \ldots & \textbf{0}_{n\times n}\\
\vdots & \ddots & \vdots & \vdots & \vdots & \ddots & \vdots\\
\textbf{0}_{n\times n} & \ldots & \textbf{0}_{n\times n} & P & \textbf{0}_{n\times n} & \ldots & \textbf{0}_{n\times n}\\
\end{array}\right].
\end{eqnarray} 

Next, since $P$ is discrete-time semistable, it follows from \cite{HH:IJC:2009} that $P$ is group invertible \cite[p.~403]{Bernstein:2009}, and hence, $P^{\#}$ exists, where $P^{\#}$ denotes the group generalized inverse of $P$ (see \cite[p.~403]{Bernstein:2009}). Note that it follows from (\ref{Wj2}) that
\begin{eqnarray}
W^{[j]}(I_{q}\otimes P^{\#})(I_{q}\otimes P)&=&\small\left[\begin{array}{ccccccc}
\textbf{0}_{n\times n} & \ldots & \textbf{0}_{n\times n} & P & \textbf{0}_{n\times n} & \ldots & \textbf{0}_{n\times n}\\
\vdots & \ddots & \vdots & \vdots & \vdots & \ddots & \vdots\\
\textbf{0}_{n\times n} & \ldots & \textbf{0}_{n\times n} & P & \textbf{0}_{n\times n} & \ldots & \textbf{0}_{n\times n}\\
\end{array}\right]\small\left[\begin{array}{ccc}
P^{\#} & \ldots & \textbf{0}_{n\times n} \\
\vdots & \ddots & \vdots \\
\textbf{0}_{n\times n} & \ldots & P^{\#} \\
\end{array}\right]\nonumber\\
&&\times\small\left[\begin{array}{ccc}
P & \ldots & \textbf{0}_{n\times n} \\
\vdots & \ddots & \vdots \\
\textbf{0}_{n\times n} & \ldots & P \\
\end{array}\right]\nonumber\\
&=&\small\left[\begin{array}{ccccccc}
\textbf{0}_{n\times n} & \ldots & \textbf{0}_{n\times n} & PP^{\#}P & \textbf{0}_{n\times n} & \ldots & \textbf{0}_{n\times n}\\
\vdots & \ddots & \vdots & \vdots & \vdots & \ddots & \vdots\\
\textbf{0}_{n\times n} & \ldots & \textbf{0}_{n\times n} & PP^{\#}P & \textbf{0}_{n\times n} & \ldots & \textbf{0}_{n\times n}\\
\end{array}\right]\nonumber\\
&=&\small\left[\begin{array}{ccccccc}
\textbf{0}_{n\times n} & \ldots & \textbf{0}_{n\times n} & P & \textbf{0}_{n\times n} & \ldots & \textbf{0}_{n\times n}\\
\vdots & \ddots & \vdots & \vdots & \vdots & \ddots & \vdots\\
\textbf{0}_{n\times n} & \ldots & \textbf{0}_{n\times n} & P & \textbf{0}_{n\times n} & \ldots & \textbf{0}_{n\times n}\\
\end{array}\right]=W^{[j]},\label{WjP1}\\
(I_{q}\otimes P)(I_{q}\otimes P^{\#})W^{[j]}&=&\small\left[\begin{array}{ccc}
P & \ldots & \textbf{0}_{n\times n} \\
\vdots & \ddots & \vdots \\
\textbf{0}_{n\times n} & \ldots & P \\
\end{array}\right]\small\left[\begin{array}{ccc}
P^{\#} & \ldots & \textbf{0}_{n\times n} \\
\vdots & \ddots & \vdots \\
\textbf{0}_{n\times n} & \ldots & P^{\#} \\
\end{array}\right]\nonumber\\
&&\times\small\left[\begin{array}{ccccccc}
\textbf{0}_{n\times n} & \ldots & \textbf{0}_{n\times n} & P & \textbf{0}_{n\times n} & \ldots & \textbf{0}_{n\times n}\\
\vdots & \ddots & \vdots & \vdots & \vdots & \ddots & \vdots\\
\textbf{0}_{n\times n} & \ldots & \textbf{0}_{n\times n} & P & \textbf{0}_{n\times n} & \ldots & \textbf{0}_{n\times n}\\
\end{array}\right]\nonumber\\
&=&\small\left[\begin{array}{ccccccc}
\textbf{0}_{n\times n} & \ldots & \textbf{0}_{n\times n} & PP^{\#}P & \textbf{0}_{n\times n} & \ldots & \textbf{0}_{n\times n}\\
\vdots & \ddots & \vdots & \vdots & \vdots & \ddots & \vdots\\
\textbf{0}_{n\times n} & \ldots & \textbf{0}_{n\times n} & PP^{\#}P & \textbf{0}_{n\times n} & \ldots & \textbf{0}_{n\times n}\\
\end{array}\right]\nonumber\\
&=&\small\left[\begin{array}{ccccccc}
\textbf{0}_{n\times n} & \ldots & \textbf{0}_{n\times n} & P & \textbf{0}_{n\times n} & \ldots & \textbf{0}_{n\times n}\\
\vdots & \ddots & \vdots & \vdots & \vdots & \ddots & \vdots\\
\textbf{0}_{n\times n} & \ldots & \textbf{0}_{n\times n} & P & \textbf{0}_{n\times n} & \ldots & \textbf{0}_{n\times n}\\
\end{array}\right]=W^{[j]},\label{WjP2}
\end{eqnarray} where we used the fact that $PP^{\#}P=P$ (see (6.2.11) in \cite[p.~403]{Bernstein:2009}). Let $M=(I_{q}\otimes P^{\#})W^{[j]}(I_{q}\otimes P^{\#})$. Then it follows from (\ref{WjP1}) and (\ref{WjP2}) that $(I_{q}\otimes P)M(I_{q}\otimes P)=(I_{q}\otimes P)(I_{q}\otimes P^{\#})W^{[j]}(I_{q}\otimes P^{\#})(I_{q}\otimes P)=W^{[j]}$. Furthermore, it follows from (\ref{WjP1}) or (\ref{WjP2}) that
\begin{eqnarray}
M(I_{q}\otimes P)M&=&(I_{q}\otimes P^{\#})W^{[j]}(I_{q}\otimes P^{\#})(I_{q}\otimes P)(I_{q}\otimes P^{\#})W^{[j]}(I_{q}\otimes P^{\#})\nonumber\\
&=&(I_{q}\otimes P^{\#})W^{[j]}(I_{q}\otimes P^{\#})W^{[j]}(I_{q}\otimes P^{\#})\nonumber\\
&=&(I_{q}\otimes P^{\#})\small\left[\begin{array}{ccccccc}
\textbf{0}_{n\times n} & \ldots & \textbf{0}_{n\times n} & P & \textbf{0}_{n\times n} & \ldots & \textbf{0}_{n\times n}\\
\vdots & \ddots & \vdots & \vdots & \vdots & \ddots & \vdots\\
\textbf{0}_{n\times n} & \ldots & \textbf{0}_{n\times n} & P & \textbf{0}_{n\times n} & \ldots & \textbf{0}_{n\times n}\\
\end{array}\right]\small\left[\begin{array}{ccc}
P^{\#} & \ldots & \textbf{0}_{n\times n} \\
\vdots & \ddots & \vdots \\
\textbf{0}_{n\times n} & \ldots & P^{\#} \\
\end{array}\right]\nonumber\\
&&\times\small\left[\begin{array}{ccccccc}
\textbf{0}_{n\times n} & \ldots & \textbf{0}_{n\times n} & P & \textbf{0}_{n\times n} & \ldots & \textbf{0}_{n\times n}\\
\vdots & \ddots & \vdots & \vdots & \vdots & \ddots & \vdots\\
\textbf{0}_{n\times n} & \ldots & \textbf{0}_{n\times n} & P & \textbf{0}_{n\times n} & \ldots & \textbf{0}_{n\times n}\\
\end{array}\right](I_{q}\otimes P^{\#})\nonumber\\
&=&(I_{q}\otimes P^{\#})\small\left[\begin{array}{ccccccc}
\textbf{0}_{n\times n} & \ldots & \textbf{0}_{n\times n} & PP^{\#}P & \textbf{0}_{n\times n} & \ldots & \textbf{0}_{n\times n}\\
\vdots & \ddots & \vdots & \vdots & \vdots & \ddots & \vdots\\
\textbf{0}_{n\times n} & \ldots & \textbf{0}_{n\times n} & PP^{\#}P & \textbf{0}_{n\times n} & \ldots & \textbf{0}_{n\times n}\\
\end{array}\right](I_{q}\otimes P^{\#})\nonumber\\
&=&(I_{q}\otimes P^{\#})\small\left[\begin{array}{ccccccc}
\textbf{0}_{n\times n} & \ldots & \textbf{0}_{n\times n} & P & \textbf{0}_{n\times n} & \ldots & \textbf{0}_{n\times n}\\
\vdots & \ddots & \vdots & \vdots & \vdots & \ddots & \vdots\\
\textbf{0}_{n\times n} & \ldots & \textbf{0}_{n\times n} & P & \textbf{0}_{n\times n} & \ldots & \textbf{0}_{n\times n}\\
\end{array}\right](I_{q}\otimes P^{\#})\nonumber\\
&=&(I_{q}\otimes P^{\#})W^{[j]}(I_{q}\otimes P^{\#})=M.
\end{eqnarray} Now it follows from Fact 2.10.30 of \cite[p.~128]{Bernstein:2009} that ${\mathrm{rank}}(I_{q}\otimes P-W^{[j]})={\mathrm{rank}}(I_{q}\otimes P)-{\mathrm{rank}}(W^{[j]})$.  Clearly it follows from (\ref{Wj2}) that ${\mathrm{rank}}(W^{[j]})={\mathrm{rank}}(P)$. 
Thus, ${\mathrm{rank}}(I_{q}\otimes P-W^{[j]})={\mathrm{rank}}(I_{q}\otimes P)-{\mathrm{rank}}(W^{[j]})=q\times{\mathrm{rank}}(P)-{\mathrm{rank}}(P)=(q-1){\mathrm{rank}}(P)$ for every $j=1,\ldots,q$. 

$ii$) It follows from (\ref{Wj2}) that for every $j=1,\ldots,q$ and every $i=1,\ldots,n$,
\begin{eqnarray*}
W^{[j]}(\textbf{1}_{q\times 1}\otimes \textbf{e}_{i})&=&\small\left[\begin{array}{ccccccc}
\textbf{0}_{n\times n} & \ldots & \textbf{0}_{n\times n} & P & \textbf{0}_{n\times n} & \ldots & \textbf{0}_{n\times n}\\
\vdots & \ddots & \vdots & \vdots & \vdots & \ddots & \vdots\\
\textbf{0}_{n\times n} & \ldots & \textbf{0}_{n\times n} & P & \textbf{0}_{n\times n} & \ldots & \textbf{0}_{n\times n}\\
\end{array}\right]\small\left[\begin{array}{c}
\textbf{e}_{i}\\
\vdots \\
\textbf{e}_{i}\\
\end{array}\right]=\small\left[\begin{array}{c}
P\textbf{e}_{i}\\
\vdots \\
P\textbf{e}_{i}\\
\end{array}\right]=\textbf{1}_{q\times 1}\otimes P\textbf{e}_{i}\nonumber\\
&=&(I_{q}\otimes P)(\textbf{1}_{q\times 1}\otimes \textbf{e}_{i}),
\end{eqnarray*} namely, $(W^{[j]}-I_{q}\otimes P)(\textbf{1}_{q\times 1}\otimes \textbf{e}_{i})=\textbf{0}_{nq\times 1}$ for every $j=1,\ldots,q$. Since by $i$), ${\mathrm{rank}}(W^{[j]}-I_{q}\otimes P)=(q-1){\mathrm{rank}}(P)$ for every $j=1,\ldots,q$, it follows from Corollary 2.5.5 of \cite[p.~105]{Bernstein:2009} that ${\mathrm{def}}(W^{[j]}-I_{q}\otimes P)=nq-{\mathrm{rank}}(W^{[j]}-I_{q}\otimes P)=nq-(q-1){\mathrm{rank}}(P)\geq n$ for every $j=1,\ldots,q$, where ${\mathrm{def}}(A)$ denotes the defect of $A$. Note that $\textbf{1}_{q\times 1}\otimes\textbf{e}_{i}$, $i=1,\ldots,n$, are linearly independent, it follows that $\textbf{1}_{q\times 1}\otimes{\mathrm{span}}\{\textbf{e}_{1},\ldots,\textbf{e}_{n}\}={\mathrm{span}}\{\textbf{1}_{q\times 1}\otimes\textbf{e}_{1},\ldots,\textbf{1}_{q\times 1}\otimes\textbf{e}_{n}\}\subseteq\ker(W^{[j]}-I_{q}\otimes P)$ for every $j=1,\ldots,q$.

Let $x=[x_{1}^{\mathrm{T}},\ldots,x_{q}^{\mathrm{T}}]^{\mathrm{T}}\in\ker(W^{[j]}-I_{q}\otimes P)$, where $x_{i}\in\mathbb{R}^{n}$, $i=1,\ldots,q$. Then it follows that $Px_{j}-Px_{i}=0$ for every $i=1,\ldots,q$, i.e., $x_{i}-x_{j}\in\ker(P)$, $i\neq j$, $i=1,\ldots,q$, where $x_{j}\in\mathbb{R}^{n}$ is arbitrary. Note that $x_{j}\in{\mathrm{span}}\{\textbf{e}_{1},\ldots,\textbf{e}_{n}\}$. Hence, $\textbf{1}_{q\times 1}\otimes{\mathrm{span}}\{\textbf{e}_{1},\ldots,\textbf{e}_{n}\}+(\textbf{1}_{q\times 1}-\textbf{g}_{j})\odot\ker(P)=\ker(W^{[j]}-I_{q}\otimes P)$.

Finally, for any $\textbf{w}=[w_{1},\ldots,w_{q}]^{\mathrm{T}}\in\mathbb{R}^{q}$, it follows from (\ref{Wj2}) that
\begin{eqnarray*}
W^{[j]}(\textbf{w}\otimes\textbf{e}_{i})&=&\small\left[\begin{array}{ccccccc}
\textbf{0}_{n\times n} & \ldots & \textbf{0}_{n\times n} & P & \textbf{0}_{n\times n} & \ldots & \textbf{0}_{n\times n}\\
\vdots & \ddots & \vdots & \vdots & \vdots & \ddots & \vdots\\
\textbf{0}_{n\times n} & \ldots & \textbf{0}_{n\times n} & P & \textbf{0}_{n\times n} & \ldots & \textbf{0}_{n\times n}\\
\end{array}\right]\small\left[\begin{array}{c}
w_{1}\textbf{e}_{i}\\
\vdots \\
w_{q}\textbf{e}_{i}\\
\end{array}\right]=\small\left[\begin{array}{c}
w_{j}P\textbf{e}_{i}\\
\vdots \\
w_{j}P\textbf{e}_{i}\\
\end{array}\right]\nonumber\\
&=&w_{j}\textbf{1}_{q\times 1}\otimes P\textbf{e}_{i}=w_{j}(I_{q}\otimes P)(\textbf{1}_{q\times 1}\otimes\textbf{e}_{i})
\end{eqnarray*} for every $j=1,\ldots,q$ and every $i=1,\ldots,n$.

$iii$) For any $\textbf{w}=[w_{1},\ldots,w_{q}]^{\mathrm{T}}\in\mathbb{R}^{q}$, $E_{n\times nq}^{[j]}(\textbf{w}\otimes\textbf{e}_{i})=[\textbf{0}_{n\times n},\ldots,\textbf{0}_{n\times n},I_{n},\textbf{0}_{n\times n},\ldots,\textbf{0}_{n\times n}][w_{1}\textbf{e}_{i}^{\mathrm{T}},\ldots,\\w_{q}\textbf{e}_{i}^{\mathrm{T}}]^{\mathrm{T}}=w_{j}\textbf{e}_{i}$ for every $j=1,\ldots,q$ and every $i=1,\ldots,n$. In particular, $E_{n\times nq}^{[j]}(\textbf{1}_{q\times 1}\otimes\textbf{e}_{i})=\textbf{e}_{i}$ for every $j=1,\ldots,q$ and every $i=1,\ldots,n$. Next, for every $j=1,\ldots,q$, $E_{n\times nq}^{[j]}(\textbf{1}_{q\times 1}\otimes A)=[\textbf{0}_{n\times n},\ldots,\textbf{0}_{n\times n},I_{n},\textbf{0}_{n\times n},\ldots,\\\textbf{0}_{n\times n}][A^{\mathrm{T}},\ldots,A^{\mathrm{T}}]^{\mathrm{T}}=A$. For every $j=1,\ldots,q$, every $s=1,\ldots,q$, and every $r=1,\ldots,n-{\mathrm{rank}}(P)$,
$E_{n\times nq}^{[j]}(\textbf{g}_{s}\otimes\textbf{j}_{r})=[\textbf{0}_{n\times n},\ldots,\textbf{0}_{n\times n},I_{n},\textbf{0}_{n\times n},\ldots,\textbf{0}_{n\times n}][\textbf{0}_{1\times q},\ldots,\textbf{j}_{r}^{\mathrm{T}},\ldots,\textbf{0}_{1\times q}]^{\mathrm{T}}=\textbf{j}_{r}$ if $s=j$. Otherwise, $E_{n\times nq}^{[j]}(\textbf{g}_{s}\otimes\textbf{j}_{r})=\textbf{0}_{n\times 1}$. Finally, $W^{[j]}(\textbf{g}_{s}\otimes\textbf{j}_{r})=(\textbf{1}_{q\times 1}\otimes P)E_{n\times nq}^{[j]}(\textbf{g}_{s}\otimes\textbf{j}_{r})=(\textbf{1}_{q\times 1}\otimes P)\textbf{j}_{r}=\textbf{1}_{q\times 1}\otimes P\textbf{j}_{r}=\textbf{0}_{nq\times 1}$ if $s=j$ and $W^{[j]}(\textbf{g}_{s}\otimes\textbf{j}_{r})=(\textbf{1}_{q\times 1}\otimes P)E_{n\times nq}^{[j]}(\textbf{g}_{s}\otimes\textbf{j}_{r})=(\textbf{1}_{q\times 1}\otimes P)\textbf{0}_{n\times 1}=\textbf{0}_{nq\times 1}$ if $s\neq j$.
\end{IEEEproof}

The following two lemmas are needed for the next result.

\begin{lemma}\label{lemma_ker}
Let $A\in\mathbb{R}^{n\times m}$ and $B\in\mathbb{R}^{l\times k}$. Then $\ker(A\otimes B)=\ker(A\otimes I_{l})+\ker(I_{n}\otimes B)$.
\end{lemma}

\begin{IEEEproof}
It follows from Equality (2.4.13) of \cite[p.~103]{Bernstein:2009} and Equality (7.1.7) of \cite[p.~440]{Bernstein:2009} that $\ker(A\otimes B)={\mathrm{ran}}((A\otimes B)^{\mathrm{T}})^{\bot}={\mathrm{ran}}(A^{\mathrm{T}}\otimes B^{\mathrm{T}})^{\bot}$, where ${\mathrm{ran}}(A)$ denotes the range space of $A$ and $S^{\bot}$ denotes the orthogonal complement of $S$. On the other hand, it follows from Fact 7.4.23 of \cite[p.~447]{Bernstein:2009} that ${\mathrm{ran}}(A^{\mathrm{T}}\otimes B^{\mathrm{T}})={\mathrm{ran}}(A^{\mathrm{T}}\otimes I_{l})\cap{\mathrm{ran}}(I_{n}\otimes B^{\mathrm{T}})$. Now it follows from Fact 2.9.16 of \cite[p.~121]{Bernstein:2009} that ${\mathrm{ran}}(A^{\mathrm{T}}\otimes B^{\mathrm{T}})^{\bot}=({\mathrm{ran}}(A^{\mathrm{T}}\otimes I_{l})\cap{\mathrm{ran}}(I_{n}\otimes B^{\mathrm{T}}))^{\bot}={\mathrm{ran}}(A^{\mathrm{T}}\otimes I_{l})^{\bot}+{\mathrm{ran}}(I_{n}\otimes B^{\mathrm{T}})^{\bot}$. Finally, it follows from Equality (2.4.13) of \cite[p.~103]{Bernstein:2009} that $\ker(A\otimes B)={\mathrm{ran}}(A^{\mathrm{T}}\otimes I_{l})^{\bot}+{\mathrm{ran}}(I_{n}\otimes B^{\mathrm{T}})^{\bot}=\ker(A\otimes I_{l})+\ker(I_{n}\otimes B)$.
\end{IEEEproof}

\begin{lemma}\label{lemma_S}
Let $S_{i}$, $i=1,2,3$, be subspaces such that $S_{1}\cup S_{2}$ or $S_{2}\cup S_{3}$ or $S_{3}\cup S_{1}$ is a subspace. Then $\dim(S_{1}+S_{2}+S_{3})=\dim S_{1}+\dim S_{2}+\dim S_{3}-\dim(S_{1}\cap S_{2})-\dim(S_{2}\cap S_{3})-\dim(S_{3}\cap S_{1})+\dim(S_{1}\cap S_{2}\cap S_{3})$, where $\dim S$ denotes the dimension of a subspace $S$.
\end{lemma}

\begin{IEEEproof}
Here we just consider the case where $S_{1}\cup S_{2}$ is a subspace. It follows from the subspace dimension theorem (Theorem 2.3.1 of \cite[p.~98]{Bernstein:2009}) that $\dim(S_{1}+S_{2}+S_{3})=\dim(S_{1}+S_{2})+\dim S_{3}-\dim[(S_{1}+S_{2})\cap S_{3}]=\dim S_{1}+\dim S_{2}-\dim(S_{1}\cap S_{2})+\dim S_{3}-\dim[(S_{1}+S_{2})\cap S_{3}]$. Since by assumption $S_{1}\cup S_{2}$ is a subspace, it follows from Fact 2.9.11 of \cite[p.~121]{Bernstein:2009} that $S_{1}+S_{2}=S_{1}\cup S_{2}$. Hence, $(S_{1}+S_{2})\cap S_{3}=(S_{1}\cup S_{2})\cap S_{3}=(S_{1}\cap S_{3})\cup(S_{2}\cap S_{3})$. On the other hand, note that $(S_{1}+S_{2})\cap S_{3}$ is a subspace, and hence, $(S_{1}\cap S_{3})\cup(S_{2}\cap S_{3})$ is a subspace as well. Thus, by Fact 2.9.11 of \cite[p.~121]{Bernstein:2009}, $(S_{1}\cap S_{3})\cup(S_{2}\cap S_{3})=S_{1}\cap S_{3}+S_{2}\cap S_{3}$. Then it follows from the subspace dimension theorem that  $\dim[(S_{1}+S_{2})\cap S_{3}]=\dim(S_{1}\cap S_{3}+S_{2}\cap S_{3})=\dim(S_{1}\cap S_{3})+\dim(S_{2}\cap S_{3})-\dim(S_{1}\cap S_{2}\cap S_{3})$. Consequently, $\dim(S_{1}+S_{2}+S_{3})=\dim S_{1}+\dim S_{2}+\dim S_{3}-\dim(S_{1}\cap S_{2})-\dim(S_{2}\cap S_{3})-\dim(S_{3}\cap S_{1})+\dim(S_{1}\cap S_{2}\cap S_{3})$.
\end{IEEEproof}

Next, we use some graph notions to state a result on the rank of certain matrices related to the matrix form of the iterative process in Algorithm~\ref{MCO}. 

\begin{lemma}\label{lemma_Arank}
Define a (possibly infinite) series of matrices $A^{[j]}_{k}$, $j=1,\ldots,q$, $k=0,1,2,\ldots$, as follows:
\begin{eqnarray}\label{Amatrix}
A_{k}^{[j]}=\small\left[\begin{array}{ccc}
\textbf{0}_{nq\times nq} & I_{nq} & \textbf{0}_{nq\times n} \\
-\mu_{k} L_{k}\otimes P_{k}-\kappa_{k} (I_{q}\otimes P_{k}) & -\eta_{k} L_{k}\otimes P_{k} & \kappa_{k} \textbf{1}_{q\times 1}\otimes P_{k} \\
\kappa_{k} E_{n\times nq}^{[j]} & \textbf{0}_{n\times nq} & -\kappa_{k} I_{n} \\
\end{array}\right],
\end{eqnarray} where $\mu_{k},\eta_{k},\kappa_{k}\geq0$, $k\in\overline{\mathbb{Z}}_{+}$, $P_{k}\in\mathbb{R}^{n\times n}$ denotes a paracontracting matrix, $L_{k}\in\mathbb{R}^{q\times q}$ denotes the Laplacian matrix of a node-fixed dynamic digraph $\mathcal{G}_{k}$, and $E_{n\times nq}^{[j]}\in\mathbb{R}^{n\times nq}$ is defined in Lemma~\ref{lemma_EW}. 
\begin{itemize}
\item[$i$)] If $\mu_{k}=0$ and $\kappa_{k}=0$, then ${\mathrm{rank}}(A_{k}^{[j]})=nq$ and $\ker(A_{k}^{[j]})=\{[\sum_{i=1}^{n}\sum_{l=1}^{q}\alpha_{il}(\textbf{e}_{i}\otimes\textbf{g}_{l})^{\mathrm{T}},\textbf{0}_{1\times nq},\sum_{i=1}^{n}\beta_{i}\\\textbf{e}_{i}^{\mathrm{T}}]^{\mathrm{T}}:\forall\alpha_{il}\in\mathbb{R},\forall\beta_{i}\in\mathbb{R},i=1,\ldots,n,l=1,\ldots,q\}$ for every $j=1,\ldots,q$, $k\in\overline{\mathbb{Z}}_{+}$.
\item[$ii$)] If $\mu_{k}=0$ and $\kappa_{k}\neq0$, then ${\mathrm{rank}}(A_{k}^{[j]})=2nq-(q-1)(n-{\mathrm{rank}}(P_{k}))$ and $\ker(A_{k}^{[j]})=\{[\sum_{i=1}^{n}\alpha_{i}(\textbf{1}_{q\times 1}\otimes\textbf{e}_{i})^{\mathrm{T}}+\sum_{s=1}^{q}\sum_{r=1}^{n-{\mathrm{rank}}(P_{k})}\beta_{sr}(\textbf{g}_{s}\otimes\textbf{j}_{r})^{\mathrm{T}}-\sum_{r=1}^{n-{\mathrm{rank}}(P_{k})}\beta_{jr}(\textbf{g}_{j}\otimes\textbf{j}_{r})^{\mathrm{T}},\textbf{0}_{1\times nq},\sum_{i=1}^{n}\alpha_{i}\textbf{e}_{i}^{\mathrm{T}}]^{\mathrm{T}}:\forall\alpha_{i},\beta_{sr}\in\mathbb{R},i=1,\ldots,n,s=1,\ldots,q,r=1,\ldots,n-{\mathrm{rank}}(P_{k})\}$ for every $j=1,\ldots,q$, $k\in\overline{\mathbb{Z}}_{+}$.
\item[$iii$)] If $\mu_{k}\neq0$ and $\kappa_{k}\neq0$, then ${\mathrm{rank}}(A_{k}^{[j]})=2nq-(q-1)(n-{\mathrm{rank}}(P_{k}))$ and $\ker(A_{k}^{[j]})=\{[\sum_{i=1}^{n}\alpha_{0i}(\textbf{w}_{0}\otimes\textbf{e}_{i})^{\mathrm{T}}+\sum_{l=1}^{q-1-{\mathrm{rank}}(L_{k})}\sum_{i=1}^{n}\sum_{m=1}^{n}\gamma_{lm}(\textbf{e}_{i}^{\mathrm{T}}(I_{n}-P_{k}^{+}P_{k})\textbf{e}_{m})(\textbf{w}_{l}\otimes\textbf{e}_{i})^{\mathrm{T}}+\sum_{s=1}^{q}\sum_{r=1}^{n-{\mathrm{rank}}(P_{k})}\beta_{sr}(\textbf{g}_{s}\otimes\textbf{j}_{r})^{\mathrm{T}},\\\textbf{0}_{1\times nq},\sum_{i=1}^{n}\alpha_{0i}\textbf{e}_{i}^{\mathrm{T}}+\sum_{l=1}^{q-1-{\mathrm{rank}}(L_{k})}\sum_{i=1}^{n}\sum_{m=1}^{n}\gamma_{lm}(\textbf{e}_{i}^{\mathrm{T}}(I_{n}-P_{k}^{+}P_{k})\textbf{e}_{m})w_{lj}\textbf{e}_{i}^{\mathrm{T}}+\sum_{r=1}^{n-{\mathrm{rank}}(P_{k})}\beta_{jr}\textbf{j}_{r}^{\mathrm{T}}]^{\mathrm{T}}:\forall\alpha_{0i},\beta_{sr},\gamma_{lm}\in\mathbb{R},i=1,\ldots,n,s=1,\ldots,q,r=1,\ldots,n-{\mathrm{rank}}(P_{k}),l=1,\ldots,q-1-{\mathrm{rank}}(L_{k}),m=1,\ldots,n\}$ for every $j=1,\ldots,q$, $k\in\overline{\mathbb{Z}}_{+}$, where $A^{+}$ denotes the Moore-Penrose generalized inverse of $A$, ${\mathrm{span}}\{\textbf{w}_{0},\textbf{w}_{1},\ldots,\textbf{w}_{q-1-{\mathrm{rank}}(L_{k})}\}=\ker(L_{k})$, $\textbf{w}_{0}=\textbf{1}_{q\times 1}$, and $\textbf{w}_{l}=[w_{l1},\ldots,w_{lq}]^{\mathrm{T}}\in\mathbb{R}^{q}$ for every $l=1,\ldots,q-1-{\mathrm{rank}}(L_{k})$.
\item[$iv$)] If $\mu_{k}\neq0$ and $\kappa_{k}=0$, then ${\mathrm{rank}}(A_{k}^{[j]})=nq+{\mathrm{rank}}(L_{k}){\mathrm{rank}}(P_{k})$ and $\ker(A_{k}^{[j]})=\{[\sum_{l=0}^{q-1-{\mathrm{rank}}(L_{k})}\sum_{i=1}^{n}\\\alpha_{li}(\textbf{w}_{l}\otimes\textbf{e}_{i})^{\mathrm{T}}+\sum_{s=1}^{q}\sum_{r=1}^{n-{\mathrm{rank}}(P_{k})}\beta_{sr}(\textbf{g}_{s}\otimes\textbf{j}_{r})^{\mathrm{T}},\textbf{0}_{1\times nq},\sum_{i=1}^{n}\gamma_{i}\textbf{e}_{i}^{\mathrm{T}}]^{\mathrm{T}}:\forall\alpha_{li},\beta_{sr},\gamma_{i}\in\mathbb{R},i=1,\ldots,n,l=0,1,\ldots,q-1-{\mathrm{rank}}(L_{k}),s=1,\ldots,q,r=1,\ldots,n-{\mathrm{rank}}(P_{k})\}$ for every $j=1,\ldots,q$, $k\in\overline{\mathbb{Z}}_{+}$.
\end{itemize}
\end{lemma}

\begin{IEEEproof}
First, it follows from (\ref{Amatrix}) that $\ker(A_{k}^{[j]})=\{[\textbf{z}_{1}^{\rm{T}},\textbf{z}_{2}^{\rm{T}},\textbf{z}_{3}^{\rm{T}}]^{\rm{T}}\in\mathbb{R}^{2nq+n}:\textbf{z}_{2}=\textbf{0}_{nq\times 1},-\mu_{k} (L_{k}\otimes P_{k})\textbf{z}_{1}-\kappa_{k}(I_{q}\otimes P_{k})\textbf{z}_{1}-\eta_{k}(L_{k}\otimes P_{k})\textbf{z}_{2}+\kappa_{k}(\textbf{1}_{q\times 1}\otimes P_{k})\textbf{z}_{3}=\textbf{0}_{nq\times 1}, \kappa_{k}E_{n\times nq}^{[j]}\textbf{z}_{1}-\kappa_{k}\textbf{z}_{3}=\textbf{0}_{n\times 1}\}$, $k\in\overline{\mathbb{Z}}_{+}$, where $\textbf{z}_{1},\textbf{z}_{2}\in\mathbb{R}^{nq}$ and $\textbf{z}_{3}\in\mathbb{R}^{n}$. 

$i$) If $\mu_{k}=0$ and $\kappa_{k}=0$, then it follows from the similar arguments as in the proof of $i$) of Lemma 4.2 of \cite{HZ:TR:2013} that the assertion holds. 

$ii$) If $\mu_{k}=0$ and $\kappa_{k}\neq0$, then 
substituting $\textbf{z}_{2}=\textbf{0}_{nq\times 1}$ and $\textbf{z}_{3}=E_{n\times nq}^{[j]}\textbf{z}_{1}$ into $-\kappa_{k}(I_{q}\otimes P_{k})\textbf{z}_{1}-\eta_{k}(L_{k}\otimes P_{k})\textbf{z}_{2}+\kappa_{k}(\textbf{1}_{q\times 1}\otimes P_{k})\textbf{z}_{3}=\textbf{0}_{nq\times 1}$ yields 
\begin{eqnarray}\label{z1p}
\kappa_{k}(W_{k}^{[j]}-I_{q}\otimes P_{k})\textbf{z}_{1}=\textbf{0}_{nq\times 1},
\end{eqnarray}
where $W_{k}^{[j]}=(\textbf{1}_{q\times 1}\otimes P_{k})E_{n\times nq}^{[j]}$. Since, by $ii$) of Lemma~\ref{lemma_EW}, $\ker(W_{k}^{[j]}-I_{q}\otimes P_{k})=\textbf{1}_{q\times 1}\otimes{\mathrm{span}}\{\textbf{e}_{1},\ldots,\textbf{e}_{n}\}+(\textbf{1}_{q\times 1}-\textbf{g}_{j})\odot\ker(P_{k})$ for every $j=1,\ldots,q$, it follows from (\ref{z1p}) and Lemma~\ref{lemma_odot} that $\textbf{z}_{1}$ can be represented as $\textbf{z}_{1}=\sum_{i=1}^{n}\alpha_{i}\textbf{1}_{q\times 1}\otimes\textbf{e}_{i}+\sum_{s=1}^{q}\sum_{r=1}^{n-{\mathrm{rank}}(P_{k})}\beta_{sr}\textbf{g}_{s}\otimes\textbf{j}_{r}-\sum_{r=1}^{n-{\mathrm{rank}}(P_{k})}\beta_{jr}\textbf{g}_{j}\otimes\textbf{j}_{r}$, where $\alpha_{i},\beta_{sr}\in\mathbb{R}$. Furthermore, it follows from $iii$) of Lemma 4.1 of \cite{HZ:TR:2013} and $iii$) of Lemma~\ref{lemma_EW} that $\textbf{z}_{3}=E_{n\times nq}^{[j]}\textbf{z}_{1}=\sum_{i=1}^{n}\alpha_{i}E_{n\times nq}^{[j]}(\textbf{1}_{q\times 1}\otimes\textbf{e}_{i})+\sum_{s=1}^{q}\sum_{r=1}^{n-{\mathrm{rank}}(P_{k})}\beta_{sr}E_{n\times nq}^{[j]}(\textbf{g}_{s}\otimes\textbf{j}_{r})-\sum_{r=1}^{n-{\mathrm{rank}}(P_{k})}\beta_{jr}E_{n\times nq}^{[j]}(\textbf{g}_{j}\otimes\textbf{j}_{r})=\sum_{i=1}^{n}\alpha_{i}\textbf{e}_{i}$ for every $j=1,\ldots,q$. Thus, $\ker(A_{k}^{[j]})=\{[\sum_{i=1}^{n}\alpha_{i}(\textbf{1}_{q\times 1}\otimes\textbf{e}_{i})^{\mathrm{T}}+\sum_{s=1}^{q}\sum_{r=1}^{n-{\mathrm{rank}}(P_{k})}\beta_{sr}(\textbf{g}_{s}\otimes\textbf{j}_{r})^{\mathrm{T}}-\sum_{r=1}^{n-{\mathrm{rank}}(P_{k})}\beta_{jr}(\textbf{g}_{j}\otimes\textbf{j}_{r})^{\mathrm{T}},\textbf{0}_{1\times nq},\sum_{i=1}^{n}\alpha_{i}\textbf{e}_{i}^{\mathrm{T}}]^{\mathrm{T}}:\forall\alpha_{i},\beta_{sr}\in\mathbb{R},i=1,\ldots,n,s=1,\ldots,q,r=1,\ldots,n-{\mathrm{rank}}(P_{k})\}$ for every $j=1,\ldots,q$, $k\in\overline{\mathbb{Z}}_{+}$. 

Let $\mathcal{S}_{1}=\{[\sum_{i=1}^{n}\alpha_{i}(\textbf{1}_{q\times 1}\otimes\textbf{e}_{i})^{\mathrm{T}},\textbf{0}_{1\times nq},\sum_{i=1}^{n}\alpha_{i}\textbf{e}_{i}^{\mathrm{T}}]^{\mathrm{T}}:\forall\alpha_{i}\in\mathbb{R},i=1,\ldots,n\}$ and $\mathcal{S}_{2}=\{[\sum_{s=1}^{q}\sum_{r=1}^{n-{\mathrm{rank}}(P_{k})}\\\beta_{sr}(\textbf{g}_{s}\otimes\textbf{j}_{r})^{\mathrm{T}}-\sum_{r=1}^{n-{\mathrm{rank}}(P_{k})}\beta_{jr}(\textbf{g}_{j}\otimes\textbf{j}_{r})^{\mathrm{T}},\textbf{0}_{1\times nq},\textbf{0}_{1\times n}]^{\mathrm{T}}:\forall\beta_{sr}\in\mathbb{R},s=1,\ldots,q,r=1,\ldots,n-{\mathrm{rank}}(P_{k})\}$. Clearly $\ker(A_{k}^{[j]})=\mathcal{S}_{1}+\mathcal{S}_{2}$ and $\mathcal{S}_{1}$ and $\mathcal{S}_{2}$ are subspaces. Now it follows from the subspace dimension theorem (Theorem 2.3.1 of \cite[p.~98]{Bernstein:2009}) that $\dim\ker(A_{k}^{[j]})=\dim\mathcal{S}_{1}+\dim\mathcal{S}_{2}-\dim(\mathcal{S}_{1}\cap\mathcal{S}_{2})=n+(q-1)(n-{\mathrm{rank}}(P_{k}))-\dim(\mathcal{S}_{1}\cap\mathcal{S}_{2})$. Since $\mathcal{S}_{1}\cap\mathcal{S}_{2}=\{\textbf{0}_{(2nq+n)\times 1}\}$, it follows that $\dim(\mathcal{S}_{1}\cap\mathcal{S}_{2})=0$, which implies that ${\mathrm{def}}(A_{k}^{[j]})=\dim\ker(A_{k}^{[j]})=nq-(q-1){\mathrm{rank}}(P_{k})$ for every $j=1,\ldots,q$, $k\in\overline{\mathbb{Z}}_{+}$. Therefore, in this case ${\mathrm{rank}}(A_{k}^{[j]})=2nq+n-{\mathrm{def}}(A_{k}^{[j]})=nq+n+(q-1){\mathrm{rank}}(P_{k})=2nq-(q-1)(n-{\mathrm{rank}}(P_{k}))$.

$iii$) If $\mu_{k}\neq0$ and $\kappa_{k}\neq0$, then we claim that $\kappa_{k}/\mu_{k}\not\in{\mathrm{spec}}(-L_{k}\otimes I_{n})$. To see this, it follows from Proposition 1 of \cite{AC:LAA:2005} that for any $\lambda_{k}\in{\mathrm{spec}}(-L_{k})$, ${\mathrm{Re}}\,\lambda_{k}\leq0$, where ${\mathrm{Re}}\,\lambda_{k}$ denotes the real part of $\lambda_{k}$. Furthermore, note that ${\mathrm{spec}}(-L_{k}\otimes I_{n})={\mathrm{spec}}(-L_{k})$. Thus, if $\kappa_{k}\neq0$, then $0<\kappa_{k}/\mu_{k}\not\in{\mathrm{spec}}(-L_{k})={\mathrm{spec}}(-L_{k}\otimes I_{n})$. 
Now, substituting $\textbf{z}_{2}=\textbf{0}_{nq\times 1}$ and $\textbf{z}_{3}=E_{n\times nq}^{[j]}\textbf{z}_{1}$ into $-\mu_{k} (L_{k}\otimes P_{k})\textbf{z}_{1}-\kappa_{k}(I_{q}\otimes P_{k})\textbf{z}_{1}-\eta_{k}(L_{k}\otimes P_{k})\textbf{z}_{2}+\kappa_{k}(\textbf{1}_{q\times 1}\otimes P_{k})\textbf{z}_{3}=\textbf{0}_{nq\times 1}$ yields 
 \begin{eqnarray}\label{z1}
 (-\mu_{k} L_{k}\otimes P_{k}-\kappa_{k} I_{q}\otimes P_{k}+\kappa_{k} W_{k}^{[j]})\textbf{z}_{1}=\textbf{0}_{nq\times 1}, \quad k\in\overline{\mathbb{Z}}_{+}.
 \end{eqnarray}
Note that $(L_{k}\otimes I_{n})W_{k}^{[j]}=(L_{k}\otimes I_{n})(\textbf{1}_{q\times 1}\otimes E_{n\times nq}^{[j]})=L_{k}\textbf{1}_{q\times 1}\otimes E_{n\times nq}^{[j]}=\textbf{0}_{q\times 1}\otimes E_{n\times nq}^{[j]}=\textbf{0}_{nq\times nq}$ and $L_{k}\otimes P_{k}=(L_{k}\otimes I_{n})(I_{q}\otimes P_{k})$, $k\in\overline{\mathbb{Z}}_{+}$. Pre-multiplying $-L_{k}\otimes I_{n}$ on both sides of (\ref{z1}) yields 
 $(\mu_{k}(L_{k}\otimes I_{n})^{2}(I_{q}\otimes P_{k})+\kappa_{k}(L_{k}\otimes I_{n})(I_{q}\otimes P_{k}))\textbf{z}_{1}=(\mu_{k} L_{k}\otimes I_{n}+\kappa_{k} I_{nq})(L_{k}\otimes P_{k})\textbf{z}_{1}=\textbf{0}_{nq\times 1}$, $k\in\overline{\mathbb{Z}}_{+}$. Since $\kappa_{k}/\mu_{k}\not\in{\mathrm{spec}}(-L_{k}\otimes I_{n})$ for every $k\in\overline{\mathbb{Z}}_{+}$, it follows that $\det(\mu_{k} L_{k}\otimes I_{n}+\kappa_{k} I_{nq})\neq0$, $k\in\overline{\mathbb{Z}}_{+}$, where $\det$ denotes the determinant. Hence, $(L_{k}\otimes P_{k})\textbf{z}_{1}=\textbf{0}_{nq\times 1}$, $k\in\overline{\mathbb{Z}}_{+}$. 

Note that $L_{k}\textbf{w}_{0}=\textbf{0}_{q\times 1}$. Next, it follows from Lemma~\ref{lemma_ker} that  $\ker(L_{k}\otimes P_{k})=\ker(L_{k}\otimes I_{n})+\ker(I_{q}\otimes P_{k})$. Then it follows that $\bigcup_{i=0}^{q-1-{\mathrm{rank}}(L_{k})}{\mathrm{span}}\{\textbf{w}_{i}\otimes\textbf{e}_{1},\ldots,\textbf{w}_{i}\otimes\textbf{e}_{n}\}=\ker(L_{k}\otimes I_{n})$, $k\in\overline{\mathbb{Z}}_{+}$. Similarly, $\bigcup_{r=1}^{n-{\mathrm{rank}}(P_{k})}{\mathrm{span}}\{\textbf{g}_{1}\otimes\textbf{j}_{r},\ldots,\textbf{g}_{q}\otimes\textbf{j}_{r}\}=\ker(I_{q}\otimes P_{k})$. Consequently, $\ker(L_{k}\otimes P_{k})=\bigcup_{i=0}^{q-1-{\mathrm{rank}}(L_{k})}{\mathrm{span}}\{\textbf{w}_{i}\otimes\textbf{e}_{1},\ldots,\textbf{w}_{i}\otimes\textbf{e}_{n}\}+\bigcup_{r=1}^{n-{\mathrm{rank}}(P_{k})}{\mathrm{span}}\{\textbf{g}_{1}\otimes\textbf{j}_{r},\ldots,\textbf{g}_{q}\otimes\textbf{j}_{r}\}$. Hence, $\textbf{z}_{1}=\sum_{l=0}^{q-1-{\mathrm{rank}}(L_{k})}\sum_{i=1}^{n}\alpha_{li}\textbf{w}_{l}\otimes\textbf{e}_{i}+\sum_{s=1}^{q}\sum_{r=1}^{n-{\mathrm{rank}}(P_{k})}\beta_{sr}\textbf{g}_{s}\otimes\textbf{j}_{r}$, where $\alpha_{li},\beta_{sr}\in\mathbb{R}$ and $\alpha_{li}=\beta_{sr}=0$ for every $i=1,\ldots,n$ and every $s=1,\ldots,q$ if $\textbf{w}_{l}=\textbf{0}_{q\times 1}$ and $\textbf{j}_{r}=\textbf{0}_{n\times 1}$ for some $l\in\{1,\ldots,q-1-{\mathrm{rank}}(L_{k})\}$ and some $r\in\{1,\ldots,n-{\mathrm{rank}}(P_{k})\}$. Substituting this $\textbf{z}_{1}$ into the left-hand side of (\ref{z1}) yields $(-\mu_{k} L_{k}\otimes P_{k}-\kappa_{k} I_{q}\otimes P_{k}+\kappa_{k} W_{k}^{[j]})\textbf{z}_{1}=\kappa_{k}(W_{k}^{[j]}-I_{q}\otimes P_{k})\textbf{z}_{1}=\kappa_{k}(W_{k}^{[j]}-I_{q}\otimes P_{k})(\sum_{l=0}^{q-1-{\mathrm{rank}}(L_{k})}\sum_{i=1}^{n}\alpha_{li}\textbf{w}_{l}\otimes\textbf{e}_{i}+\sum_{s=1}^{q}\sum_{r=1}^{n-{\mathrm{rank}}(P_{k})}\beta_{sr}\textbf{g}_{s}\otimes\textbf{j}_{r})=\kappa_{k}\sum_{l=0}^{q-1-{\mathrm{rank}}(L_{k})}\sum_{i=1}^{n}\alpha_{li}W_{k}^{[j]}(\textbf{w}_{l}\otimes\textbf{e}_{i})-\kappa_{k}\sum_{l=0}^{q-1-{\mathrm{rank}}(L_{k})}\sum_{i=1}^{n}\alpha_{li}(I_{q}\otimes P_{k})(\textbf{w}_{l}\otimes\textbf{e}_{i})+\kappa_{k}\sum_{s=1}^{q}\sum_{r=1}^{n-{\mathrm{rank}}(P_{k})}\\\beta_{sr}W_{k}^{[j]}(\textbf{g}_{s}\otimes\textbf{j}_{r})-\kappa_{k}\sum_{s=1}^{q}\sum_{r=1}^{n-{\mathrm{rank}}(P_{k})}\beta_{sr}(I_{q}\otimes P_{k})(\textbf{g}_{s}\otimes\textbf{j}_{r})$. Note that it follows from $ii$) of Lemma~\ref{lemma_EW} that $W_{k}^{[j]}(\textbf{w}_{0}\otimes\textbf{e}_{i})=(I_{q}\otimes P_{k})(\textbf{w}_{0}\otimes\textbf{e}_{i})$ for every $j=1,\ldots,q$ and every $i=1,\ldots,n$. Let $P_{k}(i,j)$ denote the $(i,j)$th entry of $P_{k}$, then it follows from $ii$) of Lemma~\ref{lemma_EW} that
\begin{eqnarray*}
&&\sum_{l=0}^{q-1-{\mathrm{rank}}(L_{k})}\sum_{i=1}^{n}\alpha_{li}W_{k}^{[j]}(\textbf{w}_{l}\otimes\textbf{e}_{i})-\sum_{l=0}^{q-1-{\mathrm{rank}}(L_{k})}\sum_{i=1}^{n}\alpha_{li}(I_{q}\otimes P_{k})(\textbf{w}_{l}\otimes\textbf{e}_{i})\nonumber\\
&&=\sum_{l=1}^{q-1-{\mathrm{rank}}(L_{k})}\sum_{i=1}^{n}\alpha_{li}(W_{k}^{[j]}(\textbf{w}_{l}\otimes\textbf{e}_{i})-(I_{q}\otimes P_{k})(\textbf{w}_{l}\otimes\textbf{e}_{i}))\nonumber\\
&&=\sum_{l=1}^{q-1-{\mathrm{rank}}(L_{k})}\sum_{i=1}^{n}\alpha_{li}(w_{lj}(I_{q}\otimes P_{k})(\textbf{w}_{0}\otimes\textbf{e}_{i})-(I_{q}\otimes P_{k})(\textbf{w}_{l}\otimes\textbf{e}_{i}))\nonumber\\
&&=\sum_{l=1}^{q-1-{\mathrm{rank}}(L_{k})}\sum_{i=1}^{n}\alpha_{li}(I_{q}\otimes P_{k})((w_{lj}\textbf{w}_{0}-\textbf{w}_{l})\otimes\textbf{e}_{i})\nonumber\\
&&=\sum_{l=1}^{q-1-{\mathrm{rank}}(L_{k})}\sum_{i=1}^{n}\alpha_{li}(w_{lj}\textbf{w}_{0}-\textbf{w}_{l})\otimes P_{k}\textbf{e}_{i}\nonumber\\
&&=\sum_{l=1}^{q-1-{\mathrm{rank}}(L_{k})}\sum_{i=1}^{n}(w_{lj}\textbf{w}_{0}-\textbf{w}_{l})\otimes\small\left[\begin{array}{c}
\alpha_{li}P_{k}(1,i)\\
\vdots\\
\alpha_{li}P_{k}(n,i)\\
\end{array}\right]\nonumber\\
&&=\sum_{l=1}^{q-1-{\mathrm{rank}}(L_{k})}(w_{lj}\textbf{w}_{0}-\textbf{w}_{l})\otimes\small\left[\begin{array}{c}
\sum_{i=1}^{n}\alpha_{li}P_{k}(1,i)\\
\vdots\\
\sum_{i=1}^{n}\alpha_{li}P_{k}(n,i)\\
\end{array}\right]\nonumber\\
&&=\sum_{l=1}^{q-1-{\mathrm{rank}}(L_{k})}(w_{lj}\textbf{w}_{0}-\textbf{w}_{l})\otimes\Big(\sum_{s=1}^{n}\Big(\sum_{i=1}^{n}\alpha_{li}P_{k}(s,i)\Big)\textbf{e}_{s}\Big)\nonumber\\
&&=\sum_{l=1}^{q-1-{\mathrm{rank}}(L_{k})}\sum_{s=1}^{n}\Big(\sum_{i=1}^{n}\alpha_{li}P_{k}(s,i)\Big)(w_{lj}\textbf{w}_{0}-\textbf{w}_{l})\otimes\textbf{e}_{s}\nonumber\\
&&=\sum_{l=1}^{q-1-{\mathrm{rank}}(L_{k})}\sum_{s=1}^{n}\Big(\sum_{i=1}^{n}\alpha_{li}P_{k}(s,i)\Big)w_{lj}\textbf{w}_{0}\otimes\textbf{e}_{s}+\sum_{l=1}^{q-1-{\mathrm{rank}}(L_{k})}\sum_{s=1}^{n}\Big(-\sum_{i=1}^{n}\alpha_{li}P_{k}(s,i)\Big)\textbf{w}_{l}\otimes\textbf{e}_{s}.
\end{eqnarray*} Moreover, it follows from $iii$) of Lemma~\ref{lemma_EW} that
\begin{eqnarray*}
&&\kappa_{k}\sum_{s=1}^{q}\sum_{r=1}^{n-{\mathrm{rank}}(P_{k})}\beta_{sr}W_{k}^{[j]}(\textbf{g}_{s}\otimes\textbf{j}_{r})-\kappa_{k}\sum_{s=1}^{q}\sum_{r=1}^{n-{\mathrm{rank}}(P_{k})}\beta_{sr}(I_{q}\otimes P_{k})(\textbf{g}_{s}\otimes\textbf{j}_{r})\nonumber\\
&&=-\kappa_{k}\sum_{s=1}^{q}\sum_{r=1}^{n-{\mathrm{rank}}(P_{k})}\beta_{sr}\textbf{g}_{s}\otimes P_{k}\textbf{j}_{r}=\textbf{0}_{nq\times 1}.
\end{eqnarray*} Note that $\textbf{w}_{l}\otimes\textbf{e}_{s}$, $l=0,1,\ldots,q-1-{\mathrm{rank}}(L_{k})$, $s=1,\ldots,n$, are linearly independent. Hence, $\textbf{z}_{1}$ satisfies (\ref{z1}) if and only if $\sum_{i=1}^{n}\alpha_{li}P_{k}(s,i)=0$ for every $s=1,\ldots,q$ and every $l=1,\ldots,q-1-{\mathrm{rank}}(L_{k})$, which is equivalent to 
\begin{eqnarray}\label{Pka}
P_{k}\small\left[\begin{array}{c}
\alpha_{l1}\\
\vdots\\
\alpha_{ln}
\end{array}\right]=\textbf{0}_{n\times 1}
\end{eqnarray}
for every $l=1,\ldots,q-1-{\mathrm{rank}}(L_{k})$. Thus, $\small\left[\begin{array}{c}
\alpha_{l1}\\
\vdots\\
\alpha_{ln}
\end{array}\right]=(I_{n}-P_{k}^{+}P_{k})(\sum_{i=1}^{n}\gamma_{li}\textbf{e}_{i})$, where $\gamma_{li}\in\mathbb{R}$, $i=1,\ldots,n$, $l=1,\ldots,q-1-{\mathrm{rank}}(L_{k})$, are arbitrary. In other words, $\alpha_{li}=\sum_{m=1}^{n}\gamma_{lm}\textbf{e}_{i}^{\mathrm{T}}(I_{n}-P_{k}^{+}P_{k})\textbf{e}_{m}$ for every $i=1,\ldots,n$ and every $l=1,\ldots,q-1-{\mathrm{rank}}(L_{k})$. In this case, we have $\textbf{z}_{1}=\sum_{i=1}^{n}\alpha_{0i}\textbf{w}_{0}\otimes\textbf{e}_{i}+\sum_{l=1}^{q-1-{\mathrm{rank}}(L_{k})}\sum_{i=1}^{n}\sum_{m=1}^{n}\gamma_{lm}(\textbf{e}_{i}^{\mathrm{T}}(I_{n}-P_{k}^{+}P_{k})\textbf{e}_{m})(\textbf{w}_{l}\otimes\textbf{e}_{i})+\sum_{s=1}^{q}\sum_{r=1}^{n-{\mathrm{rank}}(P_{k})}\beta_{sr}\textbf{g}_{s}\otimes\textbf{j}_{r}$, where $\alpha_{0i},\beta_{sr},\gamma_{lm}\in\mathbb{R}$ are arbitrary.

Note that by $iii$) of Lemma~\ref{lemma_EW}, $\textbf{z}_{3}=E_{n\times nq}^{[j]}\textbf{z}_{1}=\sum_{i=1}^{n}\alpha_{0i}E_{n\times nq}^{[j]}(\textbf{1}_{q\times 1}\otimes\textbf{e}_{i})+\sum_{l=1}^{q-1-{\mathrm{rank}}(L_{k})}\sum_{i=1}^{n}\sum_{m=1}^{n}\\\gamma_{lm}(\textbf{e}_{i}^{\mathrm{T}}(I_{n}-P_{k}^{+}P_{k})\textbf{e}_{m})E_{n\times nq}^{[j]}(\textbf{w}_{l}\otimes\textbf{e}_{i})+\sum_{s=1}^{q}\sum_{r=1}^{n-{\mathrm{rank}}(P_{k})}\beta_{sr}E_{n\times nq}^{[j]}(\textbf{g}_{s}\otimes\textbf{j}_{r})=\sum_{i=1}^{n}\alpha_{0i}\textbf{e}_{i}+\sum_{l=1}^{q-1-{\mathrm{rank}}(L_{k})}\\\sum_{i=1}^{n}\sum_{m=1}^{n}\gamma_{lm}(\textbf{e}_{i}^{\mathrm{T}}(I_{n}-P_{k}^{+}P_{k})\textbf{e}_{m})w_{lj}\textbf{e}_{i}+\sum_{r=1}^{n-{\mathrm{rank}}(P_{k})}\beta_{jr}\textbf{j}_{r}$ for every $j=1,\ldots,q$. Thus, $\ker(A_{k}^{[j]})=\{[\sum_{i=1}^{n}\alpha_{0i}(\textbf{w}_{0}\otimes\textbf{e}_{i})^{\mathrm{T}}+\sum_{l=1}^{q-1-{\mathrm{rank}}(L_{k})}\sum_{i=1}^{n}\sum_{m=1}^{n}\gamma_{lm}(\textbf{e}_{i}^{\mathrm{T}}(I_{n}-P_{k}^{+}P_{k})\textbf{e}_{m})(\textbf{w}_{l}\otimes\textbf{e}_{i})^{\mathrm{T}}+\sum_{s=1}^{q}\sum_{r=1}^{n-{\mathrm{rank}}(P_{k})}\\\beta_{sr}(\textbf{g}_{s}\otimes\textbf{j}_{r})^{\mathrm{T}},\textbf{0}_{1\times nq},\sum_{i=1}^{n}\alpha_{0i}\textbf{e}_{i}^{\mathrm{T}}+\sum_{l=1}^{q-1-{\mathrm{rank}}(L_{k})}\sum_{i=1}^{n}\sum_{m=1}^{n}\gamma_{lm}(\textbf{e}_{i}^{\mathrm{T}}(I_{n}-P_{k}^{+}P_{k})\textbf{e}_{m})w_{lj}\textbf{e}_{i}^{\mathrm{T}}+\sum_{r=1}^{n-{\mathrm{rank}}(P_{k})}\beta_{jr}\\\textbf{j}_{r}^{\mathrm{T}}]^{\mathrm{T}}:\forall\alpha_{0i},\beta_{sr},\gamma_{lm}\in\mathbb{R},i=1,\ldots,n,s=1,\ldots,q,r=1,\ldots,n-{\mathrm{rank}}(P_{k}),l=1,\ldots,q-1-{\mathrm{rank}}(L_{k}),m=1,\ldots,n\}$ for every $j=1,\ldots,q$, $k\in\overline{\mathbb{Z}}_{+}$. 

Let $S_{1}=\{[\sum_{i=1}^{n}\alpha_{0i}(\textbf{w}_{0}\otimes\textbf{e}_{i})^{\mathrm{T}},\textbf{0}_{1\times nq},\sum_{i=1}^{n}\alpha_{0i}\textbf{e}_{i}^{\mathrm{T}}]^{\mathrm{T}}:\forall\alpha_{0i}\in\mathbb{R},i=1,\ldots,n\}$, $S_{2}=\{[\sum_{l=1}^{q-1-{\mathrm{rank}}(L_{k})}\sum_{i=1}^{n}\\\sum_{m=1}^{n}\gamma_{lm}(\textbf{e}_{i}^{\mathrm{T}}(I_{n}-P_{k}^{+}P_{k})\textbf{e}_{m})(\textbf{w}_{l}\otimes\textbf{e}_{i})^{\mathrm{T}},\textbf{0}_{1\times nq},\sum_{l=1}^{q-1-{\mathrm{rank}}(L_{k})}\sum_{i=1}^{n}\sum_{m=1}^{n}\gamma_{lm}(\textbf{e}_{i}^{\mathrm{T}}(I_{n}-P_{k}^{+}P_{k})\textbf{e}_{m})w_{lj}\textbf{e}_{i}^{\mathrm{T}}:\forall\gamma_{lm}\in\mathbb{R},l=1,\ldots,q-1-{\mathrm{rank}}(L_{k}),m=1,\ldots,n\}$, and $S_{3}=\{[\sum_{s=1}^{q}\sum_{r=1}^{n-{\mathrm{rank}}(P_{k})}\beta_{sr}(\textbf{g}_{s}\otimes\textbf{j}_{r})^{\mathrm{T}},\textbf{0}_{1\times nq},\\\sum_{r=1}^{n-{\mathrm{rank}}(P_{k})}\beta_{jr}\textbf{j}_{r}^{\mathrm{T}}]^{\mathrm{T}}:\forall\beta_{sr}\in\mathbb{R},s=1,\ldots,q,r=1,\ldots,n-{\mathrm{rank}}(P_{k})\}$. Clearly $\ker(A_{k}^{[j]})=S_{1}+S_{2}+S_{3}$ and $S_{i}$ is a subspace for every $i=1,2,3$. Furthermore, note that $S_{1}\cup S_{2}=S_{1}+S_{2}$ is a subspace. Hence, it follows from Lemma~\ref{lemma_S} that $\dim\ker(A_{k}^{[j]})=\dim S_{1}+\dim S_{2}+\dim S_{3}-\dim(S_{1}\cap S_{2})-\dim(S_{2}\cap S_{3})-\dim(S_{3}\cap S_{1})+\dim(S_{1}\cap S_{2}\cap S_{3})$ for every $j=1,\ldots,q$, $k\in\overline{\mathbb{Z}}_{+}$. Note that $\dim S_{1}=n$ and $\dim S_{3}=q(n-{\mathrm{rank}}(P_{k}))$. To determine $\dim S_{2}$, it first follows from (\ref{Pka}) that $\dim\left\{\small\left[\begin{array}{c}
\alpha_{l1}\\
\vdots\\
\alpha_{ln}
\end{array}\right]:P_{k}\small\left[\begin{array}{c}
\alpha_{l1}\\
\vdots\\
\alpha_{ln}
\end{array}\right]=\textbf{0}_{n\times 1}\right\}=\dim\ker(P_{k})=n-{\mathrm{rank}}(P_{k})$ for every $l=1,\ldots,q-1-{\mathrm{rank}}(L_{k})$. Since $S_{2}=\{[\sum_{l=1}^{q-1-{\mathrm{rank}}(L_{k})}\sum_{i=1}^{n}\alpha_{li}(\textbf{w}_{l}\otimes\textbf{e}_{i})^{\mathrm{T}},\textbf{0}_{1\times nq},\sum_{l=1}^{q-1-{\mathrm{rank}}(L_{k})}\sum_{i=1}^{n}\alpha_{li}w_{lj}\textbf{e}_{i}^{\mathrm{T}}:\alpha_{li}\in\mathbb{R},l=1,\ldots,q-1-{\mathrm{rank}}(L_{k}),i=1,\ldots,n,\,\,{\mathrm{satisfy}}\,\,(\ref{Pka})\}$, it follows that $\dim S_{2}=\sum_{l=1}^{q-1-{\mathrm{rank}}(L_{k})}\dim\left\{\small\left[\begin{array}{c}
\alpha_{l1}\\
\vdots\\
\alpha_{ln}
\end{array}\right]:P_{k}\small\left[\begin{array}{c}
\alpha_{l1}\\
\vdots\\
\alpha_{ln}
\end{array}\right]=\textbf{0}_{n\times 1}\right\}=(q-1-{\mathrm{rank}}(L_{k}))(n-{\mathrm{rank}}(P_{k}))$.

Let $S_{4}=\{[\sum_{l=1}^{q-1-{\mathrm{rank}}(L_{k})}\sum_{i=1}^{n}\alpha_{li}(\textbf{w}_{l}\otimes\textbf{e}_{i})^{\mathrm{T}},\textbf{0}_{1\times nq},\sum_{l=1}^{q-1-{\mathrm{rank}}(L_{k})}\sum_{i=1}^{n}\alpha_{li}w_{lj}\textbf{e}_{i}^{\mathrm{T}}:\forall\alpha_{li}\in\mathbb{R},l=1,\ldots,q-1-{\mathrm{rank}}(L_{k}),i=1,\ldots,n\}$. Clearly $S_{2}\subseteq S_{4}$. Next, since $\textbf{w}_{l}\otimes\textbf{e}_{i}$, $l=0,1,\ldots,q-1-{\mathrm{rank}}(L_{k})$, $i=1,\ldots,n$, are linearly independent, it follows that $\sum_{i=1}^{n}\alpha_{0i}\textbf{w}_{0}\otimes\textbf{e}_{i}-\sum_{l=1}^{q-1-{\mathrm{rank}}(L_{k})}\sum_{i=1}^{n}\alpha_{li}\textbf{w}_{l}\otimes\textbf{e}_{i}=\textbf{0}_{nq\times 1}$ if and only if $\alpha_{li}=0$ for every $l=0,1,\ldots,q-1-{\mathrm{rank}}(L_{k})$ and every $i=1,\ldots,n$. Hence, $S_{1}\cap S_{4}=\{\textbf{0}_{(2nq+n)\times 1}\}$. Consequently, $\{\textbf{0}_{(2nq+n)\times 1}\}\subseteq S_{1}\cap S_{2}\subseteq S_{1}\cap S_{4}=\{\textbf{0}_{(2nq+n)\times 1}\}$ and $\{\textbf{0}_{(2nq+n)\times 1}\}\subseteq S_{1}\cap S_{2}\cap S_{3}\subseteq S_{1}\cap S_{2}\subseteq S_{1}\cap S_{4}=\{\textbf{0}_{(2nq+n)\times 1}\}$, which imply that $S_{1}\cap S_{2}=\{\textbf{0}_{(2nq+n)\times 1}\}$ and $S_{1}\cap S_{2}\cap S_{3}=\{\textbf{0}_{(2nq+n)\times 1}\}$. Hence, $\dim(S_{1}\cap S_{2})=0$ and $\dim(S_{1}\cap S_{2}\cap S_{3})=0$. 

Alternatively, note that  $\sum_{i=1}^{n}\alpha_{0i}\textbf{w}_{0}\otimes\textbf{e}_{i}=\sum_{s=1}^{q}\sum_{r=1}^{n-{\mathrm{rank}}(P_{k})}\beta_{sr}\textbf{g}_{s}\otimes\textbf{j}_{r}$ for some $\alpha_{0i}$ and $\beta_{sr}$ is equivalent to $\sum_{i=1}^{n}\alpha_{0i}(I_{q}\otimes P_{k})(\textbf{w}_{0}\otimes\textbf{e}_{i})=\textbf{0}_{nq\times 1}$ due to the fact that $\bigcup_{r=1}^{n-{\mathrm{rank}}(P_{k})}{\mathrm{span}}\{\textbf{g}_{1}\otimes\textbf{j}_{r},\ldots,\textbf{g}_{q}\otimes\textbf{j}_{r}\}=\ker(I_{q}\otimes P_{k})$. Thus,
$\textbf{0}_{nq\times 1}=\sum_{i=1}^{n}\alpha_{0i}(I_{q}\otimes P_{k})(\textbf{w}_{0}\otimes\textbf{e}_{i})=\sum_{i=1}^{n}\alpha_{0i}\textbf{w}_{0}\otimes P_{k}\textbf{e}_{i}=\textbf{w}_{0}\otimes P_{k}\Big(\sum_{i=1}^{n}\alpha_{0i}\textbf{e}_{i}\Big)=\textbf{w}_{0}\otimes P_{k}\small\left[\begin{array}{c}
\alpha_{01}\\
\vdots\\
\alpha_{0n}\\
\end{array}\right]$, which is equivalent to $P_{k}\small\left[\begin{array}{c}
\alpha_{01}\\
\vdots\\
\alpha_{0n}\\
\end{array}\right]=\textbf{0}_{n\times 1}$. Hence, $\dim(S_{1}\cap S_{3})=\dim\left\{\small\left[\begin{array}{c}
\alpha_{01}\\
\vdots\\
\alpha_{0n}
\end{array}\right]:P_{k}\small\left[\begin{array}{c}
\alpha_{01}\\
\vdots\\
\alpha_{0n}
\end{array}\right]=\textbf{0}_{n\times 1}\right\}=n-{\mathrm{rank}}(P_{k})$.

Likewise, $\sum_{l=1}^{q-1-{\mathrm{rank}}(L_{k})}\sum_{i=1}^{n}\alpha_{li}\textbf{w}_{l}\otimes\textbf{e}_{i}=\sum_{s=1}^{q}\sum_{r=1}^{n-{\mathrm{rank}}(P_{k})}\beta_{sr}\textbf{g}_{s}\otimes\textbf{j}_{r}$ for some $\alpha_{li}$ and $\beta_{sr}$ is equivalent to $\sum_{l=1}^{q-1-{\mathrm{rank}}(L_{k})}\sum_{i=1}^{n}\alpha_{li}(I_{q}\otimes P_{k})(\textbf{w}_{l}\otimes\textbf{e}_{i})=\textbf{0}_{nq\times 1}$. Thus, $\textbf{0}_{nq\times 1}=\sum_{l=1}^{q-1-{\mathrm{rank}}(L_{k})}\sum_{i=1}^{n}\alpha_{li}\textbf{w}_{l}\otimes P_{k}\textbf{e}_{i}=\sum_{l=1}^{q-1-{\mathrm{rank}}(L_{k})}\textbf{w}_{l}\otimes P_{k}(\sum_{i=1}^{n}\alpha_{li}\textbf{e}_{i})=\sum_{l=1}^{q-1-{\mathrm{rank}}(L_{k})}\textbf{w}_{l}\otimes P_{k}\small\left[\begin{array}{c}
\alpha_{l1}\\
\vdots\\
\alpha_{ln}
\end{array}\right]=\sum_{l=1}^{q-1-{\mathrm{rank}}(L_{k})}\sum_{s=1}^{n}(\sum_{i=1}^{n}\alpha_{li}\\P_{k}(s,i))\textbf{w}_{l}\otimes\textbf{e}_{s}$, which is equivalent to $\sum_{i=1}^{n}\alpha_{li}P_{k}(s,i)=0$ for every $s=1,\ldots,q$ and every $l=1,\ldots,q-1-{\mathrm{rank}}(L_{k})$. Hence, (\ref{Pka}) holds for every $l=1,\ldots,q-1-{\mathrm{rank}}(L_{k})$. Thus, $\dim(S_{2}\cap S_{3})=\dim S_{2}=(q-1-{\mathrm{rank}}(L_{k}))(n-{\mathrm{rank}}(P_{k}))$.

Now, ${\mathrm{def}}(A_{k}^{[j]})=n+(q-1-{\mathrm{rank}}(L_{k}))(n-{\mathrm{rank}}(P_{k}))+q(n-{\mathrm{rank}}(P_{k}))-0-(q-1-{\mathrm{rank}}(L_{k}))(n-{\mathrm{rank}}(P_{k}))-(n-{\mathrm{rank}}(P_{k}))+0=n+(q-1)(n-{\mathrm{rank}}(P_{k}))$. Therefore, it follows from Corollary 2.5.5 of \cite[p.~105]{Bernstein:2009} that  ${\mathrm{rank}}(A_{k}^{[j]})=2nq+n-{\mathrm{def}}(A_{k}^{[j]})=2nq-(q-1)(n-{\mathrm{rank}}(P_{k}))$ for every $j=1,\ldots,q$, $k\in\overline{\mathbb{Z}}_{+}$.

$iv)$ If $\mu_{k}\neq0$ and $\kappa_{k}=0$, then $\textbf{z}_{2}=\textbf{0}_{nq\times 1}$, $-\mu_{k}(L_{k}\otimes P_{k})\textbf{z}_{1}=\textbf{0}_{nq\times 1}$, and $\textbf{z}_{3}$ in $\ker(A_{k}^{[j]})$ can be chosen arbitrarily in $\mathbb{R}^{n}$. Thus, $\textbf{z}_{3}$ can be represented as $\textbf{z}_{3}=\sum_{i=1}^{n}\gamma_{i}\textbf{e}_{i}$, where $\gamma_{i}\in\mathbb{R}$. In this case, since $(L_{k}\otimes P_{k})\textbf{z}_{1}=\textbf{0}_{nq\times 1}$, $k\in\overline{\mathbb{Z}}_{+}$, it follows from the similar arguments as in the proof of $iii$) that $\textbf{z}_{1}=\sum_{l=0}^{q-1-{\mathrm{rank}}(L_{k})}\sum_{i=1}^{n}\alpha_{li}\textbf{w}_{l}\otimes\textbf{e}_{i}+\sum_{s=1}^{q}\sum_{r=1}^{n-{\mathrm{rank}}(P_{k})}\beta_{sr}\textbf{g}_{s}\otimes\textbf{j}_{r}$, where $\alpha_{li},\beta_{sr}\in\mathbb{R}$. Therefore, $\ker(A_{k}^{[j]})=\{[\sum_{l=0}^{q-1-{\mathrm{rank}}(L_{k})}\sum_{i=1}^{n}\alpha_{li}(\textbf{w}_{l}\otimes\textbf{e}_{i})^{\mathrm{T}}+\sum_{s=1}^{q}\sum_{r=1}^{n-{\mathrm{rank}}(P_{k})}\beta_{sr}(\textbf{g}_{s}\otimes\textbf{j}_{r})^{\mathrm{T}},\textbf{0}_{1\times nq},\sum_{i=1}^{n}\gamma_{i}\textbf{e}_{i}^{\mathrm{T}}]^{\mathrm{T}}:\forall\alpha_{li},\beta_{sr},\gamma_{i}\in\mathbb{R},i=1,\ldots,n,l=0,1,\ldots,q-1-{\mathrm{rank}}(L_{k}),s=1,\ldots,q,r=1,\ldots,n-{\mathrm{rank}}(P_{k})\}$ for every $j=1,\ldots,q$, $k\in\overline{\mathbb{Z}}_{+}$. 

Again, let $\mathcal{S}_{1}=\{[\sum_{l=0}^{q-1-{\mathrm{rank}}(L_{k})}\sum_{i=1}^{n}\alpha_{li}(\textbf{w}_{l}\otimes\textbf{e}_{i})^{\mathrm{T}},\textbf{0}_{1\times nq},\textbf{0}_{1\times n}]^{\mathrm{T}}:\forall\alpha_{li}\in\mathbb{R},i=1,\ldots,n,l=0,1,\ldots,q-1-{\mathrm{rank}}(L_{k})\}$, $\mathcal{S}_{2}=\{[\sum_{s=1}^{q}\sum_{r=1}^{n-{\mathrm{rank}}(P_{k})}\beta_{sr}(\textbf{g}_{s}\otimes\textbf{j}_{r})^{\mathrm{T}},\textbf{0}_{1\times nq},\textbf{0}_{1\times n}]^{\mathrm{T}}:\forall\beta_{sr}\in\mathbb{R},s=1,\ldots,q,r=1,\ldots,n-{\mathrm{rank}}(P_{k})\}$, and $\mathcal{S}_{3}=\{[\textbf{0}_{1\times nq},\textbf{0}_{1\times nq},\sum_{i=1}^{n}\gamma_{i}\textbf{e}_{i}^{\mathrm{T}}]^{\mathrm{T}}:\forall\gamma_{i}\in\mathbb{R},i=1,\ldots,n\}$.
Clearly $\ker(A_{k}^{[j]})=\mathcal{S}_{1}+\mathcal{S}_{2}+\mathcal{S}_{3}$ and $\mathcal{S}_{i}$ is a subspace for every $i=1,2,3$. Next, note that $\mathcal{S}_{1}\cup\mathcal{S}_{3}=\mathcal{S}_{1}+\mathcal{S}_{3}$ is a subspace, $\dim\mathcal{S}_{1}=n(q-{\mathrm{rank}}(L_{k}))$, $\dim\mathcal{S}_{2}=q(n-{\mathrm{rank}}(P_{k}))$, and $\dim\mathcal{S}_{3}=n$. Furthermore, note that $\mathcal{S}_{1}\cap\mathcal{S}_{3}=\{\textbf{0}_{(2nq+n)\times 1}\}$, $\mathcal{S}_{2}\cap\mathcal{S}_{3}=\{\textbf{0}_{(2nq+n)\times 1}\}$, and $\mathcal{S}_{1}\cap\mathcal{S}_{2}\cap\mathcal{S}_{3}=\{\textbf{0}_{(2nq+n)\times 1}\}$. Using the similar arguments as in the proof of $iii$), it follows that $\dim(\mathcal{S}_{1}\cap\mathcal{S}_{2})=\sum_{l=0}^{q-1-{\mathrm{rank}}(L_{k})}\dim\left\{\small\left[\begin{array}{c}
\alpha_{l1}\\
\vdots\\
\alpha_{ln}
\end{array}\right]:P_{k}\small\left[\begin{array}{c}
\alpha_{l1}\\
\vdots\\
\alpha_{ln}
\end{array}\right]=\textbf{0}_{n\times 1}\right\}=(q-{\mathrm{rank}}(L_{k}))(n-{\mathrm{rank}}(P_{k}))$. Now it follows from Lemma~\ref{lemma_S} that $\dim\ker(A_{k}^{[j]})=\dim(\mathcal{S}_{1}+\mathcal{S}_{2}+\mathcal{S}_{3})=\dim\mathcal{S}_{1}+\dim\mathcal{S}_{2}+\dim\mathcal{S}_{3}-\dim(\mathcal{S}_{1}\cap\mathcal{S}_{2})-\dim(\mathcal{S}_{2}\cap\mathcal{S}_{3})-\dim(\mathcal{S}_{3}\cap\mathcal{S}_{1})+\dim(\mathcal{S}_{1}\cap\mathcal{S}_{2}\cap\mathcal{S}_{3})=n(q-{\mathrm{rank}}(L_{k}))+q(n-{\mathrm{rank}}(P_{k}))+n-(q-{\mathrm{rank}}(L_{k}))(n-{\mathrm{rank}}(P_{k}))-0-0+0=nq+n-{\mathrm{rank}}(L_{k}){\mathrm{rank}}(P_{k})$ for every $j=1,\ldots,q$, $k\in\overline{\mathbb{Z}}_{+}$. Therefore, it follows from Corollary 2.5.5 of \cite[p.~105]{Bernstein:2009} that  ${\mathrm{rank}}(A_{k}^{[j]})=2nq+n-{\mathrm{def}}(A_{k}^{[j]})=nq+{\mathrm{rank}}(L_{k}){\mathrm{rank}}(P_{k})$ for every $j=1,\ldots,q$, $k\in\overline{\mathbb{Z}}_{+}$.
\end{IEEEproof}

It follows from Lemma~\ref{lemma_Arank} that 0 is an eigenvalue of $A_{k}^{[j]}$ for every $j=1,\ldots,q$ and every $k\in\overline{\mathbb{Z}}_{+}$. Next, we further investigate some relationships of the null spaces between a row-addition transformed matrix of $A_{k}^{[j]}$ and $A_{k}^{[j]}$ itself in order to unveil an important property of this eigenvalue 0 later.

\begin{lemma}\label{lemma_Ah}
Consider the (possibly infinitely many) matrices $A_{k}^{[j]}+h_{k}A_{{\mathrm{c}}k}$, $j=1,\ldots,q$, $k=0,1,2,\ldots$, where $A_{k}^{[j]}$ is defined by (\ref{Amatrix}) in Lemma~\ref{lemma_Arank},
\begin{eqnarray}\label{Ac}
A_{{\mathrm{c}}k}=\small\left[\begin{array}{ccc}
-\mu_{k} L_{k}\otimes P_{k}-\kappa_{k} I_{q}\otimes P_{k} & -\eta_{k} L_{k}\otimes P_{k} & \kappa_{k} \textbf{1}_{q\times 1}\otimes P_{k} \\
\textbf{0}_{nq\times nq} & \textbf{0}_{nq\times nq} & \textbf{0}_{nq\times n} \\
\textbf{0}_{n\times nq} & \textbf{0}_{n\times nq} & \textbf{0}_{n\times n} \\
\end{array}\right],
\end{eqnarray} and $\mu_{k},\kappa_{k},\eta_{k},h_{k}\geq0$, $k\in\overline{\mathbb{Z}}_{+}$. Then $\ker(A_{k}^{[j]})=\ker(A_{k}^{[j]}+h_{k}A_{{\mathrm{c}}k})$ and $\ker(A_{k}^{[j]}(A_{k}^{[j]}+h_{k}A_{{\mathrm{c}}k}))=\ker((A_{k}^{[j]}+h_{k}A_{{\mathrm{c}}k})^{2})$ for every $j=1,\ldots,q$ and every $k\in\overline{\mathbb{Z}}_{+}$.
\end{lemma}

\begin{IEEEproof}
To show that $\ker(A_{k}^{[j]})=\ker(A_{k}^{[j]}+h_{k}A_{{\mathrm{c}}k})$, note that for every $j=1,\ldots,q$, $\ker(A_{k}^{[j]})=\{[\textbf{z}_{1}^{\rm{T}},\textbf{z}_{2}^{\rm{T}},\textbf{z}_{3}^{\rm{T}}]^{\rm{T}}\in\mathbb{R}^{2nq+n}:\textbf{z}_{2}=\textbf{0}_{nq\times 1},-\mu_{k} (L_{k}\otimes P_{k})\textbf{z}_{1}-\kappa_{k}(I_{q}\otimes P_{k})\textbf{z}_{1}-\eta_{k}(L_{k}\otimes P_{k})\textbf{z}_{2}+\kappa_{k}(\textbf{1}_{q\times 1}\otimes P_{k})\textbf{z}_{3}=\textbf{0}_{nq\times 1}, \kappa_{k}E_{n\times nq}^{[j]}\textbf{z}_{1}-\kappa_{k}\textbf{z}_{3}=\textbf{0}_{n\times 1}\}$, $k\in\overline{\mathbb{Z}}_{+}$. Alternatively, for every $j=1,\ldots,q$ and every $k\in\overline{\mathbb{Z}}_{+}$, let  $\textbf{y}=[\textbf{y}_{1}^{\rm{T}},\textbf{y}_{2}^{\rm{T}},\textbf{y}_{3}^{\rm{T}}]^{\rm{T}}\in\ker(A_{k}^{[j]}+h_{k}A_{{\mathrm{c}}k})$, where $\textbf{y}_{1},\textbf{y}_{2}\in\mathbb{R}^{nq}$ and $\textbf{y}_{3}\in\mathbb{R}^{n}$, we have
\begin{eqnarray}
h_{k}(-\mu_{k} L_{k}\otimes P_{k}-\kappa_{k} I_{q}\otimes P_{k})\textbf{y}_{1}+h_{k}(-\eta_{k} L_{k}\otimes P_{k})\textbf{y}_{2}+\textbf{y}_{2}+h_{k}(\kappa_{k} \textbf{1}_{q\times 1}\otimes P_{k})\textbf{y}_{3}=\textbf{0}_{nq\times 1},\label{y_1p}\\
(-\mu_{k} L_{k}\otimes P_{k}-\kappa_{k} I_{q}\otimes P_{k})\textbf{y}_{1}+(-\eta_{k} L_{k}\otimes P_{k})\textbf{y}_{2}+(\kappa_{k} \textbf{1}_{q\times 1}\otimes P_{k})\textbf{y}_{3}=\textbf{0}_{nq\times 1},\label{y_2p}\\
\kappa_{k} E_{n\times nq}^{[j]}\textbf{y}_{1}-\kappa_{k}\textbf{y}_{3}=\textbf{0}_{n\times 1}.\label{y_3p}
\end{eqnarray} Substituting (\ref{y_2p}) into (\ref{y_1p}) yields $\textbf{y}_{2}=\textbf{0}_{nq\times 1}$. Together with (\ref{y_2p}) and (\ref{y_3p}), we have $\textbf{y}\in\ker(A_{k}^{[j]})$, which implies that $\ker(A_{k}^{[j]}+h_{k}A_{{\mathrm{c}}k})\subseteq\ker(A_{k}^{[j]})$ for every $j=1,\ldots,q$ and every $k\in\overline{\mathbb{Z}}_{+}$. On the other hand, if $\textbf{y}\in\ker(A_{k}^{[j]})$, then $\textbf{y}_{2}=\textbf{0}_{nq\times 1}$, $-\mu_{k} (L_{k}\otimes P_{k})\textbf{y}_{1}-\kappa_{k}(I_{q}\otimes P_{k})\textbf{y}_{1}-\eta_{k}(L_{k}\otimes P_{k})\textbf{y}_{2}+\kappa_{k}(\textbf{1}_{q\times 1}\otimes P_{k})\textbf{y}_{3}=\textbf{0}_{nq\times 1}$, and $\kappa_{k}E_{n\times nq}^{[j]}\textbf{y}_{1}-\kappa_{k}\textbf{y}_{3}=\textbf{0}_{n\times 1}$. Clearly in this case, (\ref{y_1p})--(\ref{y_3p}) hold, i.e., $\textbf{y}\in\ker(A_{k}^{[j]}+h_{k}A_{{\mathrm{c}}k})$, which implies that $\ker(A_{k}^{[j]})\subseteq\ker(A_{k}^{[j]}+h_{k}A_{{\mathrm{c}}k})$ for every $j=1,\ldots,q$ and every $k\in\overline{\mathbb{Z}}_{+}$. Thus, $\ker(A_{k}^{[j]})=\ker(A_{k}^{[j]}+h_{k}A_{{\mathrm{c}}k})$ for every $j=1,\ldots,q$ and every $k\in\overline{\mathbb{Z}}_{+}$.

Finally, to show that $\ker(A_{k}^{[j]}(A_{k}^{[j]}+h_{k}A_{{\mathrm{c}}k}))=\ker((A_{k}^{[j]}+h_{k}A_{{\mathrm{c}}k})^{2})$, note that $\ker((A_{k}^{[j]}+h_{k}A_{{\mathrm{c}}k})^{2})=\ker((A_{k}^{[j]}+h_{k}A_{{\mathrm{c}}k})(A_{k}^{[j]}+h_{k}A_{{\mathrm{c}}k}))$ for every $j=1,\ldots,q$ and every $k\in\overline{\mathbb{Z}}_{+}$. Let $\textbf{y}\in\ker((A_{k}^{[j]}+h_{k}A_{{\mathrm{c}}k})(A_{k}^{[j]}+h_{k}A_{{\mathrm{c}}k}))$, then $(A_{k}^{[j]}+h_{k}A_{{\mathrm{c}}k})\textbf{y}\in\ker(A_{k}^{[j]}+h_{k}A_{{\mathrm{c}}k})=\ker(A_{k}^{[j]})$, and hence, $\textbf{y}\in\ker((A_{k}^{[j]}+h_{k}A_{{\mathrm{c}}k})^{2})$, which implies that $\ker(A_{k}^{[j]}(A_{k}^{[j]}+h_{k}A_{{\mathrm{c}}k}))\subseteq\ker(A_{k}^{[j]}(A_{k}^{[j]}+h_{k}A_{{\mathrm{c}}k}))$ for every $j=1,\ldots,q$ and every $k\in\overline{\mathbb{Z}}_{+}$. Alternatively, let $\textbf{z}\in\ker(A_{k}^{[j]}(A_{k}^{[j]}+h_{k}A_{{\mathrm{c}}k}))$, then $(A_{k}^{[j]}+h_{k}A_{{\mathrm{c}}k})\textbf{z}\in\ker(A_{k}^{[j]})=\ker(A_{k}^{[j]}+h_{k}A_{{\mathrm{c}}k})$, and hence, $\textbf{z}\in\ker((A_{k}^{[j]}+h_{k}A_{{\mathrm{c}}k})^{2})$, which implies that $\ker(A_{k}^{[j]}(A_{k}^{[j]}+h_{k}A_{{\mathrm{c}}k}))\subseteq\ker((A_{k}^{[j]}+h_{k}A_{{\mathrm{c}}k})^{2})$ for every $j=1,\ldots,q$ and every $k\in\overline{\mathbb{Z}}_{+}$. Thus, $\ker(A_{k}^{[j]}(A_{k}^{[j]}+h_{k}A_{{\mathrm{c}}k}))=\ker((A_{k}^{[j]}+h_{k}A_{{\mathrm{c}}k})^{2})$ for every $j=1,\ldots,q$ and every $k\in\overline{\mathbb{Z}}_{+}$.
\end{IEEEproof}

Next, we assert that 0 is semisimple for $A_{k}^{[j]}+h_{k}A_{{\mathrm{c}}k}$. Recall from Definition 5.5.4 of \cite[p.~322]{Bernstein:2009} that $0$ is semisimple if its geometric multiplicity and algebraic multiplicity are equal.

\begin{lemma}\label{lemma_semisimple}
Consider the (possibly infinitely many) matrices $A_{k}^{[j]}+h_{k}A_{{\mathrm{c}}k}$, $j=1,\ldots,q$, $k=0,1,2,\ldots$, defined in Lemma~\ref{lemma_Ah}, where $\mu_{k},\kappa_{k},\eta_{k},h_{k}\geq0$, $k\in\overline{\mathbb{Z}}_{+}$.
\begin{itemize}
\item[$i$)] If $\kappa_{k}=0$ and $\mu_{k}=0$, then ${\mathrm{rank}}(A_{k}^{[j]}+h_{k}A_{{\mathrm{c}}k})=nq$ and 0 is not a semisimple eigenvalue of $A_{k}^{[j]}+h_{k}A_{{\mathrm{c}}k}$ for every $j=1,\ldots,q$, $k\in\overline{\mathbb{Z}}_{+}$.
\item[$ii$)] If $\kappa_{k}=0$ and $\mu_{k}\neq0$, then ${\mathrm{rank}}(A_{k}^{[j]}+h_{k}A_{{\mathrm{c}}k})=nq+{\mathrm{rank}}(L_{k}){\mathrm{rank}}(P_{k})$ and 0 is not a semisimple eigenvalue of $A_{k}^{[j]}+h_{k}A_{{\mathrm{c}}k}$ for every $j=1,\ldots,q$, $k\in\overline{\mathbb{Z}}_{+}$. 
\item[$iii$)] If $\kappa_{k}\neq0$, then ${\mathrm{rank}}(A_{k}^{[j]}+h_{k}A_{{\mathrm{c}}k})=2nq-(q-1)(n-{\mathrm{rank}}(P_{k}))$ for every $j=1,\ldots,q$, $k\in\overline{\mathbb{Z}}_{+}$. In this case, for every $j=1,\ldots,q$ and every $k\in\overline{\mathbb{Z}}_{+}$, 0 is a semisimple eigenvalue of $A_{k}^{[j]}+h_{k}A_{{\mathrm{c}}k}$ if and only if ${\mathrm{rank}}(P_{k})=n$.
\end{itemize}
\end{lemma}

\begin{IEEEproof}
First, it follows from Lemma~\ref{lemma_Ah} that $\ker(A_{k}^{[j]}+h_{k}A_{{\mathrm{c}}k})=\ker(A_{k}^{[j]})$, and hence ${\mathrm{def}}(A_{k}^{[j]}+h_{k}A_{{\mathrm{c}}k})={\mathrm{def}}(A_{k}^{[j]})$ for every $j=1,\ldots,q$ and every $k\in\overline{\mathbb{Z}}_{+}$. Thus,  ${\mathrm{rank}}(A_{k}^{[j]}+h_{k}A_{{\mathrm{c}}k})=2nq+n-{\mathrm{def}}(A_{k}^{[j]}+h_{k}A_{{\mathrm{c}}k})=2nq+n-{\mathrm{def}}(A_{k}^{[j]})={\mathrm{rank}}(A_{k}^{[j]})$ for every $j=1,\ldots,q$ and every $k\in\overline{\mathbb{Z}}_{+}$.
Therefore, all the rank conclusions on $A_{k}^{[j]}+h_{k}A_{{\mathrm{c}}k}$ in $i$)--$iii$) directly follow from Lemma~\ref{lemma_Arank}.

Next, it follows from these rank conclusions on $A_{k}^{[j]}+h_{k}A_{{\mathrm{c}}k}$ that $A_{k}^{[j]}+h_{k}A_{{\mathrm{c}}k}$ has an eigenvalue 0 for every $j=1,\ldots,q$ and every $k\in\overline{\mathbb{Z}}_{+}$. Now we want to further investigate whether 0 is a semisimple eigenvalue of $A_{k}^{[j]}+h_{k}A_{{\mathrm{c}}k}$ or not for every $j=1,\ldots,q$, $k\in\overline{\mathbb{Z}}_{+}$. To this end, we need to study the relationship between $\ker(A_{k}^{[j]})$ and $\ker(A_{k}^{[j]}(A_{k}^{[j]}+h_{k}A_{{\mathrm{c}}k}))$ for every $j=1,\ldots,q$, $k\in\overline{\mathbb{Z}}_{+}$.
  
Noting that $(L_{k}\otimes P_{k})(\textbf{1}_{q\times 1}\otimes P_{k})=(L_{k}\textbf{1}_{q\times 1})\otimes P_{k}^{2}=\textbf{0}_{nq\times n}$ and by $iii$) of Lemma~\ref{lemma_EW}, $E_{n\times nq}^{[j]}(\textbf{1}_{q\times 1}\otimes P_{k})=P_{k}$, we have 
\begin{eqnarray*}
&&\hspace{-2em}(A_{k}^{[j]})^{2}=\small\left[\begin{array}{ccc}
-\mu_{k} L_{k}\otimes P_{k}-\kappa_{k} I_{q}\otimes P_{k} & -\eta_{k} L_{k}\otimes P_{k} & \kappa_{k} \textbf{1}_{q\times 1}\otimes P_{k} \\
\eta_{k}\mu_{k}(L_{k}\otimes P_{k})^{2}+\eta_{k}\kappa_{k} L_{k}\otimes P_{k}^{2}+\kappa_{k}^{2}W_{k}^{[j]} & \eta_{k}^{2}(L_{k}\otimes P_{k})^{2}-\mu_{k} L_{k}\otimes P_{k}-\kappa_{k} I_{q}\otimes P_{k} & -\kappa_{k}^{2} \textbf{1}_{q\times 1}\otimes P_{k} \\
-\kappa_{k}^{2} E_{n\times nq}^{[j]} & \kappa_{k} E_{n\times nq}^{[j]} & \kappa_{k}^{2} I_{n} \\
\end{array}\right],\\
&&\hspace{-2em}A_{k}^{[j]}A_{{\mathrm{c}}k}=\small\left[\begin{array}{ccc}
\textbf{0}_{nq\times nq} & \textbf{0}_{nq\times nq} & \textbf{0}_{nq\times n} \\
\mu_{k}^{2} (L_{k}\otimes P_{k})^{2}+2\mu_{k}\kappa_{k}(L_{k}\otimes P_{k}^{2})+\kappa_{k}^{2} (I_{q}\otimes P_{k})^{2} & \mu_{k}\eta_{k}(L_{k}\otimes P_{k})^{2}+\kappa_{k}\eta_{k}L_{k}\otimes P_{k}^{2} & -\kappa_{k}^{2} \textbf{1}_{q\times 1}\otimes P_{k}^{2} \\
-\kappa_{k}\mu_{k}E_{n\times nq}^{[j]}(L_{k}\otimes P_{k})-\kappa_{k}^{2}E_{n\times nq} ^{[j]}(I_{q}\otimes P_{k}) & -\kappa_{k}\eta_{k}E_{n\times nq}^{[j]}(L_{k}\otimes P_{k}) & \kappa_{k}^{2}P_{k} \\
\end{array}\right].
\end{eqnarray*} Thus, for every $j=1,\ldots,q$ and every $k\in\overline{\mathbb{Z}}_{+}$, let $\textbf{y}=[\textbf{y}_{1}^{\rm{T}},\textbf{y}_{2}^{\rm{T}},\textbf{y}_{3}^{\rm{T}}]^{\rm{T}}\in\ker(A_{k}^{[j]}(A_{k}^{[j]}+h_{k}A_{{\mathrm{c}}k}))$, where $\textbf{y}_{1},\textbf{y}_{2}\in\mathbb{R}^{nq}$ and $\textbf{y}_{3}\in\mathbb{R}^{n}$, we have 
\begin{eqnarray}
(-\mu_{k} L_{k}\otimes P_{k}-\kappa_{k} I_{q}\otimes P_{k})\textbf{y}_{1}-(\eta_{k} L_{k}\otimes P_{k})\textbf{y}_{2}+(\kappa_{k} \textbf{1}_{q\times 1}\otimes P_{k})\textbf{y}_{3}=\textbf{0}_{nq\times 1},\label{y_1}\\
(\eta_{k}\mu_{k}(L_{k}\otimes P_{k})^{2}+\eta_{k}\kappa_{k} L_{k}\otimes P_{k}^{2}+\kappa_{k}^{2}W_{k}^{[j]})\textbf{y}_{1}+(\eta_{k}^{2}(L_{k}\otimes P_{k})^{2}-\mu_{k} L_{k}\otimes P_{k}-\kappa_{k} I_{q}\otimes P_{k})\textbf{y}_{2}\nonumber\\
+(-\kappa_{k}^{2} \textbf{1}_{q\times 1}\otimes P_{k})\textbf{y}_{3}\nonumber\\
+h_{k}(\mu_{k}^{2} (L_{k}\otimes P_{k})^{2}+2\mu_{k}\kappa_{k}(L_{k}\otimes P_{k}^{2})+\kappa_{k}^{2} (I_{q}\otimes P_{k})^{2})\textbf{y}_{1}+h_{k}(\mu_{k}\eta_{k}(L_{k}\otimes P_{k})^{2}+\kappa_{k}\eta_{k}L_{k}\otimes P_{k}^{2})\textbf{y}_{2}\nonumber\\
+h_{k}(-\kappa_{k}^{2} \textbf{1}_{q\times 1}\otimes P_{k}^{2})\textbf{y}_{3}=\textbf{0}_{nq\times 1},\label{y_2}\\
-\kappa_{k}^{2} E_{n\times nq}^{[j]}\textbf{y}_{1}+\kappa_{k}E_{n\times nq}^{[j]}\textbf{y}_{2}+\kappa_{k}^{2}\textbf{y}_{3}+h_{k}(-\kappa_{k}\mu_{k}E_{n\times nq}^{[j]}(L_{k}\otimes P_{k})-\kappa_{k}^{2}E_{n\times nq} ^{[j]}(I_{q}\otimes P_{k}))\textbf{y}_{1}\nonumber\\
+h_{k}(-\kappa_{k}\eta_{k}E_{n\times nq}^{[j]}(L_{k}\otimes P_{k}))\textbf{y}_{2}+h_{k}\kappa_{k}^{2}P_{k}\textbf{y}_{3}=\textbf{0}_{n\times 1}.\label{y_3}
\end{eqnarray} Now we consider two cases on $\kappa_{k}$.

\textit{Case 1.} $\kappa_{k}=0$. In this case, (\ref{y_3}) becomes trivial and (\ref{y_1}) and (\ref{y_2}) become
\begin{eqnarray}
(-\mu_{k} L_{k}\otimes P_{k})\textbf{y}_{1}-(\eta_{k} L_{k}\otimes P_{k})\textbf{y}_{2}=\textbf{0}_{nq\times 1},\label{y_1k}\\
\eta_{k}\mu_{k}(L_{k}\otimes P_{k})^{2}\textbf{y}_{1}+(\eta_{k}^{2}(L_{k}\otimes P_{k})^{2}-\mu_{k} L_{k}\otimes P_{k})\textbf{y}_{2}\nonumber\\
+h_{k}\mu_{k}^{2} (L_{k}\otimes P_{k})^{2}\textbf{y}_{1}+h_{k}\mu_{k}\eta_{k}(L_{k}\otimes P_{k})^{2}\textbf{y}_{2}=\textbf{0}_{nq\times 1}.\label{y_2k}
\end{eqnarray} 

$i$) If $\mu_{k}=0$, then it follows from (\ref{y_1k}) and (\ref{y_2k}) that $-(\eta_{k}L_{k}\otimes P_{k})\textbf{y}_{2}=\textbf{0}_{nq\times 1}$ and $\eta_{k}^{2}(L_{k}\otimes P_{k})^{2}\textbf{y}_{2}=\textbf{0}_{nq\times 1}$. Hence, either $\eta_{k}=0$ or $(L_{k}\otimes P_{k})\textbf{y}_{2}=\textbf{0}_{nq\times 1}$. If $\eta_{k}=0$, then $\textbf{y}_{1},\textbf{y}_{2}\in\mathbb{R}^{nq}$ and $\textbf{y}_{3}\in\mathbb{R}^{n}$ can be chosen arbitrarily. Thus, $\ker(A_{k}^{[j]}(A_{k}^{[j]}+h_{k}A_{{\mathrm{c}}k}))=\mathbb{R}^{2nq+n}$, and it follows from $i$) of Lemma~\ref{lemma_Arank} that $\ker(A_{k}^{[j]}(A_{k}^{[j]}+h_{k}A_{{\mathrm{c}}k}))\neq\ker(A_{k}^{[j]})$. By Lemma~\ref{lemma_Ah}, we have $\ker((A_{k}^{[j]}+h_{k}A_{{\mathrm{c}}k})^{2})\neq\ker(A_{k}^{[j]}+h_{k}A_{{\mathrm{c}}k})$. Now, by Proposition 5.5.8 of \cite[p.~323]{Bernstein:2009}, 0 is not semisimple. Alternatively, if $\eta_{k}\neq0$, then $(L_{k}\otimes P_{k})\textbf{y}_{2}=\textbf{0}_{nq\times 1}$ and $\textbf{y}_{1}\in\mathbb{R}^{nq}$ and $\textbf{y}_{3}\in\mathbb{R}^{n}$ can be chosen arbitrarily. Using the similar arguments as in the proof of $iii$) of Lemma~\ref{lemma_Arank}, it follows that $\textbf{y}_{2}=\sum_{l=0}^{q-1-{\mathrm{rank}}(L_{k})}\sum_{i=1}^{n}\alpha_{li}\textbf{w}_{l}\otimes\textbf{e}_{i}+\sum_{s=1}^{q}\sum_{r=1}^{n-{\mathrm{rank}}(P_{k})}\beta_{sr}\textbf{g}_{s}\otimes\textbf{j}_{r}$, where $\alpha_{li},\beta_{sr}\in\mathbb{R}$.
Hence, $\ker(A_{k}^{[j]}(A_{k}^{[j]}+h_{k}A_{{\mathrm{c}}k}))=\{[\sum_{i=1}^{n}\sum_{s=1}^{q}\delta_{is}(\textbf{e}_{i}\otimes\textbf{e}_{s})^{\mathrm{T}},\sum_{l=0}^{q-1-{\mathrm{rank}}(L_{k})}\sum_{i=1}^{n}\alpha_{li}(\textbf{w}_{l}\otimes\textbf{e}_{i})^{\mathrm{T}}+\sum_{s=1}^{q}\sum_{r=1}^{n-{\mathrm{rank}}(P_{k})}\beta_{sr}(\textbf{g}_{s}\otimes\textbf{j}_{r})^{\mathrm{T}},\sum_{i=1}^{n}\gamma_{i}\textbf{e}_{i}^{\mathrm{T}}]^{\mathrm{T}}:\forall\alpha_{li}\in\mathbb{R},\forall\delta_{is}\in\mathbb{R},\forall\beta_{sr}\in\mathbb{R},\forall\gamma_{i}\in\mathbb{R},i=1,\ldots,n,s=1,\ldots,q,l=0,\ldots,q-1-{\mathrm{rank}}(L_{k}),r=1,\ldots,n-{\mathrm{rank}}(P_{k})\}$ for every $j=1,\ldots,q$, $k\in\overline{\mathbb{Z}}_{+}$. Clearly it follows from $i$) of Lemma~\ref{lemma_Arank} that $\ker(A_{k}^{[j]}(A_{k}^{[j]}+h_{k}A_{{\mathrm{c}}k}))\neq\ker(A_{k}^{[j]})$. By Lemma~\ref{lemma_Ah}, we have $\ker((A_{k}^{[j]}+h_{k}A_{{\mathrm{c}}k})^{2})\neq\ker(A_{k}^{[j]}+h_{k}A_{{\mathrm{c}}k})$. Now, by Proposition 5.5.8 of \cite[p.~323]{Bernstein:2009}, 0 is not semisimple.

$ii$) If $\mu_{k}\neq0$, then substituting (\ref{y_1k}) into (\ref{y_2k}) yields $-\mu_{k}(L_{k}\otimes P_{k})\textbf{y}_{2}=\textbf{0}_{nq\times 1}$. Substituting this equation into (\ref{y_1k}) yields $-\mu_{k}(L_{k}\otimes P_{k})\textbf{y}_{1}=\textbf{0}_{nq\times 1}$. Using the similar arguments as in the proof of $iii$) of Lemma~\ref{lemma_Arank}, it follows that $\textbf{y}_{1}=\sum_{l=0}^{q-1-{\mathrm{rank}}(L_{k})}\sum_{i=1}^{n}\alpha_{li}\textbf{w}_{l}\otimes\textbf{e}_{i}+\sum_{s=1}^{q}\sum_{r=1}^{n-{\mathrm{rank}}(P_{k})}\beta_{sr}\textbf{g}_{s}\otimes\textbf{j}_{r}$ and $\textbf{y}_{2}=\sum_{l=0}^{q-1-{\mathrm{rank}}(L_{k})}\sum_{i=1}^{n}\gamma_{li}\textbf{w}_{l}\otimes\textbf{e}_{i}+\sum_{s=1}^{q}\sum_{r=1}^{n-{\mathrm{rank}}(P_{k})}\delta_{sr}\textbf{g}_{s}\otimes\textbf{j}_{r}$, where $\alpha_{li},\beta_{sr},\gamma_{li},\delta_{sr}\in\mathbb{R}$. Note that $\textbf{y}_{3}\in\mathbb{R}^{n}$ can be chosen arbitrarily, and hence, $\ker(A_{k}^{[j]}(A_{k}^{[j]}+h_{k}A_{{\mathrm{c}}k}))=\{[\sum_{l=0}^{q-1-{\mathrm{rank}}(L_{k})}\sum_{i=1}^{n}\alpha_{li}(\textbf{w}_{l}\otimes\textbf{e}_{i})^{\mathrm{T}}+\sum_{s=1}^{q}\sum_{r=1}^{n-{\mathrm{rank}}(P_{k})}\beta_{sr}(\textbf{g}_{s}\otimes\textbf{j}_{r})^{\mathrm{T}},\sum_{l=0}^{q-1-{\mathrm{rank}}(L_{k})}\sum_{i=1}^{n}\gamma_{li}(\textbf{w}_{l}\otimes\textbf{e}_{i})^{\mathrm{T}}+\sum_{s=1}^{q}\sum_{r=1}^{n-{\mathrm{rank}}(P_{k})}\delta_{sr}(\textbf{g}_{s}\otimes\textbf{j}_{r})^{\mathrm{T}},\sum_{i=1}^{n}\zeta_{i}\textbf{e}_{i}^{\mathrm{T}}]^{\mathrm{T}}:\forall\alpha_{li}\in\mathbb{R},\forall\beta_{sr}\in\mathbb{R},\forall\gamma_{li}\in\mathbb{R},\forall\delta_{sr}\in\mathbb{R},\forall\zeta_{i}\in\mathbb{R},i=1,\ldots,n,l=0,\ldots,q-1-{\mathrm{rank}}(L_{k}),s=1,\ldots,q,r=1,\ldots,n-{\mathrm{rank}}(P_{k})\}$ for every $j=1,\ldots,q$, $k\in\overline{\mathbb{Z}}_{+}$. Clearly it follows from $iv$) of Lemma~\ref{lemma_Arank} that $\ker(A_{k}^{[j]}(A_{k}^{[j]}+h_{k}A_{{\mathrm{c}}k}))\neq\ker(A_{k}^{[j]})$. By Lemma~\ref{lemma_Ah}, we have $\ker((A_{k}^{[j]}+h_{k}A_{{\mathrm{c}}k})^{2})\neq\ker(A_{k}^{[j]}+h_{k}A_{{\mathrm{c}}k})$. Now, by Proposition 5.5.8 of \cite[p.~323]{Bernstein:2009}, 0 is not semisimple.

\textit{Case 2.} $\kappa_{k}\neq0$. In this case, substituting (\ref{y_1}) into (\ref{y_2}) and (\ref{y_3}) yields
\begin{eqnarray}
(\eta_{k}\mu_{k}(L_{k}\otimes P_{k})^{2}+\eta_{k}\kappa_{k} L_{k}\otimes P_{k}^{2}+\kappa_{k}^{2}W_{k}^{[j]})\textbf{y}_{1}+(\eta_{k}^{2}(L_{k}\otimes P_{k})^{2}-\mu_{k} L_{k}\otimes P_{k}-\kappa_{k} I_{q}\otimes P_{k})\textbf{y}_{2}\nonumber\\
+(-\kappa_{k}^{2} \textbf{1}_{q\times 1}\otimes P_{k})\textbf{y}_{3}\nonumber\\
+h_{k}(\mu_{k}^{2} (L_{k}\otimes P_{k})^{2}+\mu_{k}\kappa_{k}(L_{k}\otimes P_{k}^{2}))\textbf{y}_{1}+h_{k}\mu_{k}\eta_{k}(L_{k}\otimes P_{k})^{2}\textbf{y}_{2}=\textbf{0}_{nq\times 1},\label{y12p}\\
-\kappa_{k}^{2} E_{n\times nq}^{[j]}\textbf{y}_{1}+\kappa_{k}E_{n\times nq}^{[j]}\textbf{y}_{2}+\kappa_{k}^{2}\textbf{y}_{3}=\textbf{0}_{n\times 1}.\label{y13p}
\end{eqnarray}
Note that $(L_{k}\otimes P_{k})W_{k}^{[j]}=(L_{k}\otimes P_{k})(\textbf{1}_{q\times 1}\otimes P_{k}E_{n\times nq}^{[j]})=L_{k}\textbf{1}_{q\times 1}\otimes P_{k}^{2}E_{n\times nq}^{[j]}=\textbf{0}_{q\times 1}\otimes P_{k}^{2}E_{n\times nq}^{[j]}=\textbf{0}_{nq\times nq}$. Pre-multiplying $-L_{k}\otimes P_{k}$ on both sides of (\ref{y_1}) yields $(\mu_{k}(L_{k}\otimes P_{k})^{2}+\kappa_{k} L_{k}\otimes P_{k}^{2})\textbf{y}_{1}+\eta_{k}(L_{k}\otimes P_{k})^{2}\textbf{y}_{2}=\textbf{0}_{nq\times 1}$. Substituting this equation into (\ref{y12p}) yields
\begin{eqnarray}
\kappa_{k}^{2}W_{k}^{[j]}\textbf{y}_{1}+(-\mu_{k} L_{k}\otimes P_{k}-\kappa_{k} I_{q}\otimes P_{k})\textbf{y}_{2}+(-\kappa_{k}^{2} \textbf{1}_{q\times 1}\otimes P_{k})\textbf{y}_{3}=\textbf{0}_{nq\times 1}.\label{y12pp}
\end{eqnarray} Finally, substituting (\ref{y13p}) into (\ref{y_1}) and (\ref{y12pp}) by eliminating $\textbf{y}_{3}$ yields
\begin{eqnarray}
(-\mu_{k} L_{k}\otimes P_{k}-\kappa_{k} I_{q}\otimes P_{k}+\kappa_{k} W_{k}^{[j]})\textbf{y}_{1}-(\eta_{k} L_{k}\otimes P_{k}+W_{k}^{[j]})\textbf{y}_{2}=\textbf{0}_{nq\times 1},\label{y12}\\
(-\mu_{k} L_{k}\otimes P_{k}-\kappa_{k}I_{q}\otimes P_{k}+\kappa_{k}W_{k}^{[j]})\textbf{y}_{2}=\textbf{0}_{nq\times 1}.\label{y21}
\end{eqnarray} 

$iii$) To show that 0 is a semisimple eigenvalue of $A_{k}^{[j]}+h_{k}A_{{\mathrm{c}}k}$ if ${\mathrm{rank}}(P_{k})=n$, we consider two cases on $\mu_{k}$. If $\mu_{k}=0$, then note that (\ref{y21}) is identical to (\ref{z1p}). Then it follows from the similar arguments as in the proof of $ii$) of Lemma~\ref{lemma_Arank} that $\textbf{y}_{2}=\sum_{i=1}^{n}\alpha_{i}\textbf{1}_{q\times 1}\otimes\textbf{e}_{i}+\sum_{s=1}^{q}\sum_{r=1}^{n-{\mathrm{rank}}(P_{k})}\beta_{sr}\textbf{g}_{s}\otimes\textbf{j}_{r}-\sum_{r=1}^{n-{\mathrm{rank}}(P_{k})}\beta_{jr}\textbf{g}_{j}\otimes\textbf{j}_{r}$ for some $\alpha_{i},\beta_{sr}\in\mathbb{R}$. Clearly $\textbf{y}_{2}\in\ker(L_{k}\otimes I_{n})+\ker(I_{q}\otimes P_{k})=\ker(L_{k}\otimes P_{k})$. Next, using this property of $\textbf{y}_{2}$ in $(\mu_{k}(L_{k}\otimes P_{k})^{2}+\kappa_{k} L_{k}\otimes P_{k}^{2})\textbf{y}_{1}+\eta_{k}(L_{k}\otimes P_{k})^{2}\textbf{y}_{2}=\textbf{0}_{nq\times 1}$ yields $(\mu_{k}(L_{k}\otimes P_{k})^{2}+\kappa_{k} L_{k}\otimes P_{k}^{2})\textbf{y}_{1}=(\mu_{k}L_{k}\otimes I_{n}+\kappa_{k}I_{nq})(L_{k}\otimes P_{k}^{2})\textbf{y}_{1}=\textbf{0}_{nq\times 1}$, i.e., $(L_{k}\otimes P_{k}^{2})\textbf{y}_{1}=(I_{q}\otimes P_{k})^{2}(L_{k}\otimes I_{n})\textbf{y}_{1}=\textbf{0}_{nq\times 1}$. Since by assumption $P_{k}$ is a full rank matrix, it follows that $I_{q}\otimes P_{k}$ is nonsingular. Hence, $(L_{k}\otimes I_{n})\textbf{y}_{1}=\textbf{0}_{nq\times 1}$. Substituting this relationship into (\ref{y12}) yields $(-\kappa_{k} I_{q}\otimes P_{k}+\kappa_{k} W_{k}^{[j]})\textbf{y}_{1}-W_{k}^{[j]}\textbf{y}_{2}=\textbf{0}_{nq\times 1}$.  Clearly it follows from (\ref{y21}) that $W_{k}^{[j]}\textbf{y}_{2}=(I_{q}\otimes P_{k})\textbf{y}_{2}$ for every $j=1,\ldots,q$. Then
\begin{eqnarray}\label{y1y2}
(-\kappa_{k}I_{q}\otimes P_{k}+\kappa_{k}W_{k}^{[j]})\textbf{y}_{1}-(I_{q}\otimes P_{k})\textbf{y}_{2}=\textbf{0}_{nq\times 1}. 
\end{eqnarray}
Since $(L_{k}\otimes I_{n})\textbf{y}_{1}=\textbf{0}_{nq\times 1}$, it follows that $\textbf{y}_{1}=\sum_{l=0}^{q-1-{\mathrm{rank}}(L_{k})}\sum_{i=1}^{n}\gamma_{li}\textbf{w}_{l}\otimes\textbf{e}_{i}$ for some $\gamma_{li}\in\mathbb{R}$. Now substituting these explicit expressions $\textbf{y}_{1}$ and $\textbf{y}_{2}$ into (\ref{y1y2}) together with $iii$) of Lemma~\ref{lemma_EW} yields
\begin{eqnarray}\label{ldwe}
(I_{q}\otimes P_{k})\Big(-\kappa_{k}\sum_{l=0}^{q-1-{\mathrm{rank}}(L_{k})}\sum_{i=1}^{n}\gamma_{li}\textbf{w}_{l}\otimes \textbf{e}_{i}+\kappa_{k}\sum_{l=0}^{q-1-{\mathrm{rank}}(L_{k})}\sum_{i=1}^{n}\gamma_{li}w_{lj}\textbf{w}_{0}\otimes \textbf{e}_{i}-\sum_{i=1}^{n}\alpha_{i}\textbf{w}_{0}\otimes\textbf{e}_{i}\Big)=\textbf{0}_{nq\times 1}.
\end{eqnarray} Note that $\textbf{w}_{l}\otimes \textbf{e}_{i}$, $l=0,1,\ldots,q-1-{\mathrm{rank}}(L_{k})$, $i=1,\ldots,n$, are linearly independent and $w_{0j}=1$, it follows from (\ref{ldwe}) that $\gamma_{li}=0$ and $\alpha_{i}=0$ for every $l=1,2,\ldots,q-1-{\mathrm{rank}}(L_{k})$ and every $i=1,\ldots,n$. Finally, since $\ker(P_{k})=\{\textbf{0}_{n\times 1}\}$, it follows that $\textbf{j}_{r}=\textbf{0}_{n\times 1}$. Hence, $\textbf{y}_{2}=\textbf{0}_{nq\times 1}$ and $\textbf{y}_{1}=\sum_{i=1}^{n}\gamma_{0i}\textbf{w}_{0}\otimes\textbf{e}_{i}$. Now it follows from (\ref{y13p}) and $iii$) of Lemma~\ref{lemma_EW} that $\textbf{y}_{3}=E_{n\times nq}^{[j]}\textbf{y}_{1}=\sum_{i=1}^{n}\gamma_{0i}E_{n\times nq}^{[j]}(\textbf{1}_{q\times 1}\otimes\textbf{e}_{i})=\sum_{i=1}^{n}\gamma_{0i}\textbf{e}_{i}$. Clearly such $\textbf{y}_{1}=\sum_{i=1}^{n}\gamma_{0i}\textbf{1}_{q\times 1}\otimes\textbf{e}_{i}$, $\textbf{y}_{2}=\textbf{0}_{nq\times 1}$, and $\textbf{y}_{3}=\sum_{i=1}^{n}\gamma_{0i}\textbf{e}_{i}$ satisfy (\ref{y_1})--(\ref{y_3}). Thus, $\ker(A_{k}^{[j]}(A_{k}^{[j]}+h_{k}A_{{\mathrm{c}}k}))=\{[\sum_{i=1}^{n}\gamma_{0i}(\textbf{1}_{q\times 1}\otimes\textbf{e}_{i})^{\mathrm{T}},\textbf{0}_{1\times nq},\sum_{i=1}^{n}\gamma_{0i}\textbf{e}_{i}^{\mathrm{T}}]^{\mathrm{T}}:\forall\gamma_{0i}\in\mathbb{R},i=1,\ldots,n\}=\ker(A_{k}^{[j]})$, where the last step follows from $ii$) of Lemma~\ref{lemma_Arank} with ${\mathrm{rank}}(P_{k})=n$. By Lemma~\ref{lemma_Ah}, we have $\ker((A_{k}^{[j]}+h_{k}A_{{\mathrm{c}}k})^{2})=\ker(A_{k}^{[j]}+h_{k}A_{{\mathrm{c}}k})$. Now, by Proposition 5.5.8 of \cite[p.~323]{Bernstein:2009}, 0 is semisimple.

If $\mu_{k}\neq0$, then note that (\ref{y21}) is identical to (\ref{z1}). Next, it follows from the similar arguments as in the proof of $iii$) of Lemma~\ref{lemma_Arank} that $\textbf{y}_{2}=\sum_{l=0}^{q-1-{\mathrm{rank}}(L_{k})}\sum_{i=1}^{n}\alpha_{li}\textbf{w}_{l}\otimes\textbf{e}_{i}+\sum_{s=1}^{q}\sum_{r=1}^{n-{\mathrm{rank}}(P_{k})}\beta_{sr}\textbf{g}_{s}\otimes\textbf{j}_{r}$, where $\alpha_{li},\beta_{sr}\in\mathbb{R}$ and (\ref{Pka}) holds. Since $P_{k}$ is a full rank matrix, it follows from (\ref{Pka}) that $\alpha_{li}=0$ and $\textbf{j}_{r}=\textbf{0}_{n\times 1}$ for every $l=0,1,\ldots,q-1-{\mathrm{rank}}(L_{k})$, every $i=1,\ldots,n$, and every $r=1,\ldots,n-{\mathrm{rank}}(P_{k})$, which implies that $\textbf{y}_{2}=\textbf{0}_{nq\times 1}$. Again, it follows from the similar arguments as above that $(L_{k}\otimes I_{n})\textbf{y}_{1}=\textbf{0}_{nq\times 1}$ and hence, $\textbf{y}_{1}=\sum_{i=1}^{n}\gamma_{i}\textbf{1}_{q\times 1}\otimes\textbf{e}_{i}$, where $\gamma_{i}\in\mathbb{R}$. Then it follows from (\ref{y13p}) and $iii$) of Lemma~\ref{lemma_EW} that $\textbf{y}_{3}=E_{n\times nq}^{[j]}\textbf{y}_{1}=\sum_{i=1}^{n}\gamma_{i}E_{n\times nq}^{[j]}(\textbf{1}_{q\times 1}\otimes\textbf{e}_{i})=\sum_{i=1}^{n}\gamma_{i}\textbf{e}_{i}$. Clearly such $\textbf{y}_{1}=\sum_{i=1}^{n}\gamma_{i}\textbf{1}_{q\times 1}\otimes\textbf{e}_{i}$, $\textbf{y}_{2}=\textbf{0}_{nq\times 1}$, and $\textbf{y}_{3}=\sum_{i=1}^{n}\gamma_{i}\textbf{e}_{i}$ satisfy (\ref{y_1})--(\ref{y_3}). Thus, $\ker(A_{k}^{[j]}(A_{k}^{[j]}+h_{k}A_{{\mathrm{c}}k}))=\{[\sum_{i=1}^{n}\gamma_{i}(\textbf{1}_{q\times 1}\otimes\textbf{e}_{i})^{\mathrm{T}},\textbf{0}_{1\times nq},\sum_{i=1}^{n}\gamma_{i}\textbf{e}_{i}^{\mathrm{T}}]^{\mathrm{T}}:\forall\gamma_{i}\in\mathbb{R},i=1,\ldots,n\}=\ker(A_{k}^{[j]})$, where the last step follows from $iii$) of Lemma~\ref{lemma_Arank} with ${\mathrm{rank}}(P_{k})=n$. By Lemma~\ref{lemma_Ah}, we have $\ker((A_{k}^{[j]}+h_{k}A_{{\mathrm{c}}k})^{2})=\ker(A_{k}^{[j]}+h_{k}A_{{\mathrm{c}}k})$. Now, by Proposition 5.5.8 of \cite[p.~323]{Bernstein:2009}, 0 is semisimple.

To show that 0 is a semisimple eigenvalue of $A_{k}^{[j]}+h_{k}A_{{\mathrm{c}}k}$ only if ${\mathrm{rank}}(P_{k})=n$, conversely we assume that this is not true, that is, ${\mathrm{rank}}(P_{k})<n$. We first claim that a specific solution $\textbf{y}_{2}$ to (\ref{y21}) is given by the form $\textbf{y}_{2}=\sum_{i=1}^{n}\alpha_{i}\textbf{1}_{q\times 1}\otimes\textbf{e}_{i}$, where $\alpha_{i}\in\mathbb{R}$. Indeed this is clear from $ii$) of Lemma~\ref{lemma_EW}. Next, we claim that a specific solution $\textbf{y}_{1}$ and $\textbf{y}_{2}$ to (\ref{y12}) and (\ref{y21}) is given by the form $\textbf{y}_{1}=\sum_{i=1}^{n}\gamma_{i}\textbf{1}_{q\times 1}\otimes\textbf{e}_{i}$ and $\textbf{y}_{2}=\sum_{i=1}^{n}\alpha_{i}\textbf{1}_{q\times 1}\otimes\textbf{e}_{i}$, where $\gamma_{i},\alpha_{i}\in\mathbb{R}$ satisfy
\begin{eqnarray}\label{Pka1}
P_{k}\small\left[\begin{array}{c}
\alpha_{1}\\
\vdots\\
\alpha_{n}\\
\end{array}\right]=\textbf{0}_{n\times 1}.
\end{eqnarray} To see this, substituting $\textbf{y}_{1}=\sum_{i=1}^{n}\gamma_{i}\textbf{1}_{q\times 1}\otimes\textbf{e}_{i}$ and $\textbf{y}_{2}=\sum_{i=1}^{n}\alpha_{i}\textbf{1}_{q\times 1}\otimes\textbf{e}_{i}$ into (\ref{y12}) together with $(L_{k}\otimes P_{k})(\textbf{1}_{q\times 1}\otimes\textbf{e}_{i})=\textbf{0}_{nq\times 1}$ yields $\sum_{i=1}^{n}\alpha_{i}P_{k}\textbf{e}_{i}=\textbf{0}_{n\times 1}$, which is equivalent to (\ref{Pka1}). In this case, it follows from (\ref{y13p}) and $iii$) of Lemma~\ref{lemma_EW} that $\textbf{y}_{3}=E_{n\times nq}^{[j]}\textbf{y}_{1}=\sum_{i=1}^{n}\gamma_{i}E_{n\times nq}^{[j]}(\textbf{1}_{q\times 1}\otimes\textbf{e}_{i})=\sum_{i=1}^{n}\gamma_{i}\textbf{e}_{i}$. Clearly such $\textbf{y}_{1}=\sum_{i=1}^{n}\gamma_{i}\textbf{1}_{q\times 1}\otimes\textbf{e}_{i}$, $\textbf{y}_{2}=\sum_{i=1}^{n}\alpha_{i}\textbf{1}_{q\times 1}\otimes\textbf{e}_{i}$, and $\textbf{y}_{3}=\sum_{i=1}^{n}\gamma_{i}\textbf{e}_{i}$ satisfy (\ref{y_1})--(\ref{y_3}). Since by assumption, ${\mathrm{rank}}(P_{k})<n$, it follows that (\ref{Pka1}) has nontrivial solutions, which implies that $\textbf{y}_{2}\not\equiv\textbf{0}_{nq\times 1}$. Thus, it follows from $ii$) and $iii$) of Lemma~\ref{lemma_Arank} that $\ker(A_{k}^{[j]})\neq\ker(A_{k}^{[j]}(A_{k}^{[j]}+h_{k}A_{{\mathrm{c}}k}))$, which implies that $\ker((A_{k}^{[j]}+h_{k}A_{{\mathrm{c}}k})^{2})\neq\ker(A_{k}^{[j]}+h_{k}A_{{\mathrm{c}}k})$. Now, by Proposition 5.5.8 of \cite[p.~323]{Bernstein:2009}, 0 is not semisimple, which contradicts the original condition that 0 is a semisimple eigenvalue of $A_{k}^{[j]}+h_{k}A_{{\mathrm{c}}k}$. Hence in this case, if 0 is a semisimple eigenvalue of $A_{k}^{[j]}+h_{k}A_{{\mathrm{c}}k}$, then ${\mathrm{rank}}(P_{k})=n$.
\end{IEEEproof}

It follows from Lemma~\ref{lemma_semisimple} that for every $j=1,\ldots,q$, 0 is a semisimple eigenvalue of $A_{k}^{[j]}+h_{k}A_{{\mathrm{c}}k}$ defined in Lemma~\ref{lemma_Ah}, where $\mu_{k},\kappa_{k},\eta_{k},h_{k}\geq0$, if and only if $\kappa_{k}\neq0$ and ${\mathrm{rank}}(P_{k})=n$, $k\in\overline{\mathbb{Z}}_{+}$.
To proceed, let $\mathbb{C}^{n}$ (respectively $\mathbb{C}^{m\times n}$) denote the set of complex vectors (respectively matrices). Using Lemmas \ref{lemma_EW}--\ref{lemma_semisimple}, one can show the following complete result about the nonzero eigenvalue and eigenspace structures of $A_{k}^{[j]}+h_{k}A_{{\mathrm{c}}k}$. 

\begin{lemma}\label{lemma_A}
Consider the (possibly infinitely many) matrices $A_{k}^{[j]}+h_{k}A_{{\mathrm{c}}k}$, $j=1,\ldots,q$, $k=0,1,2,\ldots$, defined by (\ref{Amatrix}) in Lemma~\ref{lemma_Arank} and (\ref{Ac}) in Lemma~\ref{lemma_semisimple}, where $\mu_{k},\kappa_{k},\eta_{k}\geq0$ and $h_{k}>0$, $k\in\overline{\mathbb{Z}}_{+}$. Assume that ${\mathrm{rank}}(P_{k})=n$, $k\in\overline{\mathbb{Z}}_{+}$.
\begin{itemize}
\item[$i$)] Then for every $j=1,\ldots,q$, ${\mathrm{spec}}(A_{k}^{[j]}+h_{k}A_{{\mathrm{c}}k})\subseteq\{0,-\kappa_{k},-\frac{\kappa_{k}(1+h_{k})}{2}\pm\frac{1}{2}\sqrt{\kappa_{k}^{2}(1+h_{k})^{2}-4\kappa_{k}},\lambda\in\mathbb{C}:\forall \frac{\lambda^{2}+\kappa_{k} h_{k}\lambda+\kappa_{k}}{\eta_{k}\lambda+\mu_{k} h_{k}\lambda+\mu_{k}}\in{\mathrm{spec}}(-L_{k})\}=\{0,-\kappa_{k},-\frac{\kappa_{k}(1+h_{k})}{2}\pm\frac{1}{2}\sqrt{\kappa_{k}^{2}(1+h_{k})^{2}-4\kappa_{k}},-\frac{\kappa_{k}h_{k}}{2}\pm\frac{1}{2}\sqrt{\kappa_{k}^{2}h_{k}^{2}-4\kappa_{k}},\lambda\\\in\mathbb{C}:\forall \frac{\lambda^{2}+\kappa_{k} h_{k}\lambda+\kappa_{k}}{\eta_{k}\lambda+\mu_{k} h_{k}\lambda+\mu_{k}}\in{\mathrm{spec}}(-L_{k})\backslash\{0\}\}$. 
\item[$ii$)] If $1\not\in{\mathrm{spec}}((\frac{\mu_{k}}{\lambda_{1,2}\kappa_{k}}+\frac{\mu_{k} h_{k}}{\kappa_{k}}+\frac{\eta_{k}}{\kappa_{k}})L_{k})$, then $\lambda_{1,2}=-\frac{\kappa_{k}(1+h_{k})}{2}\pm\frac{1}{2}\sqrt{\kappa_{k}^{2}(1+h_{k})^{2}-4\kappa_{k}}$ are the eigenvalues of $A_{k}^{[j]}+h_{k}A_{{\mathrm{c}}k}$. The corresponding eigenspace is given by 
\begin{eqnarray}\label{egns1}
&&\hspace{-2em}\ker\Big(A_{k}^{[j]}+h_{k}A_{{\mathrm{c}}k}-\lambda_{1,2} I_{2nq+n}\Big)\nonumber\\
&&\hspace{-2em}=\Big\{\Big[\frac{1+h_{k}\lambda_{1,2}^{*}}{\lambda_{1,2}^{*}}\lambda\sum_{l=0}^{q-1-{\mathrm{rank}}(L_{k})}\sum_{i=1}^{n}\omega_{li}(\textbf{w}_{l}\otimes\textbf{e}_{i})^{\mathrm{T}},\lambda\sum_{l=0}^{q-1-{\mathrm{rank}}(L_{k})}\sum_{i=1}^{n}\omega_{li}(\textbf{w}_{l}\otimes\textbf{e}_{i})^{\mathrm{T}},\nonumber\\
&&-\lambda\sum_{l=0}^{q-1-{\mathrm{rank}}(L_{k})}\sum_{i=1}^{n}\omega_{li}w_{lj}\textbf{e}_{i}^{\mathrm{T}}\Big]^{*}:\forall\omega_{li}\in\mathbb{C},i=1,\ldots,n,l=0,1,\ldots,q-1-{\mathrm{rank}}(L_{k})\Big\},
\end{eqnarray} where $\textbf{x}^{*}$ denotes the complex conjugate transpose of $\textbf{x}\in\mathbb{C}^{n}$.
\item[$iii$)] If $1\in{\mathrm{spec}}((\frac{\mu_{k}}{\lambda_{1,2}\kappa_{k}}+\frac{\mu_{k} h_{k}}{\kappa_{k}}+\frac{\eta_{k}}{\kappa_{k}})L_{k})$, and $h_{k}\kappa_{k}\neq 1$, then $\lambda_{1,2}=-\frac{\kappa_{k}(1+h_{k})}{2}\pm\frac{1}{2}\sqrt{\kappa_{k}^{2}(1+h_{k})^{2}-4\kappa_{k}}$ are the eigenvalues of $A_{k}^{[j]}+h_{k}A_{{\mathrm{c}}k}$. The corresponding eigenspace is given by 
\begin{eqnarray}\label{egns2}
&&\ker\Big(A_{k}^{[j]}+h_{k}A_{{\mathrm{c}}k}-\lambda_{1,2} I_{2nq+n}\Big)\nonumber\\
&&=\Big\{\Big[\frac{1+h_{k}\lambda_{1,2}^{*}}{\lambda_{1,2}^{*}}\sum_{i=1}^{n}\sum_{l=1}^{q}\varpi_{li}((\textbf{g}_{l}-G_{k}^{+}G_{k}\textbf{g}_{l})\otimes \textbf{e}_{i})^{\mathrm{T}}-\frac{1+h_{k}\lambda_{1,2}^{*}}{\kappa_{k}\lambda_{1,2}^{*}}\sum_{i=1}^{n}\omega_{0i}(\textbf{1}_{q\times 1}\otimes\textbf{e}_{i})^{\mathrm{T}},\nonumber\\
&&\sum_{i=1}^{n}\sum_{l=1}^{q}\varpi_{li}((\textbf{g}_{l}-G_{k}^{+}G_{k}\textbf{g}_{l})\otimes \textbf{e}_{i})^{\mathrm{T}}-\frac{1}{\kappa_{k}}\sum_{i=1}^{n}\omega_{0i}(\textbf{1}_{q\times 1}\otimes\textbf{e}_{i})^{\mathrm{T}},\nonumber\\
&&\frac{\kappa_{k}+\kappa_{k}h_{k}\lambda_{1,2}^{*}}{\lambda_{1,2}^{*}(\lambda_{1,2}^{*}+\kappa_{k})}\sum_{i=1}^{n}\sum_{l=1}^{q}\varpi_{li}(\textbf{g}_{j}^{\mathrm{T}}\textbf{g}_{l}-\textbf{g}_{j}^{\mathrm{T}}G_{k}^{+}G_{k}\textbf{g}_{l})\textbf{e}_{i}^{\mathrm{T}}-\frac{1+h_{k}\lambda_{1,2}^{*}}{\lambda_{1,2}^{*}(\lambda_{1,2}^{*}+\kappa_{k})}\sum_{i=1}^{n}\omega_{0i}\textbf{e}_{i}^{\mathrm{T}}\Big]^{*}:\nonumber\\
&&\forall\omega_{0i}\in\mathbb{C},\forall\varpi_{li}\in\mathbb{C},i=1,\ldots,n,l=1,\ldots,q\Big\},
\end{eqnarray} where $G_{k}=(\frac{\mu_{k}}{\lambda_{1,2}}+\mu_{k} h_{k}+\eta_{k})L_{k}-\kappa_{k}I_{q}$.
\item[$iv$)] If $\frac{\kappa_{k}}{\lambda_{4}}+\lambda_{4}+\kappa_{k} h_{k}\neq0$, $\lambda_{4}\neq-\kappa_{k}$, $\frac{\mu_{k}}{\lambda_{4}}+\mu_{k} h_{k}+\eta_{k}\neq0$, and $\frac{\lambda_{4}^{2}+\kappa_{k} h_{k}\lambda_{4}+\kappa_{k}}{\eta_{k}\lambda+\mu_{k} h_{k}\lambda_{4}+\mu_{k}}\in{\mathrm{spec}}(-L_{k})$, then $\lambda_{4}$ are the eigenvalues of $A_{k}^{[j]}+h_{k}A_{{\mathrm{c}}k}$. The corresponding eigenspace is given by 
\begin{eqnarray}\label{egns3}
&&\hspace{-3em}\ker\Big(A_{k}^{[j]}+h_{k}A_{{\mathrm{c}}k}-\lambda_{4} I_{2nq+n}\Big)\nonumber\\
&&\hspace{-3em}=\Big\{\Big[\frac{1+h_{k}\lambda_{4}^{*}}{\lambda_{4}^{*}}\sum_{i=1}^{n}\sum_{l=1}^{q}\varpi_{li}\Big(\textbf{g}_{l}-F_{k}^{+}F_{k}\textbf{g}_{l}+\frac{\kappa_{k}^{2}(1+h_{k}\lambda_{4})}{\lambda_{4}(\lambda_{4}+\kappa_{k})}(\textbf{g}_{j}^{\mathrm{T}}F_{k}\textbf{g}_{l})F_{k}^{+}\psi_{k}-\frac{\kappa_{k}^{2}(1+h_{k}\lambda_{4})}{\lambda_{4}(\lambda_{4}+\kappa_{k})}(\textbf{g}_{j}^{\mathrm{T}}\textbf{g}_{l})\psi_{k}\Big)^{*}\otimes \textbf{e}_{i}^{\mathrm{T}},\nonumber\\
&&\hspace{-3em}\sum_{i=1}^{n}\sum_{l=1}^{q}\varpi_{li}\Big(\textbf{g}_{l}-F_{k}^{+}F_{k}\textbf{g}_{l}+\frac{\kappa_{k}^{2}(1+h_{k}\lambda_{4})}{\lambda_{4}(\lambda_{4}+\kappa_{k})}(\textbf{g}_{j}^{\mathrm{T}}F_{k}\textbf{g}_{l})F_{k}^{+}\psi_{k}-\frac{\kappa_{k}^{2}(1+h_{k}\lambda_{4})}{\lambda_{4}(\lambda_{4}+\kappa_{k})}(\textbf{g}_{j}^{\mathrm{T}}\textbf{g}_{l})\psi_{k}\Big)^{*}\otimes \textbf{e}_{i}^{\mathrm{T}},\nonumber\\
&&\hspace{-3em}\frac{\kappa_{k}+\kappa_{k}h_{k}\lambda_{4}^{*}}{\lambda_{4}^{*}(\lambda_{4}^{*}+\kappa_{k})}\sum_{i=1}^{n}\sum_{l=1}^{q}\varpi_{li}\Big(\textbf{g}_{j}^{\mathrm{T}}\textbf{g}_{l}-\textbf{g}_{j}^{\mathrm{T}}F_{k}^{+}F_{k}\textbf{g}_{l}+\frac{\kappa_{k}^{2}(1+h_{k}\lambda_{4})}{\lambda_{4}(\lambda_{4}+\kappa_{k})}(\textbf{g}_{j}^{\mathrm{T}}F_{k}\textbf{g}_{l})\textbf{g}_{j}^{\mathrm{T}}F_{k}^{+}\psi_{k}\nonumber\\
&&\hspace{-3em}-\frac{\kappa_{k}^{2}(1+h_{k}\lambda_{4})}{\lambda_{4}(\lambda_{4}+\kappa_{k})}(\textbf{g}_{j}^{\mathrm{T}}\textbf{g}_{l})\textbf{g}_{j}^{\mathrm{T}}\psi_{k}\Big)^{*}\otimes \textbf{e}_{i}^{\mathrm{T}}\Big]^{*}:\varpi_{li}\in\mathbb{C},i=1,\ldots,n,l=1,\ldots,q\Big\},
\end{eqnarray} where $F_{k}=(\frac{\mu_{k}}{\lambda_{4}}+\mu_{k} h_{k}+\eta_{k})L_{k}+(\frac{\kappa_{k}}{\lambda_{4}}+\lambda_{4}+\kappa_{k} h_{k})I_{q}$ and
\begin{eqnarray}\label{psik}
\psi_{k}=\small\left\{\begin{array}{ll}
(\frac{\kappa_{k}^{2}(1+h_{k}\lambda_{4})}{\lambda_{4}(\lambda_{4}+\kappa_{k})}\textbf{g}_{j}^{\mathrm{T}}-\frac{\kappa_{k}^{2}(1+h_{k}\lambda_{4})}{\lambda_{4}(\lambda_{4}+\kappa_{k})}\textbf{g}_{j}^{\mathrm{T}}F_{k}^{+}F_{k})^{+}, & \frac{\kappa_{k}^{2}(1+h_{k}\lambda_{4}^{*})}{\lambda_{4}^{*}(\lambda_{4}^{*}+\kappa_{k})}\textbf{g}_{j}\neq \frac{\kappa_{k}^{2}(1+h_{k}\lambda_{4}^{*})}{\lambda_{4}^{*}(\lambda_{4}^{*}+\kappa_{k})}F_{k}^{+}F_{k}\textbf{g}_{j},\\
\frac{\kappa_{k}^{2}(1+h_{k}\lambda_{4})}{\lambda_{4}(\lambda_{4}+\kappa_{k})}(1+|\frac{\kappa_{k}^{2}(1+h_{k}\lambda_{4})}{\lambda_{4}(\lambda_{4}+\kappa_{k})}|^{2}\textbf{g}_{j}^{\mathrm{T}}(F_{k}^{\mathrm{T}}F_{k})^{+}\textbf{g}_{j})^{-1}(F_{k}^{\mathrm{T}}F_{k})^{+}\textbf{g}_{j}, & \frac{\kappa_{k}^{2}(1+h_{k}\lambda_{4}^{*})}{\lambda_{4}^{*}(\lambda_{4}^{*}+\kappa_{k})}\textbf{g}_{j}= \frac{\kappa_{k}^{2}(1+h_{k}\lambda_{4}^{*})}{\lambda_{4}^{*}(\lambda_{4}^{*}+\kappa_{k})}F_{k}^{+}F_{k}\textbf{g}_{j}. \\
\end{array}\right.
\end{eqnarray}
\item[$v$)] If $\frac{\mu_{k}}{\lambda_{5,6}}+\mu_{k} h_{k}+\eta_{k}\neq0$, $\lambda_{5,6}\neq-\kappa_{k}$, and $\frac{\kappa_{k}}{\lambda_{5,6}}+\lambda_{5,6}+\kappa_{k} h_{k}=0$, then $\lambda_{5,6}=-\frac{\kappa_{k}h_{k}}{2}\pm\frac{1}{2}\sqrt{\kappa_{k}^{2}h_{k}^{2}-4\kappa_{k}}$ are the eigenvalues of $A_{k}^{[j]}+h_{k}A_{{\mathrm{c}}k}$. The corresponding eigenspace is given by the form (\ref{egns3}) with $\lambda_{4}$ being replaced by $\lambda_{5,6}$.
\item[$vi$)] If $\frac{\mu_{k}}{\lambda_{5,6}}+\mu_{k} h_{k}+\eta_{k}=0$, $\lambda_{5,6}\neq-\kappa_{k}$, $\mu_{k}=0$, and $\frac{\kappa_{k}}{\lambda_{5,6}}+\lambda_{5,6}+\kappa_{k} h_{k}=0$, then $\lambda_{5,6}$ are the eigenvalues of $A_{k}^{[j]}+h_{k}A_{{\mathrm{c}}k}$. The corresponding eigenspace is given by
\begin{eqnarray}\label{egns4}
&&\hspace{-1em}\ker\Big(A_{k}^{[j]}+h_{k}A_{{\mathrm{c}}k}-\lambda_{5,6} I_{2nq+n}\Big)\nonumber\\
&&\hspace{-1em}=\Big\{\Big[\frac{1+h_{k}\lambda_{5,6}^{*}}{\lambda_{5,6}^{*}}\sum_{i=1}^{n}\sum_{l=1}^{q}\varpi_{li}(\textbf{g}_{l}-(\textbf{g}_{j}^{\mathrm{T}}\textbf{g}_{l})\textbf{g}_{j})^{\mathrm{T}}\otimes \textbf{e}_{i}^{\mathrm{T}},\sum_{i=1}^{n}\sum_{l=1}^{q}\varpi_{li}(\textbf{g}_{l}-(\textbf{g}_{j}^{\mathrm{T}}\textbf{g}_{l})\textbf{g}_{j})^{\mathrm{T}}\otimes \textbf{e}_{i}^{\mathrm{T}},\textbf{0}_{1\times n}\Big]^{*}:\nonumber\\
&&\hspace{-1em}\varpi_{li}\in\mathbb{C},i=1,\ldots,n,l=1,\ldots,q\Big\}.
\end{eqnarray}
\item[$vii$)] If $1\in{\mathrm{spec}}(\frac{\eta_{k}}{\kappa_{k}}L_{k})$ and $\kappa_{k}h_{k}=1$, then $\lambda_{3}=-\kappa_{k}$ is an eigenvalue of $A_{k}^{[j]}+h_{k}A_{{\mathrm{c}}k}$. The corresponding eigenspace is given by
\begin{eqnarray}\label{egns5}
&&\ker\Big(A_{k}^{[j]}+h_{k}A_{{\mathrm{c}}k}-\lambda_{3} I_{2nq+n}\Big)\nonumber\\
&&=\Big\{\Big[\textbf{0}_{1\times nq},\sum_{i=1}^{n}\sum_{l=1}^{q}\alpha_{li}(\textbf{g}_{l}\otimes\textbf{e}_{i})^{\mathrm{T}},\sum_{i=1}^{n}\sum_{l=1}^{q}\frac{\eta_{k}}{\kappa_{k}}\alpha_{li}(L_{k}\textbf{g}_{l}\otimes\textbf{e}_{i})^{\mathrm{T}}-\sum_{i=1}^{n}\sum_{l=1}^{q}\alpha_{li}(\textbf{g}_{l}\otimes\textbf{e}_{i})^{\mathrm{T}}\Big]^{*}:\nonumber\\
&&\forall\alpha_{li}\in\mathbb{C},i=1,\ldots,n,l=1,\ldots,q\Big\}.
\end{eqnarray}
\item[$viii$)] If $\frac{\mu_{k}}{\kappa_{k}}(\kappa_{k} h_{k}-1)+\eta_{k}=0$ and $h_{k}=1+\frac{1}{\kappa_{k}}$, then $\lambda_{3}=-\kappa_{k}$ is an eigenvalue of $A_{k}^{[j]}+h_{k}A_{{\mathrm{c}}k}$. The corresponding eigenspace is given by
\begin{eqnarray}\label{egns6}
&&\ker\Big(A_{k}^{[j]}+h_{k}A_{{\mathrm{c}}k}-\lambda_{3} I_{2nq+n}\Big)=\Big\{\Big[\textbf{0}_{1\times nq},\sum_{i=1}^{n}\sum_{l=1}^{q}\alpha_{li}(\textbf{g}_{l}-(\textbf{g}_{j}^{\mathrm{T}}\textbf{g}_{l})\textbf{g}_{j})^{\mathrm{T}}\otimes \textbf{e}_{i}^{\mathrm{T}},\textbf{0}_{1\times n}\Big]^{*}:\nonumber\\
&&\forall\alpha_{li}\in\mathbb{C},i=1,\ldots,n,l=1,\ldots,q\Big\}.
\end{eqnarray}
\item[$ix$)] If $1\in{\mathrm{spec}}(\frac{\mu_{k}+\eta_{k}}{\kappa_{k}}L_{k})$ and $h_{k}=1+\frac{1}{\kappa_{k}}$, then $\lambda_{3}=-\kappa_{k}$ is an eigenvalue of $A_{k}^{[j]}+h_{k}A_{{\mathrm{c}}k}$. The corresponding eigenspace is given by
\begin{eqnarray}\label{egns7}
&&\ker\Big(A_{k}^{[j]}+h_{k}A_{{\mathrm{c}}k}-\lambda_{3} I_{2nq+n}\Big)\nonumber\\
&&=\Big\{\Big[\textbf{0}_{1\times nq},\frac{\kappa_{k}}{\mu_{k}+\eta_{k}}\sum_{i=1}^{n}\beta_{i}(L_{k}^{+}\textbf{1}_{q\times 1}\otimes\textbf{e}_{i})^{\mathrm{T}}-\frac{\kappa_{k}}{\mu_{k}+\eta_{k}}\sum_{i=1}^{n}\beta_{i}(L_{k}^{+}\varphi_{k}\otimes\textbf{e}_{i})^{\mathrm{T}}\nonumber\\
&&+\sum_{l=1}^{q}\sum_{i=1}^{n}\gamma_{li}(\textbf{g}_{l}-L_{k}^{+}L_{k}\textbf{g}_{l}+(\textbf{g}_{j}^{\mathrm{T}}L_{k}\textbf{g}_{l})L_{k}^{+}\varphi_{k}-(\textbf{g}_{j}^{\mathrm{T}}\textbf{g}_{l})\varphi_{k})^{\mathrm{T}}\otimes \textbf{e}_{i}^{\mathrm{T}},\sum_{i=1}^{n}\beta_{i}\textbf{e}_{i}^{\mathrm{T}}\Big]^{*}:\nonumber\\
&&\beta_{i}\in\mathbb{C},\gamma_{li}\in\mathbb{C},i=1,\ldots,n,l=1,\ldots,q\Big\},
\end{eqnarray} where 
\begin{eqnarray}\label{varphik}
\varphi_{k}=\small\left\{\begin{array}{ll}
(\textbf{g}_{j}^{\mathrm{T}}-\textbf{g}_{j}^{\mathrm{T}}L_{k}^{+}L_{k})^{+}, & \textbf{g}_{j}\neq L_{k}^{+}L_{k}\textbf{g}_{j},\\
(1+\textbf{g}_{j}^{\mathrm{T}}(L_{k}^{\mathrm{T}}L_{k})^{+}\textbf{g}_{j})^{-1}(L_{k}^{\mathrm{T}}L_{k})^{+}\textbf{g}_{j}, & \textbf{g}_{j}= L_{k}^{+}L_{k}\textbf{g}_{j}. \\
\end{array}\right.
\end{eqnarray}
\item[$x$)] If $1\in{\mathrm{spec}}(\frac{\mu_{k}(\kappa_{k}h_{k}-1)+\eta_{k}\kappa_{k}}{\kappa_{k}(-\kappa_{k} h_{k}+1+\kappa_{k})}L_{k})$ and $\kappa_{k}h_{k}\neq 1$, then $\lambda_{3}=-\kappa_{k}$ is an eigenvalue of $A_{k}^{[j]}+h_{k}A_{{\mathrm{c}}k}$. The corresponding eigenspace is given by
\begin{eqnarray}\label{egns8}
&&\ker\Big(A_{k}^{[j]}+h_{k}A_{{\mathrm{c}}k}-\lambda_{3} I_{2nq+n}\Big)\nonumber\\
&&=\Big\{\Big[\textbf{0}_{1\times nq},\sum_{i=1}^{n}\sum_{l=1}^{q}\varpi_{li}\Big(\textbf{g}_{l}-M_{k}^{+}M_{k}\textbf{g}_{l}+(\textbf{g}_{j}^{\mathrm{T}}M_{k}\textbf{g}_{l})M_{k}^{+}\phi_{k}-(\textbf{g}_{j}^{\mathrm{T}}\textbf{g}_{l})\phi_{k}\Big)^{\mathrm{T}}\otimes\textbf{e}_{i}^{\mathrm{T}},\textbf{0}_{1\times n}\Big]^{*}:\nonumber\\
&&\varpi_{li}\in\mathbb{C},i=1,\ldots,n,l=1,\ldots,q\Big\},
\end{eqnarray} where $M_{k}=(\frac{\mu_{k}}{\kappa_{k}}(\kappa_{k} h_{k}-1)+\eta_{k})L_{k}+(\kappa_{k} h_{k}-1-\kappa_{k})I_{q}$ and 
\begin{eqnarray}\label{phik}
\phi_{k}=\small\left\{\begin{array}{ll}
(\textbf{g}_{j}^{\mathrm{T}}-\textbf{g}_{j}^{\mathrm{T}}M_{k}^{+}M_{k})^{+}, & \textbf{g}_{j}\neq M_{k}^{+}M_{k}\textbf{g}_{j},\\
(1+\textbf{g}_{j}^{\mathrm{T}}(M_{k}^{\mathrm{T}}M_{k})^{+}\textbf{g}_{j})^{-1}(M_{k}^{\mathrm{T}}M_{k})^{+}\textbf{g}_{j}, & \textbf{g}_{j}= M_{k}^{+}M_{k}\textbf{g}_{j}. \\
\end{array}\right.
\end{eqnarray}
\end{itemize}
\end{lemma}

\begin{IEEEproof}
For a fixed $j\in\{1,\ldots,q\}$ and a fixed $k\in\overline{\mathbb{Z}}_{+}$, let $\textbf{x}\in\mathbb{C}^{2nq+n}$ be an eigenvector of the
corresponding eigenvalue $\lambda\in\mathbb{C}$ for $A_{k}^{[j]}+h_{k}A_{{\mathrm{c}}k}$. We partition
$\textbf{x}$ into
$\textbf{x}=[\textbf{x}_{1}^{*},\textbf{x}_{2}^{*},\textbf{x}_{3}^{*}]^{*}\neq\textbf{0}_{(2nq+n)\times 1}$,
where $\textbf{x}_{1},\textbf{x}_{2}\in\mathbb{C}^{nq}$, and $\textbf{x}_{3}\in\mathbb{C}^{n}$. It follows from
$(A_{k}^{[j]}+h_{k}A_{{\mathrm{c}}k})\textbf{x}=\lambda\textbf{x}$ that
\begin{eqnarray}
h_{k}(-\mu_{k} L_{k}\otimes P_{k}-\kappa_{k} I_{q}\otimes P_{k})\textbf{x}_{1}+h_{k}(-\eta_{k} L_{k}\otimes P_{k})\textbf{x}_{2}+\textbf{x}_{2}+h_{k}(\kappa_{k} \textbf{1}_{q\times 1}\otimes P_{k})\textbf{x}_{3}=\lambda\textbf{x}_{1},\label{Aeig_1}\\
(-\mu_{k} L_{k}\otimes P_{k}-\kappa_{k} I_{q}\otimes P_{k})\textbf{x}_{1}+(-\eta_{k} L_{k}\otimes P_{k})\textbf{x}_{2}+(\kappa_{k} \textbf{1}_{q\times 1}\otimes P_{k})\textbf{x}_{3}=\lambda \textbf{x}_{2},\label{x2}\\
\kappa_{k} E_{n\times nq}^{[j]}\textbf{x}_{1}-\kappa_{k}\textbf{x}_{3}=\lambda \textbf{x}_{3}.\label{Aeig_3}
\end{eqnarray} Note that it follows from Lemma~\ref{lemma_semisimple} that $A_{k}^{[j]}+h_{k}A_{{\mathrm{c}}k}$ has an eigenvalue 0. Now we assume that $\lambda\neq0$. 

Substituting (\ref{x2}) into (\ref{Aeig_1}) yields
$\textbf{x}_{1}=\frac{1+h_{k}\lambda}{\lambda}\textbf{x}_{2}$. Replacing $\textbf{x}_{1}$ in (\ref{x2}) and (\ref{Aeig_3}) with $\textbf{x}_{1}=\frac{1+h_{k}\lambda}{\lambda}\textbf{x}_{2}$ yields
\begin{eqnarray}
-\Big[\Big(\frac{\mu_{k}}{\lambda}+\mu_{k} h_{k}+\eta_{k}\Big)(L_{k}\otimes P_{k})+\Big(\frac{\kappa_{k}}{\lambda}+\lambda+\kappa_{k} h_{k}\Big)(I_{q}\otimes P_{k})\Big]\textbf{x}_{2}+\kappa_{k}(\textbf{1}_{q\times 1}\otimes P_{k})\textbf{x}_{3}=\textbf{0}_{nq\times 1},\label{hx2}\\
\Big(\frac{\kappa_{k}}{\lambda}+\kappa_{k} h_{k}\Big)E_{n\times nq}^{[j]}\textbf{x}_{2}-(\lambda+\kappa_{k})\textbf{x}_{3}=\textbf{0}_{n\times 1}.\label{hx3}
\end{eqnarray} Clearly $[\textbf{x}_{2}^{*},\textbf{x}_{3}^{*}]^{*}\neq\textbf{0}_{2nq\times 1}$. Thus, (\ref{hx2}) and (\ref{hx3}) have nontrivial solutions if and only if
\begin{eqnarray}\label{detcon}
\det\small\left[\begin{array}{cc}
\Big(\frac{\mu_{k}}{\lambda}+\mu_{k} h_{k}+\eta_{k}\Big)(L_{k}\otimes P_{k})+\Big(\frac{\kappa_{k}}{\lambda}+\lambda+\kappa_{k} h_{k}\Big)(I_{q}\otimes P_{k}) & -\kappa_{k}(\textbf{1}_{q\times 1}\otimes P_{k})\\
\Big(\frac{\kappa_{k}}{\lambda}+\kappa_{k} h_{k}\Big)E_{n\times nq}^{[j]} & -(\lambda+\kappa_{k})I_{n}
\end{array}\right]=0.
\end{eqnarray} 

If $\det\Big[\Big(\frac{\mu_{k}}{\lambda}+\mu_{k} h_{k}+\eta_{k}\Big)(L_{k}\otimes P_{k})+\Big(\frac{\kappa_{k}}{\lambda}+\lambda+\kappa_{k} h_{k}\Big)(I_{q}\otimes P_{k})\Big]\neq0$, then pre-multiplying $-L_{k}\otimes I_{n}$ on both sides of (\ref{hx2}) yields
$\Big[\Big(\frac{\mu_{k}}{\lambda}+\mu_{k} h_{k}+\eta_{k}\Big)(L_{k}\otimes I_{n})+\Big(\frac{\kappa_{k}}{\lambda}+\lambda+\kappa_{k} h_{k}\Big)I_{nq}\Big](L_{k}\otimes P_{k})\textbf{x}_{2}=\textbf{0}_{nq\times 1}$,
which implies that $(L_{k}\otimes P_{k})\textbf{x}_{2}=\textbf{0}_{nq\times 1}$. Now following the similar arguments as in the proof of $iii$) in Lemma~\ref{lemma_Arank}, we have $\textbf{x}_{2}=\sum_{l=0}^{q-1-{\mathrm{rank}}(L_{k})}\sum_{i=1}^{n}\varpi_{li}\textbf{w}_{l}\otimes\textbf{e}_{i}+\sum_{s=1}^{q}\sum_{r=1}^{n-{\mathrm{rank}}(P_{k})}\beta_{sr}\textbf{g}_{s}\otimes\textbf{j}_{r}$, where $\varpi_{li},\beta_{sr}\in\mathbb{C}$ and not all $\varpi_{li},\beta_{sr}$ are zero. Substituting this expression of $\textbf{x}_{2}$ into (\ref{hx2}) and (\ref{hx3}) by using $iii$) of Lemma~\ref{lemma_EW} yields
\begin{eqnarray} 
\kappa_{k}P_{k}\textbf{x}_{3}=\left(\frac{\kappa_{k}}{\lambda}+\lambda+\kappa_{k}h_{k}\right)\sum_{l=0}^{q-1-{\mathrm{rank}}(L_{k})}\sum_{i=1}^{n}\varpi_{li}w_{lj}P_{k}\textbf{e}_{i}.\label{x3e1}\\
(\lambda+\kappa_{k})\textbf{x}_{3}=\left(\frac{\kappa_{k}}{\lambda}+\kappa_{k} h_{k}\right)\sum_{l=0}^{q-1-{\mathrm{rank}}(L_{k})}\sum_{i=1}^{n}\varpi_{li}w_{lj}\textbf{e}_{i}.\label{x3e2}
\end{eqnarray}
Furthermore, substituting (\ref{x3e1}) into $P_{k}$(\ref{x3e2}) yields
$\lambda P_{k}\textbf{x}_{3}=-\lambda\sum_{l=0}^{q-1-{\mathrm{rank}}(L_{k})}\sum_{i=1}^{n}\varpi_{li}w_{lj}P_{k}\textbf{e}_{i}$,
which implies that $P_{k}\textbf{x}_{3}=-\sum_{l=0}^{q-1-{\mathrm{rank}}(L_{k})}\sum_{i=1}^{n}\varpi_{li}w_{lj}P_{k}\textbf{e}_{i}$ since $\lambda\neq0$. Hence, $P_{k}(\textbf{x}_{3}+\sum_{l=0}^{q-1-{\mathrm{rank}}(L_{k})}\sum_{i=1}^{n}\varpi_{li}w_{lj}\textbf{e}_{i})=\textbf{0}_{n\times 1}$, which further implies that $\textbf{x}_{3}+\sum_{l=0}^{q-1-{\mathrm{rank}}(L_{k})}\sum_{i=1}^{n}\varpi_{li}w_{lj}\textbf{e}_{i}\in\ker(P_{k})$. Consequently, $\textbf{x}_{3}=-\sum_{l=0}^{q-1-{\mathrm{rank}}(L_{k})}\\\sum_{i=1}^{n}\varpi_{li}w_{lj}\textbf{e}_{i}+\sum_{r=1}^{n-{\mathrm{rank}}(P_{k})}\delta_{r}\textbf{j}_{r}$, where $\delta_{r}\in\mathbb{C}$. Finally, substituting the obtained expression for $\textbf{x}_{3}$ into (\ref{x3e2}) yields
\begin{eqnarray}\label{eqn_ei} 
\left(\frac{\kappa_{k}}{\lambda}+\kappa_{k} h_{k}+\lambda+\kappa_{k}\right)\sum_{l=0}^{q-1-{\mathrm{rank}}(L_{k})}\sum_{i=1}^{n}\varpi_{li}w_{lj}\textbf{e}_{i}-(\lambda+\kappa_{k})\sum_{r=1}^{n-{\mathrm{rank}}(P_{k})}\delta_{r}\textbf{j}_{r}=\textbf{0}_{n\times 1}.
\end{eqnarray}
In this case, (\ref{hx2}) and (\ref{hx3}) have nontrivial solutions if and only if (\ref{eqn_ei}) holds for not all zero $\varpi_{li},\delta_{r}\in\mathbb{C}$. Since by assumption, $P_{k}$ is a full rank matrix, it follows that $\textbf{j}_{r}=\textbf{0}_{n\times 1}$ and hence, (\ref{eqn_ei}) collapses into $\left(\frac{\kappa_{k}}{\lambda}+\kappa_{k} h_{k}+\lambda+\kappa_{k}\right)\sum_{l=0}^{q-1-{\mathrm{rank}}(L_{k})}\sum_{i=1}^{n}\varpi_{li}w_{lj}\textbf{e}_{i}=\textbf{0}_{n\times 1}$, which implies that either $\frac{\kappa_{k}}{\lambda}+\kappa_{k} h_{k}+\lambda+\kappa_{k}=0$ or $\sum_{l=0}^{q-1-{\mathrm{rank}}(L_{k})}\sum_{i=1}^{n}\varpi_{li}w_{lj}\textbf{e}_{i}=\textbf{0}_{n\times 1}$. If $\sum_{l=0}^{q-1-{\mathrm{rank}}(L_{k})}\sum_{i=1}^{n}\varpi_{li}w_{lj}\textbf{e}_{i}=\textbf{0}_{n\times 1}$, then it follows from the expression of $\textbf{x}_{3}$ that $\textbf{x}_{3}=\textbf{0}_{n\times 1}$ and by (\ref{hx2}), $\textbf{x}_{2}=\textbf{0}_{nq\times 1}$, and hence, $\textbf{x}_{1}=\frac{1+h_{k}\lambda}{\lambda}\textbf{x}_{2}=\textbf{0}_{nq\times 1}$. This is a contradiction. Thus, $\frac{\kappa_{k}}{\lambda}+\kappa_{k} h_{k}+\lambda+\kappa_{k}=0$, and hence, $\kappa_{k}\neq0$. Let $\lambda_{1,2}$ denote the two solutions to $\frac{\kappa_{k}}{\lambda}+\kappa_{k} h_{k}+\lambda+\kappa_{k}=0$. Then 
\begin{eqnarray}\label{lambda1}
\lambda_{1,2}=-\frac{\kappa_{k}(1+h_{k})}{2}\pm\frac{1}{2}\sqrt{\kappa_{k}^{2}(1+h_{k})^{2}-4\kappa_{k}}.
\end{eqnarray} In this case, note that
\begin{eqnarray}\label{detlambda}
&&\det\Big[\Big(\frac{\mu_{k}}{\lambda_{1,2}}+\mu_{k} h_{k}+\eta_{k}\Big)(L_{k}\otimes P_{k})+\Big(\frac{\kappa_{k}}{\lambda_{1,2}}+\lambda_{1,2}+\kappa_{k} h_{k}\Big)(I_{q}\otimes P_{k})\Big]\nonumber\\
&&=\det\Big[\Big(\frac{\mu_{k}}{\lambda_{1,2}}+\mu_{k} h_{k}+\eta_{k}\Big)(L_{k}\otimes I_{n})-\kappa_{k}I_{nq}\Big]\det(I_{q}\otimes P_{k})\nonumber\\
&&=\kappa_{k}^{nq}\det\Big[\Big(\frac{\mu_{k}}{\lambda_{1,2}\kappa_{k}}+\frac{\mu_{k} h_{k}}{\kappa_{k}}+\frac{\eta_{k}}{\kappa_{k}}\Big)(L_{k}\otimes I_{n})-I_{nq}\Big](\det(P_{k}))^{q}.
\end{eqnarray} Hence, $\det\Big[\Big(\frac{\mu_{k}}{\lambda_{1,2}}+\mu_{k} h_{k}+\eta_{k}\Big)(L_{k}\otimes P_{k})+\Big(\frac{\kappa_{k}}{\lambda_{1,2}}+\lambda_{1,2}+\kappa_{k} h_{k}\Big)(I_{q}\otimes P_{k})\Big]\neq0$ if and only if $1\not\in{\mathrm{spec}}((\frac{\mu_{k}}{\lambda_{1,2}\kappa_{k}}+\frac{\mu_{k} h_{k}}{\kappa_{k}}+\frac{\eta_{k}}{\kappa_{k}})L_{k})$. Thus, if 
$1\not\in{\mathrm{spec}}((\frac{\mu_{k}}{\lambda_{1,2}\kappa_{k}}+\frac{\mu_{k} h_{k}}{\kappa_{k}}+\frac{\eta_{k}}{\kappa_{k}})L_{k})$, then $\lambda_{1,2}$ given by (\ref{lambda1}) are indeed the eigenvalues of $A_{k}^{[j]}+h_{k}A_{{\mathrm{c}}k}$ and the corresponding eigenvectors for $\lambda_{1,2}$ are given by 
$\textbf{x}=\Big[\frac{1+h_{k}\lambda_{1,2}^{*}}{\lambda_{1,2}^{*}}\sum_{l=0}^{q-1-{\mathrm{rank}}(L_{k})}\sum_{i=1}^{n}\varpi_{li}(\textbf{w}_{l}\otimes\textbf{e}_{i})^{\mathrm{T}},\sum_{l=0}^{q-1-{\mathrm{rank}}(L_{k})}\sum_{i=1}^{n}\varpi_{li}(\textbf{w}_{l}\otimes\textbf{e}_{i})^{\mathrm{T}},-\sum_{l=0}^{q-1-{\mathrm{rank}}(L_{k})}\sum_{i=1}^{n}\varpi_{li}w_{lj}\textbf{e}_{i}^{\mathrm{T}}\Big]^{*}$,
where $\varpi_{li}\in\mathbb{C}$ and not all of $\varpi_{li}$ are zero. Therefore,
$\ker\Big(A_{k}^{[j]}+h_{k}A_{{\mathrm{c}}k}-\lambda_{1,2} I_{2nq+n}\Big)$ is given by (\ref{egns1}). 

Alternatively, if $\det\Big[\Big(\frac{\mu_{k}}{\lambda}+\mu_{k} h_{k}+\eta_{k}\Big)(L_{k}\otimes P_{k})+\Big(\frac{\kappa_{k}}{\lambda}+\lambda+\kappa_{k} h_{k}\Big)(I_{q}\otimes P_{k})\Big]=0$, then in this case, we consider two additional cases for (\ref{detcon}): 

\textit{Case 1.} If $\lambda\neq-\kappa_{k}$, then it follows from Proposition 2.8.4 of \cite[p.~116]{Bernstein:2009} that (\ref{detcon}) is equivalent to $\det\Big(\Big(\frac{\mu_{k}}{\lambda}+\mu_{k} h_{k}+\eta_{k}\Big)(L_{k}\otimes P_{k})+\Big(\frac{\kappa_{k}}{\lambda}+\lambda+\kappa_{k} h_{k}\Big)(I_{q}\otimes P_{k})-\frac{\kappa_{k}^{2}(1+h_{k}\lambda)}{\lambda(\lambda+\kappa_{k})}W_{k}^{[j]}\Big)=0$, which implies that for $\lambda\neq-\kappa_{k}$, the equation
\begin{eqnarray}\label{eqn_v}
\Big(\Big(\frac{\mu_{k}}{\lambda}+\mu_{k} h_{k}+\eta_{k}\Big)(L_{k}\otimes P_{k})+\Big(\frac{\kappa_{k}}{\lambda}+\lambda+\kappa_{k} h_{k}\Big)(I_{q}\otimes P_{k})-\frac{\kappa_{k}^{2}(1+h_{k}\lambda)}{\lambda(\lambda+\kappa_{k})}W_{k}^{[j]}\Big)\textbf{v}=\textbf{0}_{nq\times 1}
\end{eqnarray} has nontrivial solutions for $\textbf{v}\in\mathbb{C}^{nq}$. It follows from (\ref{hx2}) and (\ref{hx3}) that solving this $\textbf{v}$ is equivalent to solving $\textbf{x}_{2}$. Again, note that for every $j=1,\ldots,q$, $(L_{k}\otimes I_{n})W_{k}^{[j]}=\textbf{0}_{nq\times nq}$. Pre-multiplying $L_{k}\otimes I_{n}$ on both sides of (\ref{eqn_v}) yields $\Big(\Big(\frac{\mu_{k}}{\lambda}+\mu_{k} h_{k}+\eta_{k}\Big)(L_{k}^{2}\otimes P_{k})+\Big(\frac{\kappa_{k}}{\lambda}+\lambda+\kappa_{k} h_{k}\Big)(L_{k}\otimes P_{k})\Big)\textbf{v}=(I_{q}\otimes P_{k})(L_{k}\otimes I_{n})\Big(\Big(\frac{\mu_{k}}{\lambda}+\mu_{k} h_{k}+\eta_{k}\Big)(L_{k}\otimes I_{n})+\Big(\frac{\kappa_{k}}{\lambda}+\lambda+\kappa_{k} h_{k}\Big)I_{nq}\Big)\textbf{v}=\textbf{0}_{nq\times 1}$, which implies that $\Big(\Big(\frac{\mu_{k}}{\lambda}+\mu_{k} h_{k}+\eta_{k}\Big)(L_{k}\otimes I_{n})+\Big(\frac{\kappa_{k}}{\lambda}+\lambda+\kappa_{k} h_{k}\Big)I_{nq}\Big)\textbf{v}\in\ker(L_{k}\otimes I_{n})$ due to the assumption that $P_{k}$ is of full rank. Since $\ker(L_{k}\otimes I_{n})=\bigcup_{l=0}^{q-1-{\mathrm{rank}}(L_{k})}{\mathrm{span}}\{\textbf{w}_{l}\otimes\textbf{e}_{1},\ldots,\textbf{w}_{l}\otimes\textbf{e}_{n}\}$, it follows that
\begin{eqnarray}\label{Az=b}
\Big(\Big(\frac{\mu_{k}}{\lambda}+\mu_{k} h_{k}+\eta_{k}\Big)(L_{k}\otimes I_{n})+\Big(\frac{\kappa_{k}}{\lambda}+\lambda+\kappa_{k} h_{k}\Big)I_{nq}\Big)\textbf{v}=\sum_{i=1}^{n}\sum_{l=0}^{q-1-{\mathrm{rank}}(L_{k})}\omega_{li}\textbf{w}_{l}\otimes\textbf{e}_{i},
\end{eqnarray} where $\omega_{li}\in\mathbb{C}$. Now it follows from (\ref{eqn_v}) and (\ref{Az=b}) that
\begin{eqnarray}\label{Wv_eqn}
\frac{\kappa_{k}^{2}(1+h_{k}\lambda)}{\lambda(\lambda+\kappa_{k})}W_{k}^{[j]}\textbf{v}=\sum_{i=1}^{n}\sum_{l=0}^{q-1-{\mathrm{rank}}(L_{k})}\omega_{li}\textbf{w}_{l}\otimes\textbf{e}_{i}.
\end{eqnarray} 

If $\frac{\kappa_{k}}{\lambda}+\lambda+\kappa_{k} h_{k}\neq0$, then (\ref{Az=b}) has a particular solution $\textbf{v}=(\frac{\kappa_{k}}{\lambda}+\lambda+\kappa_{k} h_{k})^{-1}\sum_{i=1}^{n}\sum_{l=0}^{q-1-{\mathrm{rank}}(L_{k})}\omega_{li}\textbf{w}_{l}\otimes\textbf{e}_{i}$. Let $\textbf{w}_{l}=[w_{l1}^{*},\ldots,w_{lq}^{*}]^{*}$. Substituting this particular solution into (\ref{Wv_eqn}), together with $ii$) of Lemma~\ref{lemma_EW}, yields
\begin{eqnarray} &&(I_{q}\otimes P_{k})\Big(\sum_{i=1}^{n}\sum_{l=0}^{q-1-{\mathrm{rank}}(L_{k})}\omega_{li}\textbf{w}_{l}\otimes\textbf{e}_{i}\Big)-\frac{\kappa_{k}^{2}(1+h_{k}\lambda)}{\lambda(\lambda+\kappa_{k})}W_{k}^{[j]}(\frac{\kappa_{k}}{\lambda}+\lambda+\kappa_{k} h_{k})^{-1}\sum_{i=1}^{n}\sum_{l=0}^{q-1-{\mathrm{rank}}(L_{k})}\omega_{li}\textbf{w}_{l}\otimes\textbf{e}_{i}\nonumber\\
&&=(I_{q}\otimes P_{k})\Big(\sum_{i=1}^{n}\sum_{l=0}^{q-1-{\mathrm{rank}}(L_{k})}\omega_{li}\textbf{w}_{l}\otimes\textbf{e}_{i}-\frac{\kappa_{k}^{2}(1+h_{k}\lambda)}{(\lambda+\kappa_{k})(\lambda^{2}+\kappa_{k} h_{k}\lambda+\kappa_{k})}\sum_{i=1}^{n}\sum_{l=0}^{q-1-{\mathrm{rank}}(L_{k})}\omega_{li}w_{lj}\textbf{w}_{0}\otimes\textbf{e}_{i}\Big)\nonumber\\
&&=(I_{q}\otimes P_{k})\Big(\sum_{i=1}^{n}\Big[\omega_{0i}-\frac{\kappa_{k}^{2}(1+h_{k}\lambda)}{(\lambda+\kappa_{k})(\lambda^{2}+\kappa_{k} h_{k}\lambda+\kappa_{k})}\sum_{l=0}^{q-1-{\mathrm{rank}}(L_{k})}\omega_{li}w_{lj}\Big]\textbf{w}_{0}\otimes\textbf{e}_{i}\nonumber\\
&&\hspace{1em}+\sum_{i=1}^{n}\sum_{l=1}^{q-1-{\mathrm{rank}}(L_{k})}\omega_{li}\textbf{w}_{l}\otimes\textbf{e}_{i}\Big)\nonumber\\
&&=\textbf{0}_{nq\times 1},
\end{eqnarray} which implies that 
\begin{eqnarray}\label{omega}
\omega_{0i}-\frac{\kappa_{k}^{2}(1+h_{k}\lambda)}{(\lambda+\kappa_{k})(\lambda^{2}+\kappa_{k} h_{k}\lambda+\kappa_{k})}\sum_{l=0}^{q-1-{\mathrm{rank}}(L_{k})}\omega_{li}w_{lj}=0
\end{eqnarray} and $\omega_{\ell i}=0$ for every $i=1,\ldots,n$ and every $\ell=1,\ldots,q-1-{\mathrm{rank}}(L_{k})$. Note that $w_{0j}=1$ for every $j=1,\ldots,q$. Substituting $\omega_{\ell i}=0$ into (\ref{omega}) yields
\begin{eqnarray}
\omega_{0i}-\frac{\kappa_{k}^{2}(1+h_{k}\lambda)}{(\lambda+\kappa_{k})(\lambda^{2}+\kappa_{k} h_{k}\lambda+\kappa_{k})}\omega_{0i}=0,\quad i=1,\ldots,n.
\end{eqnarray}
Then either $1-\frac{\kappa_{k}^{2}(1+h_{k}\lambda)}{(\lambda+\kappa_{k})(\lambda^{2}+\kappa_{k} h_{k}\lambda+\kappa_{k})}=0$ or $\omega_{0i}=0$ for every $i=1,\ldots,n$.

If $\frac{\kappa_{k}^{2}(1+h_{k}\lambda)}{(\lambda+\kappa_{k})(\lambda^{2}+\kappa_{k} h_{k}\lambda+\kappa_{k})}=1$, then $\lambda^{2}+\kappa_{k}(1+h_{k})\lambda+\kappa_{k}=0$. Hence, $\lambda=\lambda_{12}$ where $\lambda_{1,2}$ are given by (\ref{lambda1}). In this case, note that $\frac{\kappa_{k}}{\lambda_{1,2}}+\lambda_{1,2}+\kappa_{k} h_{k}=-\kappa_{k}\neq0$. Then it follows that 
(\ref{detlambda}) holds. Hence,
$\det\Big[\Big(\frac{\mu_{k}}{\lambda_{1,2}}+\mu_{k} h_{k}+\eta_{k}\Big)(L_{k}\otimes P_{k})+\Big(\frac{\kappa_{k}}{\lambda_{1,2}}+\lambda_{1,2}+\kappa_{k} h_{k}\Big)(I_{q}\otimes P_{k})\Big]=0$ if and only if $1\in{\mathrm{spec}}((\frac{\mu_{k}}{\lambda_{1,2}\kappa_{k}}+\frac{\mu_{k} h_{k}}{\kappa_{k}}+\frac{\eta_{k}}{\kappa_{k}})L_{k})$. Furthermore, $\lambda_{1,2}\neq-\kappa_{k}$ if and only if $h_{k}\kappa_{k}\neq 1$. Thus, if 
$1\in{\mathrm{spec}}((\frac{\mu_{k}}{\lambda_{1,2}\kappa_{k}}+\frac{\mu_{k} h_{k}}{\kappa_{k}}+\frac{\eta_{k}}{\kappa_{k}})L_{k})$ and $h_{k}\kappa_{k}\neq 1$, then $\lambda_{1,2}$ given by (\ref{lambda1}) are indeed the eigenvalues of $A_{k}^{[j]}+h_{k}A_{{\mathrm{c}}k}$. In this case, 
(\ref{Az=b}) becomes
\begin{eqnarray}\label{Az=bx}
\Big(\Big(\frac{\mu_{k}}{\lambda_{1,2}}+\mu_{k} h_{k}+\eta_{k}\Big)(L_{k}\otimes I_{n})-\kappa_{k}I_{nq}\Big)\textbf{v}=\sum_{i=1}^{n}\omega_{0i}\textbf{w}_{0}\otimes\textbf{e}_{i}
\end{eqnarray} and a specific solution is given by $\textbf{v}=-\frac{1}{\kappa_{k}}\sum_{i=1}^{n}\omega_{0i}\textbf{w}_{0}\otimes\textbf{e}_{i}$. To find the general solution to (\ref{Az=bx}), let $G_{k}=(\frac{\mu_{k}}{\lambda_{1,2}}+\mu_{k} h_{k}+\eta_{k})L_{k}-\kappa_{k}I_{q}$ and consider 
\begin{eqnarray}\label{Gx0}
(G_{k}\otimes I_{n})\hat{\textbf{v}}=\textbf{0}_{nq\times 1}.
\end{eqnarray} It follows from $vi$) of Proposition 6.1.7 of \cite[p.~400]{Bernstein:2009} and $viii$) of Proposition 6.1.6 of \cite[p.~399]{Bernstein:2009} that the general solution $\hat{\textbf{v}}$ to (\ref{Gx0}) is given by the form
\begin{eqnarray}
\hat{\textbf{v}}&=&\Big[I_{nq}-(G_{k}\otimes I_{n})^{+}(G_{k}\otimes I_{n})\Big]\sum_{i=1}^{n}\sum_{l=1}^{q}\varpi_{li}\textbf{g}_{l}\otimes\textbf{e}_{i}\nonumber\\
&=&\Big[I_{nq}-(G_{k}^{+}\otimes I_{n})(G_{k}\otimes I_{n})\Big]\sum_{i=1}^{n}\sum_{l=1}^{q}\varpi_{li}\textbf{g}_{l}\otimes\textbf{e}_{i}\nonumber\\
&=&\Big[I_{q}\otimes I_{n}-((G_{k}^{+}G_{k})\otimes I_{n})\Big]\sum_{i=1}^{n}\sum_{l=1}^{q}\varpi_{li}\textbf{g}_{l}\otimes\textbf{e}_{i}\nonumber\\
&=&\Big[(I_{q}-G_{k}^{+}G_{k})\otimes I_{n}\Big]\sum_{i=1}^{n}\sum_{l=1}^{q}\varpi_{li}\textbf{g}_{l}\otimes\textbf{e}_{i}\nonumber\\
&=&\sum_{i=1}^{n}\sum_{l=1}^{q}\varpi_{li}(\textbf{g}_{l}-G_{k}^{+}G_{k}\textbf{g}_{l})\otimes \textbf{e}_{i},
\end{eqnarray} where $\varpi_{li}\in\mathbb{C}$, $j=1,\ldots,q$, and we used the facts that $(A\otimes B)^{+}=A^{+}\otimes B^{+}$, $A\otimes B-C\otimes B=(A-C)\otimes B$, and $(A\otimes B)(C\otimes D)=AC\otimes BD$ for compatible matrices $A,B,C,D$. Then the general solution to (\ref{Az=bx}) is given by
\begin{eqnarray}
\textbf{v}&=&\hat{\textbf{v}}-\frac{1}{\kappa_{k}}\sum_{i=1}^{n}\omega_{0i}\textbf{w}_{0}\otimes\textbf{e}_{i}\nonumber\\
&=&\sum_{i=1}^{n}\sum_{l=1}^{q}\varpi_{li}(\textbf{g}_{l}-G_{k}^{+}G_{k}\textbf{g}_{l})\otimes \textbf{e}_{i}-\frac{1}{\kappa_{k}}\sum_{i=1}^{n}\omega_{0i}\textbf{w}_{0}\otimes\textbf{e}_{i},
\end{eqnarray} and hence, $\textbf{x}_{2}=\textbf{v}\neq\textbf{0}_{nq\times 1}$ and  $\textbf{x}_{1}=\frac{1+h_{k}\lambda_{1,2}}{\lambda_{1,2}}\textbf{v}$. Furthermore,  note that $\textbf{g}_{j}^{\mathrm{T}}\textbf{w}_{0}=1$ for every $j=1,\ldots,q$, it follows that  
\begin{eqnarray}
\textbf{x}_{3}&=&\frac{\kappa_{k}+\kappa_{k}h_{k}\lambda_{1,2}}{\lambda_{1,2}(\lambda_{1,2}+\kappa_{k})}E_{n\times nq}^{[j]}\textbf{v}\nonumber\\
&=&\frac{\kappa_{k}+\kappa_{k}h_{k}\lambda_{1,2}}{\lambda_{1,2}(\lambda_{1,2}+\kappa_{k})}(\textbf{g}_{j}^{\mathrm{T}}\otimes I_{n})\textbf{v}\nonumber\\
&=&\frac{\kappa_{k}+\kappa_{k}h_{k}\lambda_{1,2}}{\lambda_{1,2}(\lambda_{1,2}+\kappa_{k})}\sum_{i=1}^{n}\sum_{l=1}^{q}\varpi_{li}(\textbf{g}_{j}^{\mathrm{T}}\otimes I_{n})((\textbf{g}_{l}-G_{k}^{+}G_{k}\textbf{g}_{l})\otimes \textbf{e}_{i})\nonumber\\
&&-\frac{1+h_{k}\lambda_{1,2}}{\lambda_{1,2}(\lambda_{1,2}+\kappa_{k})}\sum_{i=1}^{n}\omega_{0i}(\textbf{g}_{j}^{\mathrm{T}}\otimes I_{n})(\textbf{w}_{0}\otimes\textbf{e}_{i})\nonumber\\
&=&\frac{\kappa_{k}+\kappa_{k}h_{k}\lambda_{1,2}}{\lambda_{1,2}(\lambda_{1,2}+\kappa_{k})}\sum_{i=1}^{n}\sum_{l=1}^{q}\varpi_{li}(\textbf{g}_{j}^{\mathrm{T}}\textbf{g}_{l}-\textbf{g}_{j}^{\mathrm{T}}G_{k}^{+}G_{k}\textbf{g}_{l})\textbf{e}_{i}-\frac{1+h_{k}\lambda_{1,2}}{\lambda_{1,2}(\lambda_{1,2}+\kappa_{k})}\sum_{i=1}^{n}\omega_{0i}\textbf{e}_{i}.
\end{eqnarray} Hence,
the corresponding eigenvectors for $\lambda_{1,2}$ are given by 
\begin{eqnarray}
\textbf{x}&=&\Big[\frac{1+h_{k}\lambda_{1,2}^{*}}{\lambda_{1,2}^{*}}\sum_{i=1}^{n}\sum_{l=1}^{q}\varpi_{li}((\textbf{g}_{l}-G_{k}^{+}G_{k}\textbf{g}_{l})\otimes \textbf{e}_{i})^{\mathrm{T}}-\frac{1+h_{k}\lambda_{1,2}^{*}}{\kappa_{k}\lambda_{1,2}^{*}}\sum_{i=1}^{n}\omega_{0i}(\textbf{w}_{0}\otimes\textbf{e}_{i})^{\mathrm{T}},\nonumber\\
&&\sum_{i=1}^{n}\sum_{l=1}^{q}\varpi_{li}((\textbf{g}_{l}-G_{k}^{+}G_{k}\textbf{g}_{l})\otimes \textbf{e}_{i})^{\mathrm{T}}-\frac{1}{\kappa_{k}}\sum_{i=1}^{n}\omega_{0i}(\textbf{w}_{0}\otimes\textbf{e}_{i})^{\mathrm{T}},\nonumber\\
&&\frac{\kappa_{k}+\kappa_{k}h_{k}\lambda_{1,2}^{*}}{\lambda_{1,2}^{*}(\lambda_{1,2}^{*}+\kappa_{k})}\sum_{i=1}^{n}\sum_{l=1}^{q}\varpi_{li}(\textbf{g}_{j}^{\mathrm{T}}\textbf{g}_{l}-\textbf{g}_{j}^{\mathrm{T}}G_{k}^{+}G_{k}\textbf{g}_{l})\textbf{e}_{i}^{\mathrm{T}}-\frac{1+h_{k}\lambda_{1,2}^{*}}{\lambda_{1,2}^{*}(\lambda_{1,2}^{*}+\kappa_{k})}\sum_{i=1}^{n}\omega_{0i}\textbf{e}_{i}^{\mathrm{T}}\Big]^{*},
\end{eqnarray} where $\varpi_{li}\in\mathbb{C}$, $\omega_{0i}\in\mathbb{C}$, and not all of them are zero. Therefore,
$\ker\Big(A_{k}^{[j]}+h_{k}A_{{\mathrm{c}}k}-\lambda_{1,2} I_{2nq+n}\Big)$ is given by (\ref{egns2}).

If $\omega_{0i}=0$ for every $i=1,\ldots,n$, then it follows from (\ref{eqn_v}) and (\ref{Az=b}) that
\begin{eqnarray} 
\frac{\kappa_{k}^{2}(1+h_{k}\lambda)}{\lambda(\lambda+\kappa_{k})}W_{k}^{[j]}\textbf{v}=\textbf{0}_{nq\times 1},\label{v-1}\\
\Big(\Big(\frac{\mu_{k}}{\lambda}+\mu_{k} h_{k}+\eta_{k}\Big)(L_{k}\otimes I_{n})+\Big(\frac{\kappa_{k}}{\lambda}+\lambda+\kappa_{k} h_{k}\Big)I_{nq}\Big)\textbf{v}=\textbf{0}_{nq\times 1}.\label{v-2}
\end{eqnarray} In this case, since $\frac{\kappa_{k}}{\lambda}+\lambda+\kappa_{k} h_{k}\neq0$ and $\lambda\neq-\kappa_{k}$,  $\det\Big[\Big(\frac{\mu_{k}}{\lambda}+\mu_{k} h_{k}+\eta_{k}\Big)(L_{k}\otimes I_{n})+\Big(\frac{\kappa_{k}}{\lambda}+\lambda+\kappa_{k} h_{k}\Big)I_{nq}\Big]=0$ if and only if $\frac{\mu_{k}}{\lambda}+\mu_{k} h_{k}+\eta_{k}\neq0$ and $\frac{\lambda^{2}+\kappa_{k} h_{k}\lambda+\kappa_{k}}{\eta_{k}\lambda+\mu_{k} h_{k}\lambda+\mu_{k}}\in{\mathrm{spec}}(-L_{k})$. Thus, if $\frac{\kappa_{k}}{\lambda}+\lambda+\kappa_{k} h_{k}\neq0$, $\lambda\neq-\kappa_{k}$, $\frac{\mu_{k}}{\lambda}+\mu_{k} h_{k}+\eta_{k}\neq0$, and $\frac{\lambda^{2}+\kappa_{k} h_{k}\lambda+\kappa_{k}}{\eta_{k}\lambda+\mu_{k} h_{k}\lambda+\mu_{k}}\in{\mathrm{spec}}(-L_{k})$, then $\lambda=\lambda_{4}$, where
\begin{eqnarray}\label{lambda4}
\frac{\lambda_{4}^{2}+\kappa_{k} h_{k}\lambda_{4}+\kappa_{k}}{\eta_{k}\lambda_{4}+\mu_{k} h_{k}\lambda_{4}+\mu_{k}}\in{\mathrm{spec}}(-L_{k}),
\end{eqnarray} are the eigenvalues of $A_{k}^{[j]}+h_{k}A_{{\mathrm{c}}k}$. To find their corresponding eigenvectors,
let $F_{k}=\Big(\frac{\mu_{k}}{\lambda_{4}}+\mu_{k} h_{k}+\eta_{k}\Big)L_{k}+\Big(\frac{\kappa_{k}}{\lambda_{4}}+\lambda_{4}+\kappa_{k} h_{k}\Big)I_{q}$. We first show that (\ref{v-1}) is equivalent to 
\begin{eqnarray}\label{v-3}
\frac{\kappa_{k}^{2}(1+h_{k}\lambda)}{\lambda(\lambda+\kappa_{k})}E_{n\times nq}^{[j]}\textbf{v}=\textbf{0}_{n\times 1}
\end{eqnarray} for every $j=1,\ldots,q$. To see this, let $\textbf{v}=[\textbf{v}_{1}^{*},\ldots,\textbf{v}_{q}^{*}]^{*}$. Then it follows from (\ref{Wj2}) that $W_{k}^{[j]}\textbf{v}=[(P_{k}\textbf{v}_{j})^{*},\ldots,(P_{k}\textbf{v}_{j})^{*}]^{*}$. Hence (\ref{v-1}) holds if and only if $\frac{\kappa_{k}^{2}(1+h_{k}\lambda)}{\lambda(\lambda+\kappa_{k})}P_{k}\textbf{v}_{j}=\textbf{0}_{n\times 1}$, i.e., $\frac{\kappa_{k}^{2}(1+h_{k}\lambda)}{\lambda(\lambda+\kappa_{k})}\textbf{v}_{j}=\textbf{0}_{n\times 1}$ since $P_{k}$ is of full rank. On the other hand, note that $E_{n\times nq}^{[j]}\textbf{v}=\textbf{v}_{j}$. Hence, (\ref{v-1}) is equivalent to (\ref{v-3}).
Then by noting that $E_{n\times nq}^{[j]}=\textbf{g}_{j}^{\mathrm{T}}\otimes I_{n}$ for every $j=1,\ldots,q$, it follows from (\ref{v-2}) and (\ref{v-3}) that
\begin{eqnarray}\label{Fv}
\small\left[\begin{array}{c}
F_{k}\otimes I_{n}\\
\frac{\kappa_{k}^{2}(1+h_{k}\lambda_{4})}{\lambda_{4}(\lambda_{4}+\kappa_{k})}(\textbf{g}_{j}^{\mathrm{T}}\otimes I_{n})
\end{array}\right]\textbf{v}=\Big(\small\left[\begin{array}{c}
F_{k}\\
\frac{\kappa_{k}^{2}(1+h_{k}\lambda_{4})}{\lambda_{4}(\lambda_{4}+\kappa_{k})}\textbf{g}_{j}^{\mathrm{T}}
\end{array}\right]\otimes I_{n}\Big)\textbf{v}=\textbf{0}_{(nq+n)\times 1}.
\end{eqnarray} Next, it follows from $vi$) of Proposition 6.1.7 of \cite[p.~400]{Bernstein:2009} and $viii$) of Proposition 6.1.6 of \cite[p.~399]{Bernstein:2009} that the general solution $\textbf{v}$ to (\ref{Fv}) is given by the form
\begin{eqnarray}
\textbf{v}&=&\Big[I_{nq}-\Big(\small\left[\begin{array}{c}
F_{k}\\
\frac{\kappa_{k}^{2}(1+h_{k}\lambda_{4})}{\lambda_{4}(\lambda_{4}+\kappa_{k})}\textbf{g}_{j}^{\mathrm{T}}
\end{array}\right]\otimes I_{n}\Big)^{+}\Big(\small\left[\begin{array}{c}
F_{k}\\
\frac{\kappa_{k}^{2}(1+h_{k}\lambda_{4})}{\lambda_{4}(\lambda_{4}+\kappa_{k})}\textbf{g}_{j}^{\mathrm{T}}
\end{array}\right]\otimes I_{n}\Big)\Big]\sum_{i=1}^{n}\sum_{l=1}^{q}\varpi_{li}\textbf{g}_{l}\otimes\textbf{e}_{i}\nonumber\\
&=&\Big[I_{nq}-\Big(\small\left[\begin{array}{c}
F_{k}\\
\frac{\kappa_{k}^{2}(1+h_{k}\lambda_{4})}{\lambda_{4}(\lambda_{4}+\kappa_{k})}\textbf{g}_{j}^{\mathrm{T}}
\end{array}\right]^{+}\otimes I_{n}\Big)\Big(\small\left[\begin{array}{c}
F_{k}\\
\frac{\kappa_{k}^{2}(1+h_{k}\lambda_{4})}{\lambda_{4}(\lambda_{4}+\kappa_{k})}\textbf{g}_{j}^{\mathrm{T}}
\end{array}\right]\otimes I_{n}\Big)\Big]\sum_{i=1}^{n}\sum_{l=1}^{q}\varpi_{li}\textbf{g}_{l}\otimes\textbf{e}_{i}\nonumber\\
&=&\Big[I_{q}\otimes I_{n}-\Big(\small\left[\begin{array}{c}
F_{k}\\
\frac{\kappa_{k}^{2}(1+h_{k}\lambda_{4})}{\lambda_{4}(\lambda_{4}+\kappa_{k})}\textbf{g}_{j}^{\mathrm{T}}
\end{array}\right]^{+}\small\left[\begin{array}{c}
F_{k}\\
\frac{\kappa_{k}^{2}(1+h_{k}\lambda_{4})}{\lambda_{4}(\lambda_{4}+\kappa_{k})}\textbf{g}_{j}^{\mathrm{T}}
\end{array}\right]\otimes I_{n}\Big)\Big]\sum_{i=1}^{n}\sum_{l=1}^{q}\varpi_{li}\textbf{g}_{l}\otimes\textbf{e}_{i}\nonumber\\
&=&\Big[\Big(I_{q}-\small\left[\begin{array}{c}
F_{k}\\
\frac{\kappa_{k}^{2}(1+h_{k}\lambda_{4})}{\lambda_{4}(\lambda_{4}+\kappa_{k})}\textbf{g}_{j}^{\mathrm{T}}
\end{array}\right]^{+}\small\left[\begin{array}{c}
F_{k}\\
\frac{\kappa_{k}^{2}(1+h_{k}\lambda_{4})}{\lambda_{4}(\lambda_{4}+\kappa_{k})}\textbf{g}_{j}^{\mathrm{T}}
\end{array}\right]\Big)\otimes I_{n}\Big]\sum_{i=1}^{n}\sum_{l=1}^{q}\varpi_{li}\textbf{g}_{l}\otimes\textbf{e}_{i}\nonumber\\
&=&\sum_{i=1}^{n}\sum_{l=1}^{q}\varpi_{li}\Big(\textbf{g}_{l}-\small\left[\begin{array}{c}
F_{k}\\
\frac{\kappa_{k}^{2}(1+h_{k}\lambda_{4})}{\lambda_{4}(\lambda_{4}+\kappa_{k})}\textbf{g}_{j}^{\mathrm{T}}
\end{array}\right]^{+}\small\left[\begin{array}{c}
F_{k}\\
\frac{\kappa_{k}^{2}(1+h_{k}\lambda_{4})}{\lambda_{4}(\lambda_{4}+\kappa_{k})}\textbf{g}_{j}^{\mathrm{T}}
\end{array}\right]\textbf{g}_{l}\Big)\otimes \textbf{e}_{i},\label{gsolution1a}
\end{eqnarray} where $\varpi_{li}\in\mathbb{C}$ and $j=1,\ldots,q$. Note that by Proposition 6.1.6 of \cite[p.~399]{Bernstein:2009}, $F_{k}^{\mathrm{T}}(F_{k}^{\mathrm{T}})^{+}=F_{k}^{\mathrm{T}}(F_{k}^{+})^{\mathrm{T}}=(F_{k}^{+}F_{k})^{\mathrm{T}}=F_{k}^{+}F_{k}$. It follows from Fact 6.5.17 of \cite[p.~427]{Bernstein:2009} that 
\begin{eqnarray}
\small\left[\begin{array}{c}
F_{k}\\
\frac{\kappa_{k}^{2}(1+h_{k}\lambda_{4})}{\lambda_{4}(\lambda_{4}+\kappa_{k})}\textbf{g}_{j}^{\mathrm{T}}
\end{array}\right]^{+}=\left[\begin{array}{cc}
F_{k}^{+}(I_{q}-\frac{\kappa_{k}^{2}(1+h_{k}\lambda_{4})}{\lambda_{4}(\lambda_{4}+\kappa_{k})}\psi_{k}\textbf{g}_{j}^{\mathrm{T}}) & \psi_{k}
\end{array}\right],
\end{eqnarray} where $\psi_{k}$ is given by (\ref{psik}). Hence, it follows that for every $j,l=1,\ldots,q$,
\begin{eqnarray}
\textbf{g}_{l}-\small\left[\begin{array}{c}
F_{k}\\
\frac{\kappa_{k}^{2}(1+h_{k}\lambda_{4})}{\lambda_{4}(\lambda_{4}+\kappa_{k})}\textbf{g}_{j}^{\mathrm{T}}
\end{array}\right]^{+}\small\left[\begin{array}{c}
F_{k}\\
\frac{\kappa_{k}^{2}(1+h_{k}\lambda_{4})}{\lambda_{4}(\lambda_{4}+\kappa_{k})}\textbf{g}_{j}^{\mathrm{T}}
\end{array}\right]\textbf{g}_{l}&=&\textbf{g}_{l}-\left[\begin{array}{cc}
F_{k}^{+}(I_{q}-\frac{\kappa_{k}^{2}(1+h_{k}\lambda_{4})}{\lambda_{4}(\lambda_{4}+\kappa_{k})}\psi_{k}\textbf{g}_{j}^{\mathrm{T}}) & \psi_{k}
\end{array}\right]\small\left[\begin{array}{c}
F_{k}\\
\frac{\kappa_{k}^{2}(1+h_{k}\lambda_{4})}{\lambda_{4}(\lambda_{4}+\kappa_{k})}\textbf{g}_{j}^{\mathrm{T}}
\end{array}\right]\textbf{g}_{l}\nonumber\\
&=&\textbf{g}_{l}-\small\left[\begin{array}{cc}
F_{k}^{+}(I_{q}-\frac{\kappa_{k}^{2}(1+h_{k}\lambda_{4})}{\lambda_{4}(\lambda_{4}+\kappa_{k})}\psi_{k}\textbf{g}_{j}^{\mathrm{T}}) & \psi_{k}
\end{array}\right]\small\left[\begin{array}{c}
F_{k}\textbf{g}_{l}\\
\frac{\kappa_{k}^{2}(1+h_{k}\lambda_{4})}{\lambda_{4}(\lambda_{4}+\kappa_{k})}\textbf{g}_{j}^{\mathrm{T}}\textbf{g}_{l}
\end{array}\right]\nonumber\\
&=&\textbf{g}_{l}-F_{k}^{+}\Big(I_{q}-\frac{\kappa_{k}^{2}(1+h_{k}\lambda_{4})}{\lambda_{4}(\lambda_{4}+\kappa_{k})}\psi_{k}\textbf{g}_{j}^{\mathrm{T}}\Big)F_{k}\textbf{g}_{l}\nonumber\\
&&-\frac{\kappa_{k}^{2}(1+h_{k}\lambda_{4})}{\lambda_{4}(\lambda_{4}+\kappa_{k})}(\textbf{g}_{j}^{\mathrm{T}}\textbf{g}_{l})\psi_{k}\nonumber\\
&=&\textbf{g}_{l}-F_{k}^{+}F_{k}\textbf{g}_{l}+\frac{\kappa_{k}^{2}(1+h_{k}\lambda_{4})}{\lambda_{4}(\lambda_{4}+\kappa_{k})}(\textbf{g}_{j}^{\mathrm{T}}F_{k}\textbf{g}_{l})F_{k}^{+}\psi_{k}\nonumber\\
&&-\frac{\kappa_{k}^{2}(1+h_{k}\lambda_{4})}{\lambda_{4}(\lambda_{4}+\kappa_{k})}(\textbf{g}_{j}^{\mathrm{T}}\textbf{g}_{l})\psi_{k}.
\end{eqnarray} Thus, (\ref{gsolution1a}) becomes 
\begin{eqnarray}\label{gsoa}
\textbf{v}=\sum_{i=1}^{n}\sum_{l=1}^{q}\varpi_{li}\Big(\textbf{g}_{l}-F_{k}^{+}F_{k}\textbf{g}_{l}+\frac{\kappa_{k}^{2}(1+h_{k}\lambda_{4})}{\lambda_{4}(\lambda_{4}+\kappa_{k})}(\textbf{g}_{j}^{\mathrm{T}}F_{k}\textbf{g}_{l})F_{k}^{+}\psi_{k}-\frac{\kappa_{k}^{2}(1+h_{k}\lambda_{4})}{\lambda_{4}(\lambda_{4}+\kappa_{k})}(\textbf{g}_{j}^{\mathrm{T}}\textbf{g}_{l})\psi_{k}\Big)\otimes \textbf{e}_{i}.
\end{eqnarray} Hence, $\textbf{x}_{1}=\frac{1+h_{k}\lambda_{4}}{\lambda_{4}}\textbf{v}$, $\textbf{x}_{2}=\textbf{v}\neq\textbf{0}_{nq\times 1}$ given by (\ref{gsoa}), and 
\begin{eqnarray}
\textbf{x}_{3}&=&\frac{\kappa_{k}+\kappa_{k}h_{k}\lambda_{4}}{\lambda_{4}(\lambda_{4}+\kappa_{k})}E_{n\times nq}^{[j]}\textbf{v}\nonumber\\
&=&\frac{\kappa_{k}+\kappa_{k}h_{k}\lambda_{4}}{\lambda_{4}(\lambda_{4}+\kappa_{k})}(\textbf{g}_{j}^{\mathrm{T}}\otimes I_{n})\textbf{v}\nonumber\\
&=&\frac{\kappa_{k}+\kappa_{k}h_{k}\lambda_{4}}{\lambda_{4}(\lambda_{4}+\kappa_{k})}\sum_{i=1}^{n}\sum_{l=1}^{q}\varpi_{li}(\textbf{g}_{j}^{\mathrm{T}}\otimes I_{n})\Big(\Big(\textbf{g}_{l}-F_{k}^{+}F_{k}\textbf{g}_{l}+\frac{\kappa_{k}^{2}(1+h_{k}\lambda_{4})}{\lambda_{4}(\lambda_{4}+\kappa_{k})}(\textbf{g}_{j}^{\mathrm{T}}F_{k}\textbf{g}_{l})F_{k}^{+}\psi_{k}\nonumber\\
&&-\frac{\kappa_{k}^{2}(1+h_{k}\lambda_{4})}{\lambda_{4}(\lambda_{4}+\kappa_{k})}(\textbf{g}_{j}^{\mathrm{T}}\textbf{g}_{l})\psi_{k}\Big)\otimes \textbf{e}_{i}\Big)\nonumber\\
&=&\frac{\kappa_{k}+\kappa_{k}h_{k}\lambda_{4}}{\lambda_{4}(\lambda_{4}+\kappa_{k})}\sum_{i=1}^{n}\sum_{l=1}^{q}\varpi_{li}\Big(\textbf{g}_{j}^{\mathrm{T}}\textbf{g}_{l}-\textbf{g}_{j}^{\mathrm{T}}F_{k}^{+}F_{k}\textbf{g}_{l}+\frac{\kappa_{k}^{2}(1+h_{k}\lambda_{4})}{\lambda_{4}(\lambda_{4}+\kappa_{k})}(\textbf{g}_{j}^{\mathrm{T}}F_{k}\textbf{g}_{l})\textbf{g}_{j}^{\mathrm{T}}F_{k}^{+}\psi_{k}\nonumber\\
&&-\frac{\kappa_{k}^{2}(1+h_{k}\lambda_{4})}{\lambda_{4}(\lambda_{4}+\kappa_{k})}(\textbf{g}_{j}^{\mathrm{T}}\textbf{g}_{l})\textbf{g}_{j}^{\mathrm{T}}\psi_{k}\Big)\otimes \textbf{e}_{i},\label{gsox3}
\end{eqnarray}
where not all of $\omega_{\ell i}$ and $\varpi_{li}$ are zero. The corresponding eigenvectors for $\lambda_{4}$ are given by
\begin{eqnarray}\label{eigv4}
&&\textbf{x}=\nonumber\\
&&\Big[\frac{1+h_{k}\lambda_{4}^{*}}{\lambda_{4}^{*}}\sum_{i=1}^{n}\sum_{l=1}^{q}\varpi_{li}\Big(\textbf{g}_{l}-F_{k}^{+}F_{k}\textbf{g}_{l}+\frac{\kappa_{k}^{2}(1+h_{k}\lambda_{4})}{\lambda_{4}(\lambda_{4}+\kappa_{k})}(\textbf{g}_{j}^{\mathrm{T}}F_{k}\textbf{g}_{l})F_{k}^{+}\psi_{k}-\frac{\kappa_{k}^{2}(1+h_{k}\lambda_{4})}{\lambda_{4}(\lambda_{4}+\kappa_{k})}(\textbf{g}_{j}^{\mathrm{T}}\textbf{g}_{l})\psi_{k}\Big)^{*}\otimes \textbf{e}_{i}^{\mathrm{T}},\nonumber\\
&&\sum_{i=1}^{n}\sum_{l=1}^{q}\varpi_{li}\Big(\textbf{g}_{l}-F_{k}^{+}F_{k}\textbf{g}_{l}+\frac{\kappa_{k}^{2}(1+h_{k}\lambda_{4})}{\lambda_{4}(\lambda_{4}+\kappa_{k})}(\textbf{g}_{j}^{\mathrm{T}}F_{k}\textbf{g}_{l})F_{k}^{+}\psi_{k}-\frac{\kappa_{k}^{2}(1+h_{k}\lambda_{4})}{\lambda_{4}(\lambda_{4}+\kappa_{k})}(\textbf{g}_{j}^{\mathrm{T}}\textbf{g}_{l})\psi_{k}\Big)^{*}\otimes \textbf{e}_{i}^{\mathrm{T}},\nonumber\\
&&\frac{\kappa_{k}+\kappa_{k}h_{k}\lambda_{4}^{*}}{\lambda_{4}^{*}(\lambda_{4}^{*}+\kappa_{k})}\sum_{i=1}^{n}\sum_{l=1}^{q}\varpi_{li}\Big(\textbf{g}_{j}^{\mathrm{T}}\textbf{g}_{l}-\textbf{g}_{j}^{\mathrm{T}}F_{k}^{+}F_{k}\textbf{g}_{l}+\frac{\kappa_{k}^{2}(1+h_{k}\lambda_{4})}{\lambda_{4}(\lambda_{4}+\kappa_{k})}(\textbf{g}_{j}^{\mathrm{T}}F_{k}\textbf{g}_{l})\textbf{g}_{j}^{\mathrm{T}}F_{k}^{+}\psi_{k}\nonumber\\
&&-\frac{\kappa_{k}^{2}(1+h_{k}\lambda_{4})}{\lambda_{4}(\lambda_{4}+\kappa_{k})}(\textbf{g}_{j}^{\mathrm{T}}\textbf{g}_{l})\textbf{g}_{j}^{\mathrm{T}}\psi_{k}\Big)^{*}\otimes \textbf{e}_{i}^{\mathrm{T}}\Big]^{*},
\end{eqnarray} where $\varpi_{li}\in\mathbb{C}$ and not all of them are zero. Therefore,
$\ker\Big(A_{k}^{[j]}+h_{k}A_{{\mathrm{c}}k}-\lambda_{4} I_{2nq+n}\Big)$ is given by (\ref{egns3}).

If $\frac{\kappa_{k}}{\lambda}+\lambda+\kappa_{k} h_{k}=0$, then $\frac{\kappa_{k}^{2}(1+h_{k}\lambda)}{\lambda(\lambda+\kappa_{k})}=-\frac{\kappa_{k}\lambda}{\lambda+\kappa_{k}}\neq0$ since $\lambda\neq0$ and $\kappa_{k}\neq0$. In this case, it follows from (\ref{eqn_v}) and (\ref{Az=b}) that
\begin{eqnarray} &&\frac{\kappa_{k}^{2}(1+h_{k}\lambda)}{\lambda(\lambda+\kappa_{k})}W_{k}^{[j]}\textbf{v}=(I_{q}\otimes P_{k})\sum_{i=1}^{n}\sum_{l=0}^{q-1-{\mathrm{rank}}(L_{k})}\omega_{li}\textbf{w}_{l}\otimes\textbf{e}_{i},\label{id_1}\\
&&\Big(\frac{\mu_{k}}{\lambda}+\mu_{k} h_{k}+\eta_{k}\Big)(L_{k}\otimes I_{n})\textbf{v}=\sum_{i=1}^{n}\sum_{l=0}^{q-1-{\mathrm{rank}}(L_{k})}\omega_{li}\textbf{w}_{l}\otimes\textbf{e}_{i}.\label{id_2}
\end{eqnarray} Since $I_{q}\otimes P_{k}$ is nonsingular, pre-multiplying $(I_{q}\otimes P_{k})^{-1}$ on both sides of (\ref{id_1}) yields
\begin{eqnarray}\label{id_3}
\frac{\kappa_{k}^{2}(1+h_{k}\lambda)}{\lambda(\lambda+\kappa_{k})}(\textbf{1}_{q\times 1}\otimes I_{n})E_{n\times nq}^{[j]}\textbf{v}=\sum_{i=1}^{n}\sum_{l=0}^{q-1-{\mathrm{rank}}(L_{k})}\omega_{li}\textbf{w}_{l}\otimes\textbf{e}_{i}
\end{eqnarray}
Since by $i$) of Lemma 4.1 of \cite{HZ:TR:2013}, $(\textbf{1}_{q\times 1}\otimes I_{n})E_{n\times nq}^{[j]}$ is idempotent, it follows from (\ref{id_3}) and $ii$) of Lemma 4.1 of \cite{HZ:TR:2013} that
 \begin{eqnarray}
 \sum_{i=1}^{n}\sum_{l=0}^{q-1-{\mathrm{rank}}(L_{k})}\omega_{li}\textbf{w}_{l}\otimes\textbf{e}_{i}=\sum_{i=1}^{n}\sum_{l=0}^{q-1-{\mathrm{rank}}(L_{k})}\omega_{li}w_{lj}\textbf{w}_{0}\otimes\textbf{e}_{i},
 \end{eqnarray}
 and hence, 
\begin{eqnarray}
\sum_{i=1}^{n}\Big(\omega_{0i}-\sum_{l=0}^{q-1-{\mathrm{rank}}(L_{k})}\omega_{li}w_{lj}\Big)\textbf{w}_{0}\otimes\textbf{e}_{i}+\sum_{i=1}^{n}\sum_{l=1}^{q-1-{\mathrm{rank}}(L_{k})}\omega_{li}\textbf{w}_{l}\otimes\textbf{e}_{i}=\textbf{0}_{nq\times 1},
\end{eqnarray} which implies that $\omega_{0i}-\sum_{l=0}^{q-1-{\mathrm{rank}}(L_{k})}\omega_{li}w_{lj}=0$ and $\omega_{\ell i}=0$ for every $i=1,\ldots,n$, $j=1,\ldots,q$, and $\ell=1,\ldots,q-1-{\mathrm{rank}}(L_{k})$. Consequently, (\ref{id_3}) and (\ref{id_2}) can be simplified as
\begin{eqnarray} &&\frac{\kappa_{k}^{2}(1+h_{k}\lambda)}{\lambda(\lambda+\kappa_{k})}(\textbf{1}_{q\times 1}\otimes I_{n})E_{n\times nq}^{[j]}\textbf{v}=\sum_{i=1}^{n}\omega_{0i}\textbf{w}_{0}\otimes\textbf{e}_{i},\label{new_1}\\
&&\Big(\frac{\mu_{k}}{\lambda}+\mu_{k} h_{k}+\eta_{k}\Big)(L_{k}\otimes I_{n})\textbf{v}=\sum_{i=1}^{n}\omega_{0i}\textbf{w}_{0}\otimes\textbf{e}_{i}.\label{new_2}
\end{eqnarray} It follows from $ii$) of Lemma 4.1 of \cite{HZ:TR:2013} that (\ref{new_1}) has a specific solution 
\begin{eqnarray}\label{vss}
\textbf{v}=\Big(\frac{\kappa_{k}^{2}(1+h_{k}\lambda)}{\lambda(\lambda+\kappa_{k})}\Big)^{-1}\sum_{i=1}^{n}\omega_{0i}\textbf{w}_{0}\otimes\textbf{e}_{i}.
\end{eqnarray} Substituting (\ref{vss}) into (\ref{new_2}) yields $\sum_{i=1}^{n}\omega_{0i}\textbf{w}_{0}\otimes\textbf{e}_{i}=\textbf{0}_{nq\times 1}$, which implies that $\omega_{0i}=0$ for every $i=1,\ldots,n$. Hence, (\ref{new_1}) and (\ref{new_2}) can be further simplified as 
\begin{eqnarray} 
&&\frac{\kappa_{k}^{2}(1+h_{k}\lambda)}{\lambda(\lambda+\kappa_{k})}(\textbf{1}_{q\times 1}\otimes I_{n})E_{n\times nq}^{[j]}\textbf{v}=\textbf{0}_{nq\times 1},\label{gs0_1}\\
&&\Big(\frac{\mu_{k}}{\lambda}+\mu_{k} h_{k}+\eta_{k}\Big)(L_{k}\otimes I_{n})\textbf{v}=\textbf{0}_{nq\times 1}.\label{gs0_2}
\end{eqnarray} If $\frac{\mu_{k}}{\lambda}+\mu_{k} h_{k}+\eta_{k}\neq0$, note that for $\frac{\kappa_{k}}{\lambda}+\lambda+\kappa_{k} h_{k}=0$, $\det\Big[\Big(\frac{\mu_{k}}{\lambda}+\mu_{k} h_{k}+\eta_{k}\Big)(L_{k}\otimes I_{n})+\Big(\frac{\kappa_{k}}{\lambda}+\lambda+\kappa_{k} h_{k}\Big)I_{nq}\Big]=\det\Big[\Big(\frac{\mu_{k}}{\lambda}+\mu_{k} h_{k}+\eta_{k}\Big)(L_{k}\otimes I_{n})\Big]=0$. Hence, the general solution $\textbf{v}$ to (\ref{gs0_1}) and (\ref{gs0_2}) is given by the form of (\ref{gsoa}) in which $\lambda_{4}$ is replaced by $\lambda_{5,6}$ satisfying $\frac{\kappa_{k}}{\lambda_{5,6}}+\lambda_{5,6}+\kappa_{k} h_{k}=0$. Thus, this case is similar to the previous case where (\ref{lambda4}) still holds for $\lambda_{4}$ being replaced by $\lambda_{5,6}$, where
\begin{eqnarray}\label{lambda5}
\lambda_{5,6}=-\frac{\kappa_{k}h_{k}}{2}\pm\frac{1}{2}\sqrt{\kappa_{k}^{2}h_{k}^{2}-4\kappa_{k}}.
\end{eqnarray} Thus, $\lambda=\lambda_{5,6}$ are indeed the eigenvalues of $A_{k}^{[j]}+h_{k}A_{{\mathrm{c}}k}$ and the corresponding eigenvectors are given by the form (\ref{eigv4}) with $\lambda_{4}$ being replaced by $\lambda_{5,6}$. 

Otherwise, if $\frac{\mu_{k}}{\lambda}+\mu_{k} h_{k}+\eta_{k}=0$ and $\frac{\kappa_{k}}{\lambda}+\lambda+\kappa_{k} h_{k}=0$, then $\mu_{k}(\frac{1}{\lambda}+h_{k})=-\eta_{k}$ and $\kappa_{k}(\frac{1}{\lambda}+h_{k})=-\lambda$. Again, since $\lambda\neq0$, it follows from $\frac{\kappa_{k}}{\lambda}+\lambda+\kappa_{k} h_{k}=0$ that $\kappa_{k}\neq0$. If $\mu_{k}=0$, then it follows from $\mu_{k}(\frac{1}{\lambda}+h_{k})=-\eta_{k}$ that $\eta_{k}=0$. In this case, $\lambda=\lambda_{5,6}$ are the eigenvalues of $A_{k}^{[j]}+h_{k}A_{{\mathrm{c}}k}$. Furthermore, (\ref{gs0_2}) becomes trivial and (\ref{gs0_1}) is equivalent to $E_{n\times nq}^{[j]}\textbf{v}=\textbf{0}_{n\times 1}$, that is, $(\textbf{g}_{j}^{\mathrm{T}}\otimes I_{n})\textbf{v}=\textbf{0}_{n\times 1}$. It follows from $vi$) of Proposition 6.1.7 of \cite[p.~400]{Bernstein:2009} and $viii$) of Proposition 6.1.6 of \cite[p.~399]{Bernstein:2009} that the general solution $\textbf{v}$ to $(\textbf{g}_{j}^{\mathrm{T}}\otimes I_{n})\textbf{v}=\textbf{0}_{n\times 1}$ is given by the form
\begin{eqnarray}
\textbf{v}&=&\Big[I_{nq}-(\textbf{g}_{j}^{\mathrm{T}}\otimes I_{n})^{+}(\textbf{g}_{j}^{\mathrm{T}}\otimes I_{n})\Big]\sum_{i=1}^{n}\sum_{l=1}^{q}\varpi_{li}\textbf{g}_{l}\otimes\textbf{e}_{i}\nonumber\\
&=&\Big[I_{nq}-((\textbf{g}_{j}^{\mathrm{T}})^{+}\otimes I_{n})(\textbf{g}_{j}^{\mathrm{T}}\otimes I_{n})\Big]\sum_{i=1}^{n}\sum_{l=1}^{q}\varpi_{li}\textbf{g}_{l}\otimes\textbf{e}_{i}\nonumber\\
&=&\Big[I_{q}\otimes I_{n}-(((\textbf{g}_{j}^{\mathrm{T}})^{+}\textbf{g}_{j}^{\mathrm{T}})\otimes I_{n})\Big]\sum_{i=1}^{n}\sum_{l=1}^{q}\varpi_{li}\textbf{g}_{l}\otimes\textbf{e}_{i}\nonumber\\
&=&\Big[(I_{q}-((\textbf{g}_{j}^{\mathrm{T}})^{+}\textbf{g}_{j}^{\mathrm{T}}))\otimes I_{n}\Big]\sum_{i=1}^{n}\sum_{l=1}^{q}\varpi_{li}\textbf{g}_{l}\otimes\textbf{e}_{i}\nonumber\\
&=&\sum_{i=1}^{n}\sum_{l=1}^{q}\varpi_{li}(\textbf{g}_{l}-((\textbf{g}_{j}^{\mathrm{T}})^{+}\textbf{g}_{j}^{\mathrm{T}})\textbf{g}_{l})\otimes \textbf{e}_{i},
\end{eqnarray} where $\varpi_{li}\in\mathbb{C}$ and $j=1,\ldots,q$. Note that it follows from Fact 6.3.2 of \cite[p.~404]{Bernstein:2009} that $\textbf{g}_{j}^{+}=\textbf{g}_{j}^{\mathrm{T}}$, and hence, $(\textbf{g}_{j}^{\mathrm{T}})^{+}=\textbf{g}_{j}$ for every $j=1,\ldots,q$. Then we have
\begin{eqnarray}\label{newv}
\textbf{v}&=&\sum_{i=1}^{n}\sum_{l=1}^{q}\varpi_{li}(\textbf{g}_{l}-(\textbf{g}_{j}\textbf{g}_{j}^{\mathrm{T}})\textbf{g}_{l})\otimes \textbf{e}_{i}\nonumber\\
&=&\sum_{i=1}^{n}\sum_{l=1}^{q}\varpi_{li}(\textbf{g}_{l}-(\textbf{g}_{j}^{\mathrm{T}}\textbf{g}_{l})\textbf{g}_{j})\otimes \textbf{e}_{i}.
\end{eqnarray} Hence, $\textbf{x}_{1}=\frac{1+h_{k}\lambda_{5,6}}{\lambda_{5,6}}\textbf{v}$, $\textbf{x}_{2}=\textbf{v}\neq\textbf{0}_{nq\times 1}$ where $\textbf{v}$ is given by (\ref{newv}), and $\textbf{x}_{3}=\textbf{0}_{n\times 1}$. The corresponding eigenvectors for $\lambda_{5,6}$ in this case are given by 
\begin{eqnarray}
\textbf{x}=\Big[\frac{1+h_{k}\lambda_{5,6}^{*}}{\lambda_{5,6}^{*}}\sum_{i=1}^{n}\sum_{l=1}^{q}\varpi_{li}(\textbf{g}_{l}-(\textbf{g}_{j}^{\mathrm{T}}\textbf{g}_{l})\textbf{g}_{j})^{\mathrm{T}}\otimes \textbf{e}_{i}^{\mathrm{T}},\sum_{i=1}^{n}\sum_{l=1}^{q}\varpi_{li}(\textbf{g}_{l}-(\textbf{g}_{j}^{\mathrm{T}}\textbf{g}_{l})\textbf{g}_{j})^{\mathrm{T}}\otimes \textbf{e}_{i}^{\mathrm{T}},\textbf{0}_{1\times n}\Big]^{*},
\end{eqnarray} where $\varpi_{li}\in\mathbb{C}$ and not all of them are zero. Consequently, in this case
$\ker\Big(A_{k}^{[j]}+h_{k}A_{{\mathrm{c}}k}-\lambda_{5,6} I_{2nq+n}\Big)$ is given by (\ref{egns4}).

Finally, if $\mu_{k}\neq0$, then it follows from $\mu_{k}(\frac{1}{\lambda}+h_{k})=-\eta_{k}$ that $\frac{1}{\lambda}+h_{k}=-\frac{\eta_{k}}{\mu_{k}}$. Together with $\kappa_{k}(\frac{1}{\lambda}+h_{k})=-\lambda$, we have $\lambda=\frac{\kappa_{k}\eta_{k}}{\mu_{k}}$. Since $\lambda\neq0$, it follows that $\eta_{k}\neq0$. Substituting this $\lambda$ into $\frac{1}{\lambda}+h_{k}=-\frac{\eta_{k}}{\mu_{k}}$ yields $h_{k}=-\frac{\eta_{k}}{\mu_{k}}-\frac{\mu_{k}}{\kappa_{k}\eta_{k}}<0$, which is a contradiction since $h_{k}\geq0$. Hence, this case is impossible.

\textit{Case 2.} If $\lambda=-\kappa_{k}$, then $\kappa_{k}\neq0$ and (\ref{detcon}) becomes
\begin{eqnarray}\label{detcase2}
\det\small\left[\begin{array}{cc}
\Big(\frac{\mu_{k}}{\kappa_{k}}(\kappa_{k} h_{k}-1)+\eta_{k}\Big)(L_{k}\otimes P_{k})+(\kappa_{k} h_{k}-1-\kappa_{k})(I_{q}\otimes P_{k}) & -\kappa_{k}(\textbf{1}_{q\times 1}\otimes P_{k})\\
(\kappa_{k} h_{k}-1)E_{n\times nq}^{[j]} & \textbf{0}_{n\times n}
\end{array}\right]=0.
\end{eqnarray} If $\kappa_{k}h_{k}=1$, then clearly (\ref{detcase2}) holds. In this case,
\begin{eqnarray*}
&&\det\Big[\Big(\frac{\mu_{k}}{\lambda}+\mu_{k} h_{k}+\eta_{k}\Big)(L_{k}\otimes P_{k})+\Big(\frac{\kappa_{k}}{\lambda}+\lambda+\kappa_{k} h_{k}\Big)(I_{q}\otimes P_{k})\Big]\nonumber\\
&&=\det\Big[\Big(-\frac{\mu_{k}}{\kappa_{k}}+\mu_{k} h_{k}+\eta_{k}\Big)(L_{k}\otimes I_{n})-\kappa_{k}I_{nq}\Big]\det(I_{q}\otimes P_{k})\nonumber\\
&&=\kappa_{k}^{nq}\det\Big[\Big(-\frac{\mu_{k}}{\kappa_{k}^{2}}+\frac{\mu_{k} h_{k}}{\kappa_{k}}+\frac{\eta_{k}}{\kappa_{k}}\Big)(L_{k}\otimes I_{n})-I_{nq}\Big]\det(I_{q}\otimes P_{k})\nonumber\\
&&=\kappa_{k}^{nq}\det\Big[\frac{\eta_{k}}{\kappa_{k}}(L_{k}\otimes I_{n})-I_{nq}\Big](\det(P_{k}))^{q}.
\end{eqnarray*} Hence, $\det\Big[\Big(\frac{\mu_{k}}{\lambda}+\mu_{k} h_{k}+\eta_{k}\Big)(L_{k}\otimes P_{k})+\Big(\frac{\kappa_{k}}{\lambda}+\lambda+\kappa_{k} h_{k}\Big)(I_{q}\otimes P_{k})\Big]=0$ if and only if $1\in{\mathrm{spec}}(\frac{\eta_{k}}{\kappa_{k}}L_{k})$. Thus, if 
$1\in{\mathrm{spec}}(\frac{\eta_{k}}{\kappa_{k}}L_{k})$ and $\kappa_{k}h_{k}=1$, then $\lambda=-\kappa_{k}$ is indeed an eigenvalue of $A_{k}^{[j]}+h_{k}A_{{\mathrm{c}}k}$. Clearly when $\kappa_{k}h_{k}=1$ and $\lambda=-\kappa_{k}$, $\textbf{x}_{1}=\frac{1+h_{k}\lambda}{\lambda}\textbf{x}_{2}=\textbf{0}_{nq\times 1}$, (\ref{hx3}) becomes trivial, and (\ref{hx2}) becomes
\begin{eqnarray}\label{x2x3}
(\eta_{k}(L_{k}\otimes P_{k})-\kappa_{k}I_{q}\otimes P_{k})\textbf{x}_{2}-\kappa_{k}(\textbf{1}_{q\times 1}\otimes P_{k})\textbf{x}_{3}=\textbf{0}_{nq\times1}.
\end{eqnarray} Pre-multiplying $E_{n\times nq}^{[j]}(I_{q}\otimes P_{k}^{-1})$ on both sides of (\ref{x2x3}) yields
\begin{eqnarray}\label{x3x2}
\textbf{x}_{3}=\Big[\frac{\eta_{k}}{\kappa_{k}}(L_{k}\otimes I_{n})-I_{nq}\Big]\textbf{x}_{2}.
\end{eqnarray} Note that $\textbf{x}_{2}$ can be chosen arbitrarily in $\mathbb{C}^{nq}$ other than $\textbf{0}_{nq\times 1}$. Then $\textbf{x}_{2}$ can be represented as $\textbf{x}_{2}=\sum_{i=1}^{n}\sum_{l=1}^{q}\\\alpha_{li}(\textbf{g}_{l}\otimes\textbf{e}_{i})$, where $\alpha_{li}\in\mathbb{C}$, not all of $\alpha_{li}$ are zero. Then it follows from (\ref{x3x2}) that $\textbf{x}_{3}=\sum_{i=1}^{n}\sum_{l=1}^{q}\frac{\eta_{k}}{\kappa_{k}}\alpha_{li}(L_{k}\otimes I_{n})(\textbf{g}_{l}\otimes\textbf{e}_{i})-\sum_{i=1}^{n}\sum_{l=1}^{q}\alpha_{li}(\textbf{g}_{l}\otimes\textbf{e}_{i})=\sum_{i=1}^{n}\sum_{l=1}^{q}\frac{\eta_{k}}{\kappa_{k}}\alpha_{li}(L_{k}\textbf{g}_{l}\otimes\textbf{e}_{i})-\sum_{i=1}^{n}\sum_{l=1}^{q}\alpha_{li}(\textbf{g}_{l}\otimes\textbf{e}_{i})$, where $\alpha_{li}\in\mathbb{C}$ and not all of $\alpha_{il}$ are zero. Clearly such $\textbf{x}_{i}$, $i=1,2,3$, satisfy (\ref{Aeig_1})--(\ref{Aeig_3}). Thus, the corresponding eigenvectors for the eigenvalue $\lambda=\lambda_{3}$ are given by
\begin{eqnarray}
\textbf{x}=\Big[\textbf{0}_{1\times nq},\sum_{i=1}^{n}\sum_{l=1}^{q}\alpha_{li}(\textbf{g}_{l}\otimes\textbf{e}_{i})^{\mathrm{T}},\sum_{i=1}^{n}\sum_{l=1}^{q}\frac{\eta_{k}}{\kappa_{k}}\alpha_{li}(L_{k}\textbf{g}_{l}\otimes\textbf{e}_{i})^{\mathrm{T}}-\sum_{i=1}^{n}\sum_{l=1}^{q}\alpha_{li}(\textbf{g}_{l}\otimes\textbf{e}_{i})^{\mathrm{T}}\Big]^{*},
\end{eqnarray} where $\alpha_{li}\in\mathbb{C}$, not all of $\alpha_{il}$ are zero, and 
\begin{eqnarray}\label{lambda3}
\lambda_{3}=-\kappa_{k}.
\end{eqnarray} Therefore,
$\ker\Big(A_{k}^{[j]}+h_{k}A_{{\mathrm{c}}k}-\lambda_{3} I_{2nq+n}\Big)$ is given by (\ref{egns5}).

Now we consider the case where $\kappa_{k}h_{k}\neq1$. Then in this case (\ref{detcase2}) holds if and only if the equation
\begin{eqnarray}
\small\left[\begin{array}{cc}
\Big(\frac{\mu_{k}}{\kappa_{k}}(\kappa_{k} h_{k}-1)+\eta_{k}\Big)(L_{k}\otimes P_{k})+(\kappa_{k} h_{k}-1-\kappa_{k})(I_{q}\otimes P_{k}) & -\kappa_{k}(\textbf{1}_{q\times 1}\otimes P_{k})\\
(\kappa_{k} h_{k}-1)E_{n\times nq}^{[j]} & \textbf{0}_{n\times n}
\end{array}\right]\textbf{u}=\textbf{0}_{(nq+n)\times 1}\label{deteqn}
\end{eqnarray} has a nontrivial solution $\textbf{u}\in\mathbb{C}^{nq+n}$. Let $\textbf{u}=[\textbf{u}_{1}^{*},\ldots,\textbf{u}_{q}^{*},\textbf{u}_{0}^{*}]^{*}$, where $\textbf{u}_{i}\in\mathbb{C}^{n}$, $i=0,1,\ldots,q$. Then it follows from (\ref{deteqn}) that 
\begin{eqnarray}
\Big(\frac{\mu_{k}}{\kappa_{k}}(\kappa_{k} h_{k}-1)+\eta_{k}\Big)(L_{k}\otimes P_{k})[\textbf{u}_{1}^{*},\ldots,\textbf{u}_{q}^{*}]^{*}+(\kappa_{k} h_{k}-1-\kappa_{k})(I_{q}\otimes P_{k})[\textbf{u}_{1}^{*},\ldots,\textbf{u}_{q}^{*}]^{*}\nonumber\\
-\kappa_{k}(\textbf{1}_{q\times 1}\otimes P_{k})\textbf{u}_{0}=\textbf{0}_{nq\times 1},\label{Enqu1}\\
(\kappa_{k} h_{k}-1)E_{n\times nq}^{[j]}[\textbf{u}_{1}^{*},\ldots,\textbf{u}_{q}^{*}]^{*}=\textbf{0}_{n\times 1}.\label{Enqu}
\end{eqnarray}   

If $\frac{\mu_{k}}{\kappa_{k}}(\kappa_{k} h_{k}-1)+\eta_{k}=0$, in this case, since $\lambda=-\kappa_{k}$, then it follows that
\begin{eqnarray*}
\det\Big[\Big(\frac{\mu_{k}}{\lambda}+\mu_{k} h_{k}+\eta_{k}\Big)(L_{k}\otimes P_{k})+\Big(\frac{\kappa_{k}}{\lambda}+\lambda+\kappa_{k} h_{k}\Big)(I_{q}\otimes P_{k})\Big]&=&\det\Big[(\kappa_{k} h_{k}-1-\kappa_{k})(I_{q}\otimes P_{k})\Big]\nonumber\\
&=&(\kappa_{k} h_{k}-1-\kappa_{k})^{nq}\det(I_{q}\otimes P_{k}).
\end{eqnarray*} Hence, $\det\Big[\Big(\frac{\mu_{k}}{\lambda}+\mu_{k} h_{k}+\eta_{k}\Big)(L_{k}\otimes P_{k})+\Big(\frac{\kappa_{k}}{\lambda}+\lambda+\kappa_{k} h_{k}\Big)(I_{q}\otimes P_{k})\Big]=0$ if and only if $\kappa_{k} h_{k}-1-\kappa_{k}=0$. If $\kappa_{k} h_{k}-1-\kappa_{k}=0$, eliminating $h_{k}$ in $\frac{\mu_{k}}{\kappa_{k}}(\kappa_{k} h_{k}-1)+\eta_{k}=0$ by using $\kappa_{k} h_{k}-1-\kappa_{k}=0$ yields $\mu_{k}+\eta_{k}=0$, and hence, $\mu_{k}=\eta_{k}=0$ since $\mu_{k},\eta_{k}\geq0$. Furthermore, $h_{k}\kappa_{k}=1+\kappa_{k}\neq 1$ due to $\kappa_{k}\neq0$. Next, since $\frac{\mu_{k}}{\kappa_{k}}(\kappa_{k} h_{k}-1)+\eta_{k}=0$ and $\kappa_{k} h_{k}-1-\kappa_{k}=0$, it follows from (\ref{Enqu1}) that $P_{k}\textbf{u}_{0}=\textbf{0}_{n\times 1}$, i.e., $\textbf{u}_{0}=\textbf{0}_{n\times 1}$. Thus in this case, (\ref{Enqu}) becomes $E_{n\times nq}^{[j]}[\textbf{u}_{1}^{*},\ldots,\textbf{u}_{q}^{*}]^{*}=\textbf{0}_{n\times 1}$, that is, $(\textbf{g}_{j}^{\mathrm{T}}\otimes I_{n})[\textbf{u}_{1}^{*},\ldots,\textbf{u}_{q}^{*}]^{*}=\textbf{0}_{n\times 1}$. Now it follows from (\ref{newv}) that $[\textbf{u}_{1}^{*},\ldots,\textbf{u}_{q}^{*}]^{*}=\sum_{i=1}^{n}\sum_{l=1}^{q}\alpha_{li}(\textbf{g}_{l}-(\textbf{g}_{j}^{\mathrm{T}}\textbf{g}_{l})\textbf{g}_{j})\otimes \textbf{e}_{i}$, where $\alpha_{li}\in\mathbb{C}$ and not all of them are zero. Clearly $\textbf{x}_{1}=\textbf{0}_{nq\times 1}$, $\textbf{x}_{2}=\sum_{i=1}^{n}\sum_{l=1}^{q}\alpha_{li}(\textbf{g}_{l}-(\textbf{g}_{j}^{\mathrm{T}}\textbf{g}_{l})\textbf{g}_{j})\otimes \textbf{e}_{i}$, and $\textbf{x}_{3}=\textbf{0}_{n\times 1}$ satisfy (\ref{Aeig_1})--(\ref{Aeig_3}). Thus, if 
$\frac{\mu_{k}}{\kappa_{k}}(\kappa_{k} h_{k}-1)+\eta_{k}=0$ and $h_{k}=1+\frac{1}{\kappa_{k}}$, then $\lambda=-\kappa_{k}$ is indeed an eigenvalue of $A_{k}^{[j]}+h_{k}A_{{\mathrm{c}}k}$ and the corresponding eigenvectors for the eigenvalue $\lambda_{3}$ of the form (\ref{lambda3}) are given by
\begin{eqnarray}
\textbf{x}=\Big[\textbf{0}_{1\times nq},\sum_{i=1}^{n}\sum_{l=1}^{q}\alpha_{li}(\textbf{g}_{l}-(\textbf{g}_{j}^{\mathrm{T}}\textbf{g}_{l})\textbf{g}_{j})^{\mathrm{T}}\otimes \textbf{e}_{i}^{\mathrm{T}},\textbf{0}_{1\times n}\Big]^{*},
\end{eqnarray} where $\alpha_{li}\in\mathbb{C}$ and not all $\alpha_{li}$ are zero. Therefore,
$\ker\Big(A_{k}^{[j]}+h_{k}A_{{\mathrm{c}}k}-\lambda_{3} I_{2nq+n}\Big)$ is given by (\ref{egns6}).

If $\frac{\mu_{k}}{\kappa_{k}}(\kappa_{k} h_{k}-1)+\eta_{k}\neq 0$ and $\kappa_{k} h_{k}-1-\kappa_{k}=0$, then $h_{k}=1+\frac{1}{\kappa_{k}}$. Clearly $h_{k}\kappa_{k}\neq 1$. In this case, since $\lambda=-\kappa_{k}$, it follows that
\begin{eqnarray*}
&&\det\Big[\Big(\frac{\mu_{k}}{\lambda}+\mu_{k} h_{k}+\eta_{k}\Big)(L_{k}\otimes P_{k})+\Big(\frac{\kappa_{k}}{\lambda}+\lambda+\kappa_{k} h_{k}\Big)(I_{q}\otimes P_{k})\Big]\nonumber\\
&&=\det\Big[\Big(-\frac{\mu_{k}}{\kappa_{k}}+\mu_{k} h_{k}+\eta_{k}\Big)(L_{k}\otimes I_{n})-\kappa_{k}I_{nq}\Big]\det(I_{q}\otimes P_{k})\nonumber\\
&&=\kappa_{k}^{nq}\det\Big[\frac{\mu_{k}+\eta_{k}}{\kappa_{k}}(L_{k}\otimes I_{n})-I_{nq}\Big](\det(P_{k}))^{q}.
\end{eqnarray*} Hence, $\det\Big[\Big(\frac{\mu_{k}}{\lambda}+\mu_{k} h_{k}+\eta_{k}\Big)(L_{k}\otimes P_{k})+\Big(\frac{\kappa_{k}}{\lambda}+\lambda+\kappa_{k} h_{k}\Big)(I_{q}\otimes P_{k})\Big]=0$ if and only if $1\in{\mathrm{spec}}(\frac{\mu_{k}+\eta_{k}}{\kappa_{k}}L_{k})$. Note that $1\in{\mathrm{spec}}(\frac{\mu_{k}+\eta_{k}}{\kappa_{k}}L_{k})$ implies that $\mu_{k}+\eta_{k}\neq0$ and hence, by using $\kappa_{k} h_{k}-1-\kappa_{k}=0$, $\frac{\mu_{k}}{\kappa_{k}}(\kappa_{k} h_{k}-1)+\eta_{k}=\mu_{k}+\eta_{k}\neq0$. Now we assume that
$1\in{\mathrm{spec}}(\frac{\mu_{k}+\eta_{k}}{\kappa_{k}}L_{k})$ and $h_{k}=1+\frac{1}{\kappa_{k}}$. Next, since $\kappa_{k} h_{k}-1-\kappa_{k}=0$ and $\mu_{k}+\eta_{k}\neq0$, it follows from (\ref{Enqu1}) that 
\begin{eqnarray}\label{u1u0uq}
(L_{k}\otimes I_{n})[\textbf{u}_{1}^{*},\ldots,\textbf{u}_{q}^{*}]^{*}=\frac{\kappa_{k}}{\mu_{k}+\eta_{k}}(\textbf{1}_{q\times 1}\otimes I_{n})\textbf{u}_{0}.
\end{eqnarray} Note that $(L_{k}\otimes I_{n})(\textbf{1}_{q\times 1}\otimes I_{n})=\textbf{0}_{nq\times n}$. Pre-multiplying $L_{k}\otimes I_{n}$ on both sides of (\ref{u1u0uq}) yields $(L_{k}\otimes I_{n})(L_{k}\otimes I_{n})[\textbf{u}_{1}^{*},\ldots,\textbf{u}_{q}^{*}]^{*}=\textbf{0}_{nq\times 1}$, which implies that $(L_{k}\otimes I_{n})[\textbf{u}_{1}^{*},\ldots,\textbf{u}_{q}^{*}]^{*}\in\ker(L_{k}\otimes I_{n})$. Hence,
\begin{eqnarray}\label{lku1}
(L_{k}\otimes I_{n})[\textbf{u}_{1}^{*},\ldots,\textbf{u}_{q}^{*}]^{*}=\sum_{l=0}^{q-1-{\mathrm{rank}}(L_{k})}\sum_{i=1}^{n}\alpha_{li}\textbf{w}_{l}\otimes\textbf{e}_{i},
\end{eqnarray} where $\alpha_{li}\in\mathbb{C}$. Let $\textbf{u}_{0}=\sum_{i=1}^{n}\beta_{i}\textbf{e}_{i}$, where $\beta_{i}\in\mathbb{C}$. Then it follows that $(\textbf{1}_{q\times 1}\otimes I_{n})\textbf{u}_{0}=\sum_{i=1}^{n}\beta_{i}(\textbf{1}_{q\times 1}\otimes I_{n})\textbf{e}_{i}=\sum_{i=1}^{n}\beta_{i}(\textbf{w}_{0}\otimes\textbf{e}_{i})$. Now it follows from (\ref{u1u0uq}) and (\ref{lku1}) that 
\begin{eqnarray*}
\sum_{i=1}^{n}\Big(\alpha_{0i}-\beta_{i}\frac{\kappa_{k}}{\mu_{k}+\eta_{k}}\Big)\textbf{w}_{0}\otimes\textbf{e}_{i}+\sum_{l=1}^{q-1-{\mathrm{rank}}(L_{k})}\sum_{i=1}^{n}\alpha_{li}\textbf{w}_{l}\otimes\textbf{e}_{i}=\textbf{0}_{nq\times 1},
\end{eqnarray*} which implies that $\alpha_{0i}-\beta_{i}\frac{\kappa_{k}}{\mu_{k}+\eta_{k}}=0$ and $\alpha_{li}=0$ for every $i=1,\ldots,n$ and every $l=1,\ldots,q-1-{\mathrm{rank}}(L_{k})$. Hence, 
\begin{eqnarray}\label{Axbeqn}
(L_{k}\otimes I_{n})[\textbf{u}_{1}^{*},\ldots,\textbf{u}_{q}^{*}]^{*}=\frac{\kappa_{k}}{\mu_{k}+\eta_{k}}\sum_{i=1}^{n}\beta_{i}\textbf{w}_{0}\otimes\textbf{e}_{i}.
\end{eqnarray} Together with $E_{n\times nq}^{[j]}[\textbf{u}_{1}^{*},\ldots,\textbf{u}_{q}^{*}]^{*}=(\textbf{g}_{j}^{\mathrm{T}}\otimes I_{n})[\textbf{u}_{1}^{*},\ldots,\textbf{u}_{q}^{*}]^{*}=\textbf{0}_{n\times 1}$, we have
\begin{eqnarray}\label{Axbeqnab}
\small\left[\begin{array}{c}
L_{k}\otimes I_{n}\\
\textbf{g}_{j}^{\mathrm{T}}\otimes I_{n}\\
\end{array}\right][\textbf{u}_{1}^{*},\ldots,\textbf{u}_{q}^{*}]^{*}=\small\left[\begin{array}{c}
\frac{\kappa_{k}}{\mu_{k}+\eta_{k}}\sum_{i=1}^{n}\beta_{i}\textbf{w}_{0}\otimes\textbf{e}_{i} \\
\textbf{0}_{n\times 1}\\
\end{array}\right].
\end{eqnarray} Now it follows from $ii$) of Theorem 2.6.4 of \cite[p.~108]{Bernstein:2009} that (\ref{Axbeqnab}) has a solution $[\textbf{u}_{1}^{*},\ldots,\textbf{u}_{q}^{*}]^{*}$ if and only if
\begin{eqnarray}\label{rankconab}
{\mathrm{rank}}\small\left[\begin{array}{c}
L_{k}\otimes I_{n}\\
\textbf{g}_{j}^{\mathrm{T}}\otimes I_{n}\\
\end{array}\right]={\mathrm{rank}}\small\left[\begin{array}{cc}
L_{k}\otimes I_{n} & \frac{\kappa_{k}}{\mu_{k}+\eta_{k}}\sum_{i=1}^{n}\beta_{i}\textbf{w}_{0}\otimes\textbf{e}_{i}\\
\textbf{g}_{j}^{\mathrm{T}}\otimes I_{n} & \textbf{0}_{n\times 1}\\
\end{array}\right].
\end{eqnarray} We claim that (\ref{rankconab}) is indeed true. First, if $\beta_{i}=0$ for every $i=1,\ldots,n$, then it is clear that ${\mathrm{rank}}\small\left[\begin{array}{c}
L_{k}\otimes I_{n}\\
\textbf{g}_{j}^{\mathrm{T}}\otimes I_{n}\\
\end{array}\right]={\mathrm{rank}}\small\left[\begin{array}{cc}
L_{k}\otimes I_{n} & \textbf{0}_{nq\times 1}\\
\textbf{g}_{j}^{\mathrm{T}}\otimes I_{n} & \textbf{0}_{n\times 1}\\
\end{array}\right]$. Alternatively, assume that $\beta_{i}\neq0$ for some $i\in\{1,\ldots,n\}$. Note that it follows from Fact 2.11.8 of \cite[p.~132]{Bernstein:2009} that ${\mathrm{rank}}\small\left[\begin{array}{c}
L_{k}\otimes I_{n}\\
\textbf{g}_{j}^{\mathrm{T}}\otimes I_{n}\\
\end{array}\right]\leq{\mathrm{rank}}\small\left[\begin{array}{cc}
L_{k}\otimes I_{n} & \frac{\kappa_{k}}{\mu_{k}+\eta_{k}}\sum_{i=1}^{n}\beta_{i}\textbf{w}_{0}\otimes\textbf{e}_{i}\\
\textbf{g}_{j}^{\mathrm{T}}\otimes I_{n} & \textbf{0}_{n\times 1}\\
\end{array}\right]$. To show (\ref{rankconab}), it suffices to show that 
\begin{eqnarray*}
{\mathrm{def}}\small\left[\begin{array}{c}
L_{k}\otimes I_{n}\\
\textbf{g}_{j}^{\mathrm{T}}\otimes I_{n}\\
\end{array}\right]\leq{\mathrm{def}}\small\left[\begin{array}{cc}
L_{k}\otimes I_{n} & \frac{\kappa_{k}}{\mu_{k}+\eta_{k}}\sum_{i=1}^{n}\beta_{i}\textbf{w}_{0}\otimes\textbf{e}_{i}\\
\textbf{g}_{j}^{\mathrm{T}}\otimes I_{n} & \textbf{0}_{n\times 1}\\
\end{array}\right],
\end{eqnarray*} or, equivalently,
\begin{eqnarray*}
\dim\ker\small\left[\begin{array}{c}
L_{k}\otimes I_{n}\\
\textbf{g}_{j}^{\mathrm{T}}\otimes I_{n}\\
\end{array}\right]\leq\dim\ker\small\left[\begin{array}{cc}
L_{k}\otimes I_{n} & \frac{\kappa_{k}}{\mu_{k}+\eta_{k}}\sum_{i=1}^{n}\beta_{i}\textbf{w}_{0}\otimes\textbf{e}_{i}\\
\textbf{g}_{j}^{\mathrm{T}}\otimes I_{n} & \textbf{0}_{n\times 1}\\
\end{array}\right].
\end{eqnarray*}
 Let $s\in\mathbb{C}$ be such that $s\in\ker\small\left[\begin{array}{c}
 \frac{\kappa_{k}}{\mu_{k}+\eta_{k}}\sum_{i=1}^{n}\beta_{i}\textbf{w}_{0}\otimes\textbf{e}_{i}\\
 \textbf{0}_{n\times 1}\\
 \end{array}\right]$. Then $s\frac{\kappa_{k}}{\mu_{k}+\eta_{k}}\beta_{i}=0$ for some $i\in\{1,\ldots,n\}$, which implies that $s=0$. Thus, $\dim\ker\small\left[\begin{array}{c}
  \frac{\kappa_{k}}{\mu_{k}+\eta_{k}}\sum_{i=1}^{n}\beta_{i}\textbf{w}_{0}\otimes\textbf{e}_{i}\\
  \textbf{0}_{n\times 1}\\
  \end{array}\right]=0$. Consequently, it follows from Fact 2.11.8 of \cite[p.~132]{Bernstein:2009} that
 \begin{eqnarray*}
 \dim\ker\small\left[\begin{array}{c}
 L_{k}\otimes I_{n}\\
 \textbf{g}_{j}^{\mathrm{T}}\otimes I_{n}\\
 \end{array}\right]&=&\dim\ker\small\left[\begin{array}{c}
 L_{k}\otimes I_{n}\\
 \textbf{g}_{j}^{\mathrm{T}}\otimes I_{n}\\
 \end{array}\right]+\dim\ker\small\left[\begin{array}{c}
  \frac{\kappa_{k}}{\mu_{k}+\eta_{k}}\sum_{i=1}^{n}\beta_{i}\textbf{w}_{0}\otimes\textbf{e}_{i}\\
  \textbf{0}_{n\times 1}\\
  \end{array}\right]\nonumber\\
 &\leq&\dim\ker\small\left[\begin{array}{cc}
 L_{k}\otimes I_{n} & \frac{\kappa_{k}}{\mu_{k}+\eta_{k}}\sum_{i=1}^{n}\beta_{i}\textbf{w}_{0}\otimes\textbf{e}_{i}\\
 \textbf{g}_{j}^{\mathrm{T}}\otimes I_{n} & \textbf{0}_{n\times 1}\\
 \end{array}\right],
 \end{eqnarray*} which implies that  ${\mathrm{rank}}\small\left[\begin{array}{c}
 L_{k}\otimes I_{n}\\
 \textbf{g}_{j}^{\mathrm{T}}\otimes I_{n}\\
 \end{array}\right]\geq{\mathrm{rank}}\small\left[\begin{array}{cc}
 L_{k}\otimes I_{n} & \frac{\kappa_{k}}{\mu_{k}+\eta_{k}}\sum_{i=1}^{n}\beta_{i}\textbf{w}_{0}\otimes\textbf{e}_{i}\\
 \textbf{g}_{j}^{\mathrm{T}}\otimes I_{n} & \textbf{0}_{n\times 1}\\
 \end{array}\right]$. Hence, (\ref{rankconab}) holds.
Next, it follows from $vi$) of Proposition 6.1.7 of \cite[p.~400]{Bernstein:2009} and $viii)$ of Proposition 6.1.6 of \cite[p.~399]{Bernstein:2009} that the general solution to (\ref{Axbeqnab}) is given by the form
\begin{eqnarray}\label{usolutionab} 
[\textbf{u}_{1}^{*},\ldots,\textbf{u}_{q}^{*}]^{*}&=&\small\left[\begin{array}{c}
 L_{k}\otimes I_{n}\\
 \textbf{g}_{j}^{\mathrm{T}}\otimes I_{n}\\
 \end{array}\right]^{+}\small\left[\begin{array}{c}
   \frac{\kappa_{k}}{\mu_{k}+\eta_{k}}\sum_{i=1}^{n}\beta_{i}\textbf{w}_{0}\otimes\textbf{e}_{i}\\
   \textbf{0}_{n\times 1}\\
   \end{array}\right]+\sum_{l=1}^{q}\sum_{i=1}^{n}\gamma_{li}\Big(I_{nq}-\small\left[\begin{array}{c}
    L_{k}\otimes I_{n}\\
    \textbf{g}_{j}^{\mathrm{T}}\otimes I_{n}\\
    \end{array}\right]^{+}\small\left[\begin{array}{c}
     L_{k}\otimes I_{n}\\
     \textbf{g}_{j}^{\mathrm{T}}\otimes I_{n}\\
     \end{array}\right]\Big)\nonumber\\
     &&(\textbf{g}_{l}\otimes\textbf{e}_{i})\nonumber\\
&=&\Big(\small\left[\begin{array}{c}
 L_{k}\\
 \textbf{g}_{j}^{\mathrm{T}}\\
 \end{array}\right]\otimes I_{n}\Big)^{+}\small\left[\begin{array}{c}
   \frac{\kappa_{k}}{\mu_{k}+\eta_{k}}\sum_{i=1}^{n}\beta_{i}\textbf{w}_{0}\otimes\textbf{e}_{i}\\
   \sum_{i=1}^{n}0\otimes\textbf{e}_{i}\\
   \end{array}\right]+\sum_{l=1}^{q}\sum_{i=1}^{n}\gamma_{li}\Big(I_{nq}-\Big(\small\left[\begin{array}{c}
    L_{k}\\
    \textbf{g}_{j}^{\mathrm{T}}\\
    \end{array}\right]\otimes I_{n}\Big)^{+}\nonumber\\
    &&\Big(\small\left[\begin{array}{c}
     L_{k}\\
     \textbf{g}_{j}^{\mathrm{T}}\\
     \end{array}\right]\otimes I_{n}\Big)\Big)(\textbf{g}_{l}\otimes\textbf{e}_{i})\nonumber\\
&=&\Big(\small\left[\begin{array}{c}
      L_{k}\\
      \textbf{g}_{j}^{\mathrm{T}}\\
      \end{array}\right]^{+}\otimes I_{n}\Big)\Big(\sum_{i=1}^{n}\small\left[\begin{array}{c}
        \frac{\kappa_{k}}{\mu_{k}+\eta_{k}}\beta_{i}\textbf{w}_{0}\\
        0\\
        \end{array}\right]\otimes\textbf{e}_{i}\Big)+\sum_{l=1}^{q}\sum_{i=1}^{n}\gamma_{li}\Big(I_{q}\otimes I_{n}-\Big(\small\left[\begin{array}{c}
         L_{k}\\
         \textbf{g}_{j}^{\mathrm{T}}\\
         \end{array}\right]^{+}\otimes I_{n}\Big)\nonumber\\
         &&\Big(\small\left[\begin{array}{c}
          L_{k}\\
          \textbf{g}_{j}^{\mathrm{T}}\\
          \end{array}\right]\otimes I_{n}\Big)\Big)(\textbf{g}_{l}\otimes\textbf{e}_{i})\nonumber\\
          &=&\sum_{i=1}^{n}\Big(\small\left[\begin{array}{c}
                L_{k}\\
                \textbf{g}_{j}^{\mathrm{T}}\\
                \end{array}\right]^{+}\small\left[\begin{array}{c}
                        \frac{\kappa_{k}}{\mu_{k}+\eta_{k}}\beta_{i}\textbf{w}_{0}\\
                        0\\
                        \end{array}\right]\Big)\otimes\textbf{e}_{i}+\sum_{l=1}^{q}\sum_{i=1}^{n}\gamma_{li}\Big(I_{q}\otimes I_{n}-\Big(\small\left[\begin{array}{c}
                                 L_{k}\\
                                 \textbf{g}_{j}^{\mathrm{T}}\\
                                 \end{array}\right]^{+}\small\left[\begin{array}{c}
                                                                   L_{k}\\
                                                                   \textbf{g}_{j}^{\mathrm{T}}\\
                                                                   \end{array}\right]\otimes I_{n}\Big)\Big)\nonumber\\
                                 &&(\textbf{g}_{l}\otimes\textbf{e}_{i})\nonumber\\
          &=&\sum_{i=1}^{n}\Big(\small\left[\begin{array}{c}
                L_{k}\\
                \textbf{g}_{j}^{\mathrm{T}}\\
                \end{array}\right]^{+}\small\left[\begin{array}{c}
                        \frac{\kappa_{k}}{\mu_{k}+\eta_{k}}\beta_{i}\textbf{w}_{0}\\
                        0\\
                        \end{array}\right]\Big)\otimes\textbf{e}_{i}+\sum_{l=1}^{q}\sum_{i=1}^{n}\gamma_{li}\Big(\textbf{g}_{l}-\small\left[\begin{array}{c}
                                 L_{k}\\
                                 \textbf{g}_{j}^{\mathrm{T}}\\
                                 \end{array}\right]^{+}\small\left[\begin{array}{c}
                                                                   L_{k}\\
                                                                   \textbf{g}_{j}^{\mathrm{T}}\\
                                                                   \end{array}\right]\textbf{g}_{l}\Big)\otimes\textbf{e}_{i},
\end{eqnarray} where $\gamma_{li}\in\mathbb{C}$. Note that by Proposition 6.1.6 of \cite[p.~399]{Bernstein:2009}, $L_{k}^{\mathrm{T}}(L_{k}^{\mathrm{T}})^{+}=L_{k}^{\mathrm{T}}(L_{k}^{+})^{\mathrm{T}}=(L_{k}^{+}L_{k})^{\mathrm{T}}=L_{k}^{+}L_{k}$. It follows from Fact 6.5.17 of \cite[p.~427]{Bernstein:2009} that 
\begin{eqnarray}
\small\left[\begin{array}{c}
L_{k}\\
\textbf{g}_{j}^{\mathrm{T}}
\end{array}\right]^{+}=\small\left[\begin{array}{cc}
L_{k}^{+}(I_{q}-\varphi_{k}\textbf{g}_{j}^{\mathrm{T}}) & \varphi_{k}
\end{array}\right],
\end{eqnarray} where $\varphi_{k}$ is given by (\ref{varphik}). Note that $\textbf{g}_{j}^{\mathrm{T}}\textbf{w}_{0}=1$ for every $j=1,\ldots,q$. Hence, it follows that for every $i=1,\ldots,n$ and every $j,l=1,\ldots,q$,
\begin{eqnarray}
\small\left[\begin{array}{cc}
L_{k}^{+}(I_{q}-\varphi_{k}\textbf{g}_{j}^{\mathrm{T}}) & \varphi_{k}
\end{array}\right]\small\left[\begin{array}{c}
                        \frac{\kappa_{k}}{\mu_{k}+\eta_{k}}\beta_{i}\textbf{w}_{0}\\
                        0\\
                        \end{array}\right]&=&\frac{\kappa_{k}}{\mu_{k}+\eta_{k}}\beta_{i}L_{k}^{+}\textbf{w}_{0}-\frac{\kappa_{k}}{\mu_{k}+\eta_{k}}\beta_{i}L_{k}^{+}\varphi_{k},\\
\textbf{g}_{l}-\small\left[\begin{array}{c}
L_{k}\\
\textbf{g}_{j}^{\mathrm{T}}
\end{array}\right]^{+}\small\left[\begin{array}{c}
L_{k}\\
\textbf{g}_{j}^{\mathrm{T}}
\end{array}\right]\textbf{g}_{l}&=&\textbf{g}_{l}-\small\left[\begin{array}{cc}
L_{k}^{+}(I_{q}-\varphi_{k}\textbf{g}_{j}^{\mathrm{T}}) & \varphi_{k}
\end{array}\right]\small\left[\begin{array}{c}
L_{k}\\
\textbf{g}_{j}^{\mathrm{T}}
\end{array}\right]\textbf{g}_{l}\nonumber\\
&=&\textbf{g}_{l}-\small\left[\begin{array}{cc}
L_{k}^{+}(I_{q}-\varphi_{k}\textbf{g}_{j}^{\mathrm{T}}) & \varphi_{k}
\end{array}\right]\small\left[\begin{array}{c}
L_{k}\textbf{g}_{l}\\
\textbf{g}_{j}^{\mathrm{T}}\textbf{g}_{l}
\end{array}\right]\nonumber\\
&=&\textbf{g}_{l}-L_{k}^{+}(I_{q}-\varphi_{k}\textbf{g}_{j}^{\mathrm{T}})L_{k}\textbf{g}_{l}-(\textbf{g}_{j}^{\mathrm{T}}\textbf{g}_{l})\varphi_{k}\nonumber\\
&=&\textbf{g}_{l}-L_{k}^{+}L_{k}\textbf{g}_{l}+(\textbf{g}_{j}^{\mathrm{T}}L_{k}\textbf{g}_{l})L_{k}^{+}\varphi_{k}-(\textbf{g}_{j}^{\mathrm{T}}\textbf{g}_{l})\varphi_{k}.
\end{eqnarray} Then (\ref{usolutionab}) becomes
\begin{eqnarray}\label{gsolutionab}
[\textbf{u}_{1}^{*},\ldots,\textbf{u}_{q}^{*}]^{*}&=&\frac{\kappa_{k}}{\mu_{k}+\eta_{k}}\sum_{i=1}^{n}\beta_{i}L_{k}^{+}\textbf{w}_{0}\otimes\textbf{e}_{i}-\frac{\kappa_{k}}{\mu_{k}+\eta_{k}}\sum_{i=1}^{n}\beta_{i}L_{k}^{+}\varphi_{k}\otimes\textbf{e}_{i}\nonumber\\
&&+\sum_{l=1}^{q}\sum_{i=1}^{n}\gamma_{li}(\textbf{g}_{l}-L_{k}^{+}L_{k}\textbf{g}_{l}+(\textbf{g}_{j}^{\mathrm{T}}L_{k}\textbf{g}_{l})L_{k}^{+}\varphi_{k}-(\textbf{g}_{j}^{\mathrm{T}}\textbf{g}_{l})\varphi_{k})\otimes \textbf{e}_{i}.
\end{eqnarray} 

In summary, if $1\in{\mathrm{spec}}(\frac{\mu_{k}+\eta_{k}}{\kappa_{k}}L_{k})$ and $h_{k}=1+\frac{1}{\kappa_{k}}$, then $\lambda=-\kappa_{k}$ is indeed an eigenvalue of $A_{k}^{[j]}+h_{k}A_{{\mathrm{c}}k}$. In this case, $\textbf{x}_{1}=\textbf{0}_{nq\times 1}$, $\textbf{x}_{2}=[\textbf{u}_{1}^{*},\ldots,\textbf{u}_{q}^{*}]^{*}$ given by (\ref{gsolutionab}), and $\textbf{x}_{3}=\sum_{i=1}^{n}\beta_{i}\textbf{e}_{i}$, where not all of $\beta_{i}$ and $\gamma_{li}$ are zero. The corresponding eigenvectors for $\lambda_{3}$ are given by
\begin{eqnarray}\label{xw0ab}
\textbf{x}&=&\Big[\textbf{0}_{1\times nq},\frac{\kappa_{k}}{\mu_{k}+\eta_{k}}\sum_{i=1}^{n}\beta_{i}(L_{k}^{+}\textbf{w}_{0}\otimes\textbf{e}_{i})^{\mathrm{T}}-\frac{\kappa_{k}}{\mu_{k}+\eta_{k}}\sum_{i=1}^{n}\beta_{i}(L_{k}^{+}\varphi_{k}\otimes\textbf{e}_{i})^{\mathrm{T}}\nonumber\\
&&+\sum_{l=1}^{q}\sum_{i=1}^{n}\gamma_{li}(\textbf{g}_{l}-L_{k}^{+}L_{k}\textbf{g}_{l}+(\textbf{g}_{j}^{\mathrm{T}}L_{k}\textbf{g}_{l})L_{k}^{+}\varphi_{k}-(\textbf{g}_{j}^{\mathrm{T}}\textbf{g}_{l})\varphi_{k})^{\mathrm{T}}\otimes \textbf{e}_{i}^{\mathrm{T}},\sum_{i=1}^{n}\beta_{i}\textbf{e}_{i}^{\mathrm{T}}\Big]^{*},
\end{eqnarray} where $\beta_{i}\in\mathbb{C}$ and $\gamma_{li}\in\mathbb{C}$ and not all of them are zero. Therefore,
$\ker\Big(A_{k}^{[j]}+h_{k}A_{{\mathrm{c}}k}-\lambda_{3} I_{2nq+n}\Big)$ is given by (\ref{egns7}).

If $\frac{\mu_{k}}{\kappa_{k}}(\kappa_{k} h_{k}-1)+\eta_{k}\neq 0$, $\kappa_{k} h_{k}-1-\kappa_{k}\neq0$, and $\kappa_{k} h_{k}-1\neq0$, in this case, since $\lambda=-\kappa_{k}$, then it follows that
\begin{eqnarray*}
&&\det\Big[\Big(\frac{\mu_{k}}{\lambda}+\mu_{k} h_{k}+\eta_{k}\Big)(L_{k}\otimes P_{k})+\Big(\frac{\kappa_{k}}{\lambda}+\lambda+\kappa_{k} h_{k}\Big)(I_{q}\otimes P_{k})\Big]\nonumber\\
&&=\det\Big[\Big(\frac{\mu_{k}}{\kappa_{k}}(\kappa_{k} h_{k}-1)+\eta_{k}\Big)(L_{k}\otimes I_{n})+(\kappa_{k} h_{k}-1-\kappa_{k})I_{nq}\Big]\det(I_{q}\otimes P_{k})\nonumber\\
&&=(-\kappa_{k} h_{k}+1+\kappa_{k})^{nq}\det\Big[\frac{\mu_{k}(\kappa_{k}h_{k}-1)+\eta_{k}\kappa_{k}}{\kappa_{k}(-\kappa_{k} h_{k}+1+\kappa_{k})}(L_{k}\otimes I_{n})-I_{nq}\Big](\det(P_{k}))^{q}.
\end{eqnarray*} Hence, $\det\Big[\Big(\frac{\mu_{k}}{\lambda}+\mu_{k} h_{k}+\eta_{k}\Big)(L_{k}\otimes I_{n})+\Big(\frac{\kappa_{k}}{\lambda}+\lambda+\kappa_{k} h_{k}\Big)I_{nq}\Big]=0$ if and only if $1\in{\mathrm{spec}}(\frac{\mu_{k}(\kappa_{k}h_{k}-1)+\eta_{k}\kappa_{k}}{\kappa_{k}(-\kappa_{k} h_{k}+1+\kappa_{k})}L_{k})$. Again, note that $1\in{\mathrm{spec}}(\frac{\mu_{k}(\kappa_{k}h_{k}-1)+\eta_{k}\kappa_{k}}{\kappa_{k}(-\kappa_{k} h_{k}+1+\kappa_{k})}L_{k})$ implies that $\frac{\mu_{k}}{\kappa_{k}}(\kappa_{k} h_{k}-1)+\eta_{k}\neq 0$ and $\kappa_{k} h_{k}-1-\kappa_{k}\neq0$. Now we assume that
$1\in{\mathrm{spec}}(\frac{\mu_{k}(\kappa_{k}h_{k}-1)+\eta_{k}\kappa_{k}}{\kappa_{k}(-\kappa_{k} h_{k}+1+\kappa_{k})}L_{k})$ and $\kappa_{k}h_{k}\neq 1$. Next, let $\textbf{u}_{0}=\sum_{i=1}^{n}\beta_{i}\textbf{e}_{i}$, where $\beta_{i}\in\mathbb{C}$ and it follows from (\ref{Enqu1}) that 
\begin{eqnarray}\label{xabs}
\Big(\Big(\frac{\mu_{k}}{\kappa_{k}}(\kappa_{k} h_{k}-1)+\eta_{k}\Big)(L_{k}\otimes I_{n})+(\kappa_{k} h_{k}-1-\kappa_{k})I_{nq}\Big)[\textbf{u}_{1}^{*},\ldots,\textbf{u}_{q}^{*}]^{*}=\kappa_{k}\sum_{i=1}^{n}\beta_{i}\textbf{1}_{q\times 1}\otimes\textbf{e}_{i}.
\end{eqnarray} Note that a specific solution $[\textbf{u}_{1}^{*},\ldots,\textbf{u}_{q}^{*}]^{*}$ to (\ref{xabs}) is given by the form
\begin{eqnarray}\label{specific}
[\textbf{u}_{1}^{*},\ldots,\textbf{u}_{q}^{*}]^{*}=\frac{\kappa_{k}}{\kappa_{k}h_{k}-1-\kappa_{k}}\sum_{i=1}^{n}\beta_{i}\textbf{1}_{q\times 1}\otimes\textbf{e}_{i}.
\end{eqnarray} Substituting (\ref{specific}) into (\ref{Enqu}) by using $iii$) of Lemma~\ref{lemma_EW} yields $\frac{\kappa_{k}(\kappa_{k}h_{k}-1)}{\kappa_{k}h_{k}-1-\kappa_{k}}\sum_{i=1}^{n}\beta_{i}E_{n\times nq}^{[j]}(\textbf{1}_{q\times 1}\otimes\textbf{e}_{i})=\frac{\kappa_{k}(\kappa_{k}h_{k}-1)}{\kappa_{k}h_{k}-1-\kappa_{k}}\\\sum_{i=1}^{n}\beta_{i}\textbf{e}_{i}=\textbf{0}_{n\times 1}$, which implies that $\beta_{i}=0$ for every $i=1,\ldots,n$, and hence, $\textbf{u}_{0}=\textbf{0}_{n\times 1}$. Thus, (\ref{xabs}) becomes
\begin{eqnarray}\label{homo}
\Big(\Big(\frac{\mu_{k}}{\kappa_{k}}(\kappa_{k} h_{k}-1)+\eta_{k}\Big)(L_{k}\otimes I_{n})+(\kappa_{k} h_{k}-1-\kappa_{k})I_{nq}\Big)[\textbf{u}_{1}^{*},\ldots,\textbf{u}_{q}^{*}]^{*}=\textbf{0}_{nq\times 1}.
\end{eqnarray} Let $M_{k}=(\frac{\mu_{k}}{\kappa_{k}}(\kappa_{k} h_{k}-1)+\eta_{k})L_{k}+(\kappa_{k} h_{k}-1-\kappa_{k})I_{q}$. Again, note that $E_{n\times nq}^{[j]}=\textbf{g}_{j}^{\mathrm{T}}\otimes I_{n}$ for every $j=1,\ldots,q$. Then it follows from  (\ref{homo}) and (\ref{Enqu}) that
\begin{eqnarray}\label{uMg}
\small\left[\begin{array}{c}
M_{k}\otimes I_{n}\\
\textbf{g}_{j}^{\mathrm{T}}\otimes I_{n}
\end{array}\right][\textbf{u}_{1}^{*},\ldots,\textbf{u}_{q}^{*}]^{*}=\Big(\small\left[\begin{array}{c}
M_{k}\\
\textbf{g}_{j}^{\mathrm{T}}
\end{array}\right]\otimes I_{n}\Big)[\textbf{u}_{1}^{*},\ldots,\textbf{u}_{q}^{*}]^{*}=\textbf{0}_{(nq+n)\times 1}.
\end{eqnarray} Next, it follows from $vi$) of Proposition 6.1.7 of \cite[p.~400]{Bernstein:2009} and $viii$) of Proposition 6.1.6 of \cite[p.~399]{Bernstein:2009} that the general solution $[\textbf{u}_{1}^{*},\ldots,\textbf{u}_{q}^{*}]^{*}$ to (\ref{uMg}) is given by the form
\begin{eqnarray}
[\textbf{u}_{1}^{*},\ldots,\textbf{u}_{q}^{*}]^{*}&=&\Big[I_{nq}-\Big(\small\left[\begin{array}{c}
M_{k}\\
\textbf{g}_{j}^{\mathrm{T}}
\end{array}\right]\otimes I_{n}\Big)^{+}\Big(\small\left[\begin{array}{c}
M_{k}\\
\textbf{g}_{j}^{\mathrm{T}}
\end{array}\right]\otimes I_{n}\Big)\Big]\sum_{i=1}^{n}\sum_{l=1}^{q}\varpi_{li}\textbf{g}_{l}\otimes\textbf{e}_{i}\nonumber\\
&=&\Big[I_{nq}-\Big(\small\left[\begin{array}{c}
M_{k}\\
\textbf{g}_{j}^{\mathrm{T}}
\end{array}\right]^{+}\otimes I_{n}\Big)\Big(\small\left[\begin{array}{c}
M_{k}\\
\textbf{g}_{j}^{\mathrm{T}}
\end{array}\right]\otimes I_{n}\Big)\Big]\sum_{i=1}^{n}\sum_{l=1}^{q}\varpi_{li}\textbf{g}_{l}\otimes\textbf{e}_{i}\nonumber\\
&=&\Big[I_{q}\otimes I_{n}-\Big(\small\left[\begin{array}{c}
M_{k}\\
\textbf{g}_{j}^{\mathrm{T}}
\end{array}\right]^{+}\small\left[\begin{array}{c}
M_{k}\\
\textbf{g}_{j}^{\mathrm{T}}
\end{array}\right]\otimes I_{n}\Big)\Big]\sum_{i=1}^{n}\sum_{l=1}^{q}\varpi_{li}\textbf{g}_{l}\otimes\textbf{e}_{i}\nonumber\\
&=&\Big[\Big(I_{q}-\small\left[\begin{array}{c}
M_{k}\\
\textbf{g}_{j}^{\mathrm{T}}
\end{array}\right]^{+}\small\left[\begin{array}{c}
M_{k}\\
\textbf{g}_{j}^{\mathrm{T}}
\end{array}\right]\Big)\otimes I_{n}\Big]\sum_{i=1}^{n}\sum_{l=1}^{q}\varpi_{li}\textbf{g}_{l}\otimes\textbf{e}_{i}\nonumber\\
&=&\sum_{i=1}^{n}\sum_{l=1}^{q}\varpi_{li}\Big(\textbf{g}_{l}-\small\left[\begin{array}{c}
M_{k}\\
\textbf{g}_{j}^{\mathrm{T}}
\end{array}\right]^{+}\small\left[\begin{array}{c}
M_{k}\\
\textbf{g}_{j}^{\mathrm{T}}
\end{array}\right]\textbf{g}_{l}\Big)\otimes \textbf{e}_{i},\label{gsolution1}
\end{eqnarray} where $\varpi_{li}\in\mathbb{C}$ and $j=1,\ldots,q$. Note that by Proposition 6.1.6 of \cite[p.~399]{Bernstein:2009}, $M_{k}^{\mathrm{T}}(M_{k}^{\mathrm{T}})^{+}=M_{k}^{\mathrm{T}}(M_{k}^{+})^{\mathrm{T}}=(M_{k}^{+}M_{k})^{\mathrm{T}}=M_{k}^{+}M_{k}$. It follows from Fact 6.5.17 of \cite[p.~427]{Bernstein:2009} that 
\begin{eqnarray}
\small\left[\begin{array}{c}
M_{k}\\
\textbf{g}_{j}^{\mathrm{T}}
\end{array}\right]^{+}=\small\left[\begin{array}{cc}
M_{k}^{+}(I_{q}-\phi_{k}\textbf{g}_{j}^{\mathrm{T}}) & \phi_{k}
\end{array}\right],
\end{eqnarray} where $\phi_{k}$ is given by (\ref{phik}). Hence, it follows that for every $j,l=1,\ldots,q$,
\begin{eqnarray}
\textbf{g}_{l}-\small\left[\begin{array}{c}
M_{k}\\
\textbf{g}_{j}^{\mathrm{T}}
\end{array}\right]^{+}\small\left[\begin{array}{c}
M_{k}\\
\textbf{g}_{j}^{\mathrm{T}}
\end{array}\right]\textbf{g}_{l}&=&\textbf{g}_{l}-\small\left[\begin{array}{cc}
M_{k}^{+}(I_{q}-\phi_{k}\textbf{g}_{j}^{\mathrm{T}}) & \phi_{k}
\end{array}\right]\small\left[\begin{array}{c}
M_{k}\\
\textbf{g}_{j}^{\mathrm{T}}
\end{array}\right]\textbf{g}_{l}\nonumber\\
&=&\textbf{g}_{l}-\small\left[\begin{array}{cc}
M_{k}^{+}(I_{q}-\phi_{k}\textbf{g}_{j}^{\mathrm{T}}) & \phi_{k}
\end{array}\right]\small\left[\begin{array}{c}
M_{k}\textbf{g}_{l}\\
\textbf{g}_{j}^{\mathrm{T}}\textbf{g}_{l}
\end{array}\right]\nonumber\\
&=&\textbf{g}_{l}-M_{k}^{+}(I_{q}-\phi_{k}\textbf{g}_{j}^{\mathrm{T}})M_{k}\textbf{g}_{l}-(\textbf{g}_{j}^{\mathrm{T}}\textbf{g}_{l})\phi_{k}\nonumber\\
&=&\textbf{g}_{l}-M_{k}^{+}M_{k}\textbf{g}_{l}+(\textbf{g}_{j}^{\mathrm{T}}M_{k}\textbf{g}_{l})M_{k}^{+}\phi_{k}-(\textbf{g}_{j}^{\mathrm{T}}\textbf{g}_{l})\phi_{k}.
\end{eqnarray} Thus, (\ref{gsolution1}) becomes 
\begin{eqnarray}\label{gso}
[\textbf{u}_{1}^{*},\ldots,\textbf{u}_{q}^{*}]^{*}=\sum_{i=1}^{n}\sum_{l=1}^{q}\varpi_{li}\Big(\textbf{g}_{l}-M_{k}^{+}M_{k}\textbf{g}_{l}+(\textbf{g}_{j}^{\mathrm{T}}M_{k}\textbf{g}_{l})M_{k}^{+}\phi_{k}-(\textbf{g}_{j}^{\mathrm{T}}\textbf{g}_{l})\phi_{k}\Big)\otimes \textbf{e}_{i}.
\end{eqnarray} 

In summary, if $1\in{\mathrm{spec}}(\frac{\mu_{k}(\kappa_{k}h_{k}-1)+\eta_{k}\kappa_{k}}{\kappa_{k}(-\kappa_{k} h_{k}+1+\kappa_{k})}L_{k})$ and $\kappa_{k}h_{k}\neq 1$, then $\lambda=-\kappa_{k}$ is indeed an eigenvalue of $A_{k}^{[j]}+h_{k}A_{{\mathrm{c}}k}$. In this case, $\textbf{x}_{1}=\textbf{0}_{nq\times 1}$, $\textbf{x}_{2}=[\textbf{u}_{1}^{*},\ldots,\textbf{u}_{q}^{*}]^{*}$ given by (\ref{gso}), and $\textbf{x}_{3}=\textbf{0}_{n\times 1}$, where not all of $\varpi_{li}$ are zero. The corresponding eigenvectors for $\lambda_{3}$ are given by
\begin{eqnarray}
\textbf{x}&=&\Big[\textbf{0}_{1\times nq},\sum_{i=1}^{n}\sum_{l=1}^{q}\varpi_{li}\Big(\textbf{g}_{l}-M_{k}^{+}M_{k}\textbf{g}_{l}+(\textbf{g}_{j}^{\mathrm{T}}M_{k}\textbf{g}_{l})M_{k}^{+}\phi_{k}-(\textbf{g}_{j}^{\mathrm{T}}\textbf{g}_{l})\phi_{k}\Big)^{\mathrm{T}}\otimes\textbf{e}_{i}^{\mathrm{T}},\textbf{0}_{1\times n}\Big]^{*},
\end{eqnarray} where $\varpi_{li}\in\mathbb{C}$ and not all of them are zero. Therefore, $\ker\Big(A_{k}^{[j]}+h_{k}A_{{\mathrm{c}}k}-\lambda_{3} I_{2nq+n}\Big)$ is given by (\ref{egns8}).
\end{IEEEproof}

\begin{lemma}\label{lemma_B}
Define a (possibly infinite) series of matrices $B^{[j]}_{k}$, $j=1,\ldots,q$, $k=0,1,2,\ldots$, as follows:
\begin{eqnarray}\label{Bmatrix}
B_{k}^{[j]}=\small\left[\begin{array}{ccc}
\textbf{0}_{nq\times nq} & h_{k}I_{nq} & \textbf{0}_{nq\times n} \\
-h_{k}\mu_{k} L_{k}\otimes P_{k}-h_{k}\kappa I_{q}\otimes P_{k} & -h_{k}\eta_{k} L_{k}\otimes P_{k} & h_{k}\kappa_{k} \textbf{1}_{q\times 1}\otimes P_{k} \\
E_{n\times nq}^{[j]} & \textbf{0}_{n\times nq} & -I_{n} \\
\end{array}\right],
\end{eqnarray} where $\mu_{k},\eta_{k},\kappa_{k}\geq0$ and $h_{k}>0$, $k\in\overline{\mathbb{Z}}_{+}$, $L_{k}\in\mathbb{R}^{q\times q}$ denotes the Laplacian matrix of a node-fixed dynamic digraph $\mathcal{G}_{k}$, $P_{k}\in\mathbb{R}^{n\times n}$ denotes a paracontracting matrix, and $E_{n\times nq}^{[j]}\in\mathbb{R}^{n\times nq}$ is defined in Lemma~\ref{lemma_EW}. Assume that ${\mathrm{rank}}(P_{k})=n$ for every $k\in\overline{\mathbb{Z}}_{+}$. Then for every $j=1,\ldots,q$, $\{0\}\subseteq{\mathrm{spec}}(B_{k}^{[j]}+h_{k}^{2}A_{{\mathrm{c}}k})\subseteq\{0,-1,-\frac{h_{k}^{2}\kappa_{k}}{2}\pm\frac{1}{2}\sqrt{(h_{k}^{2}\kappa_{k})^{2}-4h_{k}^{2}\kappa_{k}},\lambda_{1},\lambda_{2}\in\mathbb{C}:\forall\frac{\lambda_{1}^{2}+\kappa_{k} h_{k}^{2}\lambda_{1}+\kappa_{k}h_{k}^{2}}{\eta_{k}h_{k}\lambda_{1}+\mu_{k} h_{k}^{2}\lambda_{1}+\mu_{k}h_{k}^{2}}\in{\mathrm{spec}}(-L_{k})\backslash\{0\},\lambda_{2}^{3}+(1+h_{k}^{2}\kappa_{k})\lambda_{2}^{2}+(2h_{k}^{2}\kappa_{k}-h_{k}\kappa_{k})\lambda_{2}+h_{k}^{2}\kappa_{k}=0\}$, where $A_{{\mathrm{c}}k}$ is defined by (\ref{Ac}) in Lemma~\ref{lemma_semisimple}. Furthermore, if $h_{k}\kappa_{k}\neq0$, then 0 is semisimple.  
\end{lemma}

\begin{IEEEproof}
For a fixed $j\in\{1,\ldots,q\}$, let $\lambda\in{\mathrm{spec}}(B_{k}^{[j]}+h_{k}^{2}A_{{\mathrm{c}}k})$ and $\textbf{x}=[\textbf{x}_{1}^{*},\textbf{x}_{2}^{*},\textbf{x}_{3}^{*}]^{*}\in\mathbb{C}^{2nq+n}$ be the corresponding eigenvector for $\lambda$, where $\textbf{x}_{1},\textbf{x}_{2}\in\mathbb{C}^{nq}$ and $\textbf{x}_{3}\in\mathbb{C}^{n}$. Then it follows from
$(B_{k}^{[j]}+h_{k}^{2}A_{{\mathrm{c}}k})\textbf{x}=\lambda\textbf{x}$ that 
\begin{eqnarray}
h_{k}\textbf{x}_{2}+h_{k}^{2}[-\mu_{k} (L_{k}\otimes P_{k})\textbf{x}_{1}-\kappa_{k}(I_{q}\otimes P_{k})\textbf{x}_{1}-\eta_{k}(L_{k}\otimes P_{k})\textbf{x}_{2}+\kappa_{k}(\textbf{1}_{q\times 1}\otimes P_{k})\textbf{x}_{3}]=\lambda\textbf{x}_{1},\label{hx11}\\
h_{k}[-\mu_{k} (L_{k}\otimes P_{k})\textbf{x}_{1}-\kappa_{k}(I_{q}\otimes P_{k})\textbf{x}_{1}-\eta_{k}(L_{k}\otimes P_{k})\textbf{x}_{2}+\kappa_{k}(\textbf{1}_{q\times 1}\otimes P_{k})\textbf{x}_{3}]=\lambda \textbf{x}_{2},\label{x21}\\
E_{n\times nq}^{[j]}\textbf{x}_{1}-\textbf{x}_{3}=\lambda \textbf{x}_{3}.\label{Aeig_31}
\end{eqnarray} Let $\textbf{x}_{3}\neq\textbf{0}_{n\times1}$ be arbitrary, $\textbf{x}_{1}=(\textbf{1}_{q\times 1}\otimes I_{n})\textbf{x}_{3}$, and $\textbf{x}_{2}=\textbf{0}_{nq\times 1}$. Clearly such $\textbf{x}_{i}$, $i=1,2,3$, satisfy (\ref{hx11})--(\ref{Aeig_31}) with $\lambda=0$. Hence, $\lambda=0$ is always an eigenvalue of $B_{k}^{[j]}+h_{k}^{2}A_{{\mathrm{c}}k}$. Next, we assume that $\lambda\neq0$.

Substituting (\ref{x21}) into (\ref{hx11}) yields
$\textbf{x}_{1}=\frac{h_{k}(1+\lambda)}{\lambda}\textbf{x}_{2}$. Replacing $\textbf{x}_{1}$ in (\ref{x21}) and (\ref{Aeig_31}) with $\textbf{x}_{1}=\frac{h_{k}(1+\lambda)}{\lambda}\textbf{x}_{2}$ yields
\begin{eqnarray}
\Big[\Big(\frac{h_{k}^{2}\mu_{k}}{\lambda}+\mu_{k} h_{k}^{2}+\eta_{k}h_{k}\Big)(L_{k}\otimes P_{k})+\Big(\frac{h_{k}^{2}\kappa_{k}}{\lambda}+\lambda+h_{k}^{2}\kappa_{k}\Big)(I_{q}\otimes P_{k})\Big]\textbf{x}_{2}-h_{k}\kappa_{k}(\textbf{1}_{q\times 1}\otimes P_{k})\textbf{x}_{3}=\textbf{0}_{nq\times 1},\label{hx21}\\
E_{n\times nq}^{[j]}\textbf{x}_{2}-(1+\lambda)\textbf{x}_{3}=\textbf{0}_{n\times 1}.\label{hx31}
\end{eqnarray} Thus, (\ref{hx21}) and (\ref{hx31}) have nontrivial solutions if and only if
\begin{eqnarray}\label{detcon1}
\det\small\left[\begin{array}{cc}
\Big(\frac{h_{k}^{2}\mu_{k}}{\lambda}+\mu_{k} h_{k}^{2}+\eta_{k}h_{k}\Big)(L_{k}\otimes P_{k})+\Big(\frac{h_{k}^{2}\kappa_{k}}{\lambda}+\lambda+h_{k}^{2}\kappa_{k}\Big)(I_{q}\otimes P_{k}) & -h_{k}\kappa_{k}(\textbf{1}_{q\times 1}\otimes P_{k})\\
E_{n\times nq}^{[j]} & -(1+\lambda)I_{n}
\end{array}\right]=0.
\end{eqnarray} 

If $\det\Big[\Big(\frac{h_{k}^{2}\mu_{k}}{\lambda}+\mu_{k} h_{k}^{2}+\eta_{k}h_{k}\Big)(L_{k}\otimes P_{k})+\Big(\frac{h_{k}^{2}\kappa_{k}}{\lambda}+\lambda+h_{k}^{2}\kappa_{k}\Big)(I_{q}\otimes P_{k})\Big]\neq0$, then pre-multiplying $L_{k}\otimes I_{n}$ on both sides of (\ref{hx21}) and following the similar arguments as in the proof of Lemma~\ref{lemma_A}, we have $\textbf{x}_{2}=\sum_{l=0}^{q-1-{\mathrm{rank}}(L_{k})}\sum_{i=1}^{n}\varpi_{li}(\textbf{w}_{l}\otimes\textbf{e}_{i})$, where $\varpi_{li}\in\mathbb{C}$. Substituting this expression of $\textbf{x}_{2}$ into (\ref{hx21}) and (\ref{hx31}) by using $iii$) of Lemma~\ref{lemma_EW} and noting that $P_{k}$ is invertible yields
\begin{eqnarray}
\Big(\frac{h_{k}^{2}\kappa_{k}}{\lambda}+\lambda+h_{k}^{2}\kappa_{k}\Big)\sum_{l=0}^{q-1-{\mathrm{rank}}(L_{k})}\sum_{i=1}^{n}\varpi_{li}w_{lj}\textbf{e}_{i}-h_{k}\kappa_{k}\textbf{x}_{3}=\textbf{0}_{n\times 1},\label{x3_eqn1}\\
\sum_{l=0}^{q-1-{\mathrm{rank}}(L_{k})}\sum_{i=1}^{n}\varpi_{li}w_{lj}\textbf{e}_{i}-(1+\lambda)\textbf{x}_{3}=\textbf{0}_{n\times 1}.\label{x3_eqn2}
\end{eqnarray} Substituting (\ref{x3_eqn2}) into (\ref{x3_eqn1}) yields
\begin{eqnarray}
\Big[\Big(\frac{h_{k}^{2}\kappa_{k}}{\lambda}+\lambda+h_{k}^{2}\kappa_{k}\Big)(1+\lambda)-h_{k}\kappa_{k}\Big]\textbf{x}_{3}=\textbf{0}_{n\times 1}.
\end{eqnarray} If $\textbf{x}_{3}=\textbf{0}_{n\times 1}$, then it follows from (\ref{hx21}) that $\textbf{x}_{2}=\textbf{0}_{nq\times 1}$, and hence, $\textbf{x}_{1}=\textbf{0}_{nq\times 1}$, which is a contradiction since $\textbf{x}$ is an eigenvector. Thus, $\textbf{x}_{3}\neq\textbf{0}_{n\times 1}$ and consequently,
$\Big(\frac{h_{k}^{2}\kappa_{k}}{\lambda}+\lambda+h_{k}^{2}\kappa_{k}\Big)(1+\lambda)-h_{k}\kappa_{k}=0$,
i.e.,
\begin{eqnarray}\label{cubic3}
\lambda^{3}+(1+h_{k}^{2}\kappa_{k})\lambda^{2}+(2h_{k}^{2}\kappa_{k}-h_{k}\kappa_{k})\lambda+h_{k}^{2}\kappa_{k}=0.
\end{eqnarray}
Solving this cubic equation in terms of $\lambda$ gives the possible eigenvalues of $B_{k}^{[j]}+h_{k}^{2}A_{{\mathrm{c}}k}$. This can be done via Cardano's formula. If $h_{k}\kappa_{k}=0$, then $\lambda=-1$. Otherwise, if $h_{k}\kappa_{k}\neq0$, then it follows from Routh's Stability Criterion that ${\mathrm{Re}}\,\lambda<0$ if and only if $2h_{k}^{2}\kappa_{k}-h_{k}\kappa_{k}>0$ and $(1+h_{k}^{2}\kappa_{k})(2h_{k}^{2}\kappa_{k}-h_{k}\kappa_{k})>h_{k}^{2}\kappa_{k}$, that is, $h_{k}>1/2$ and $h_{k}+2h_{k}^{3}\kappa_{k}>1+h_{k}^{2}\kappa_{k}$. 

Alternatively, if $\det\Big[\Big(\frac{h_{k}^{2}\mu_{k}}{\lambda}+\mu_{k} h_{k}^{2}+\eta_{k}h_{k}\Big)(L_{k}\otimes P_{k})+\Big(\frac{h_{k}^{2}\kappa_{k}}{\lambda}+\lambda+h_{k}^{2}\kappa_{k}\Big)(I_{q}\otimes P_{k})\Big]=0$, then in this case, (\ref{detcon1}) holds if $\lambda=-1$, or $\lambda\neq-1$ and by Proposition 2.8.4 of \cite[p.~116]{Bernstein:2009}, $\det\Big(\Big(\frac{h_{k}^{2}\mu_{k}}{\lambda}+\mu_{k} h_{k}^{2}+\eta_{k}h_{k}\Big)(L_{k}\otimes P_{k})+\Big(\frac{h_{k}^{2}\kappa_{k}}{\lambda}+\lambda+h_{k}^{2}\kappa_{k}\Big)(I_{q}\otimes P_{k})-\frac{\kappa_{k}h_{k}}{1+\lambda}W_{k}^{[j]}\Big)=0$, which implies that for $\lambda\neq-1$, the equation
\begin{eqnarray}\label{eqn_v1}
\Big(\Big(\frac{h_{k}^{2}\mu_{k}}{\lambda}+\mu_{k} h_{k}^{2}+\eta_{k}h_{k}\Big)(L_{k}\otimes P_{k})+\Big(\frac{h_{k}^{2}\kappa_{k}}{\lambda}+\lambda+h_{k}^{2}\kappa_{k}\Big)(I_{q}\otimes P_{k})-\frac{\kappa_{k}h_{k}}{1+\lambda}W_{k}^{[j]}\Big)\textbf{v}=\textbf{0}_{nq\times 1}
\end{eqnarray} has nontrivial solutions for $\textbf{v}\in\mathbb{C}^{nq}$. Again, note that for every $j=1,\ldots,q$, $(L_{k}\otimes I_{n})W_{k}^{[j]}=\textbf{0}_{nq\times nq}$. Pre-multiplying $L_{k}\otimes I_{n}$ on both sides of (\ref{eqn_v1}) yields $\Big(\Big(\frac{h_{k}^{2}\mu_{k}}{\lambda}+\mu_{k} h_{k}^{2}+\eta_{k}h_{k}\Big)(L_{k}^{2}\otimes P_{k})+\Big(\frac{h_{k}^{2}\kappa_{k}}{\lambda}+\lambda+h_{k}^{2}\kappa_{k}\Big)(L_{k}\otimes P_{k})\Big)\textbf{v}=(I_{q}\otimes P_{k})(L_{k}\otimes I_{n})\Big(\Big(\frac{h_{k}^{2}\mu_{k}}{\lambda}+\mu_{k} h_{k}^{2}+\eta_{k}h_{k}\Big)(L_{k}\otimes I_{n})+\Big(\frac{h_{k}^{2}\kappa_{k}}{\lambda}+\lambda+h_{k}^{2}\kappa_{k}\Big)I_{nq}\Big)\textbf{v}=\textbf{0}_{nq\times 1}$, which implies that $\Big(\Big(\frac{h_{k}^{2}\mu_{k}}{\lambda}+\mu_{k} h_{k}^{2}+\eta_{k}h_{k}\Big)(L_{k}\otimes I_{n})+\Big(\frac{h_{k}^{2}\kappa_{k}}{\lambda}+\lambda+h_{k}^{2}\kappa_{k}\Big)I_{nq}\Big)\textbf{v}\in\ker(L_{k}\otimes I_{n})$. Since $\ker(L_{k}\otimes I_{n})=\bigcup_{l=0}^{q-1-{\mathrm{rank}}(L_{k})}{\mathrm{span}}\{\textbf{w}_{l}\otimes\textbf{e}_{1},\ldots,\textbf{w}_{l}\otimes\textbf{e}_{n}\}$, it follows that
\begin{eqnarray}\label{Az=b1}
\Big(\Big(\frac{h_{k}^{2}\mu_{k}}{\lambda}+\mu_{k} h_{k}^{2}+\eta_{k}h_{k}\Big)(L_{k}\otimes I_{n})+\Big(\frac{h_{k}^{2}\kappa_{k}}{\lambda}+\lambda+h_{k}^{2}\kappa_{k}\Big)I_{nq}\Big)\textbf{v}=\sum_{i=1}^{n}\sum_{l=0}^{q-1-{\mathrm{rank}}(L_{k})}\omega_{li}\textbf{w}_{l}\otimes\textbf{e}_{i},
\end{eqnarray} where $\omega_{li}\in\mathbb{C}$, which is similar to (\ref{Az=b}). Now it follows from (\ref{eqn_v1}) and (\ref{Az=b1}) that
\begin{eqnarray}
\frac{\kappa_{k}h_{k}}{1+\lambda}W_{k}^{[j]}\textbf{v}=\sum_{i=1}^{n}\sum_{l=0}^{q-1-{\mathrm{rank}}(L_{k})}\omega_{li}\textbf{w}_{l}\otimes\textbf{e}_{i}.
\end{eqnarray} 

If $\frac{h_{k}^{2}\kappa_{k}}{\lambda}+\lambda+h_{k}^{2}\kappa_{k}\neq0$, then it follows from the similar arguments after (\ref{Wv_eqn}) that $\omega_{\ell i}=0$ for every $i=1,\ldots,n$ and every $\ell=1,\ldots,q-1-{\mathrm{rank}}(L_{k})$. Furthermore, 
\begin{eqnarray}
\omega_{0i}-\frac{\lambda\kappa_{k}h_{k}}{(1+\lambda)(\lambda^{2}+h_{k}^{2}\kappa_{k}\lambda+h_{k}^{2}\kappa_{k})}\omega_{0i}=0,\quad i=1,\ldots,n.
\end{eqnarray}
Then either $1-\frac{\lambda\kappa_{k}h_{k}}{(1+\lambda)(\lambda^{2}+h_{k}^{2}\kappa_{k}\lambda+h_{k}^{2}\kappa_{k})}=0$ or $\omega_{0i}=0$ for every $i=1,\ldots,n$.
If $\frac{\lambda\kappa_{k}h_{k}}{(1+\lambda)(\lambda^{2}+h_{k}^{2}\kappa_{k}\lambda+h_{k}^{2}\kappa_{k})}=1$, then
\begin{eqnarray} 
\lambda^{3}+(1+h_{k}^{2}\kappa_{k})\lambda^{2}+(2h_{k}^{2}\kappa_{k}-h_{k}\kappa_{k})\lambda+h_{k}^{2}\kappa_{k}=0,
\end{eqnarray} which is the same as (\ref{cubic3}). Since $\lambda\neq -1$, in this case $\kappa_{k}h_{k}\neq0$. Then it follows from Routh's Stability Criterion that ${\mathrm{Re}}\,\lambda<0$ if and only if $h_{k}>1/2$ and $h_{k}+2h_{k}^{3}\kappa_{k}>1+h_{k}^{2}\kappa_{k}$.
If $\omega_{0i}=0$ for every $i=1,\ldots,n$, then it follows from (\ref{eqn_v1}) and (\ref{Az=b1}) that $\frac{\kappa_{k}h_{k}}{1+\lambda}W_{k}^{[j]}\textbf{v}=\textbf{0}_{nq\times 1}$ and $\Big(\Big(\frac{h_{k}^{2}\mu_{k}}{\lambda}+\mu_{k} h_{k}^{2}+\eta_{k}h_{k}\Big)(L_{k}\otimes I_{n})+\Big(\frac{h_{k}^{2}\kappa_{k}}{\lambda}+\lambda+h_{k}^{2}\kappa_{k}\Big)I_{nq}\Big)\textbf{v}=\textbf{0}_{nq\times 1}$, which implies that $\textbf{v}\in\ker\Big(\Big(\frac{h_{k}^{2}\mu_{k}}{\lambda}+\mu_{k} h_{k}^{2}+\eta_{k}h_{k}\Big)(L_{k}\otimes I_{n})+\Big(\frac{h_{k}^{2}\kappa_{k}}{\lambda}+\lambda+h_{k}^{2}\kappa_{k}\Big)I_{nq}\Big)\cap\ker(\frac{\kappa_{k}h_{k}}{1+\lambda}W_{k}^{[j]})$. Clearly $\frac{h_{k}^{2}\mu_{k}}{\lambda}+\mu_{k} h_{k}^{2}+\eta_{k}h_{k}\neq0$. In this case, $\lambda\in\{\lambda_{1}\in\mathbb{C}:\forall\frac{\lambda_{1}^{2}+\kappa_{k} h_{k}^{2}\lambda_{1}+\kappa_{k}h_{k}^{2}}{\eta_{k}h_{k}\lambda_{1}+\mu_{k} h_{k}^{2}\lambda_{1}+\mu_{k}h_{k}^{2}}\in{\mathrm{spec}}(-L_{k})\backslash\{0\}\}$.

Alternatively, if $\frac{h_{k}^{2}\kappa_{k}}{\lambda}+\lambda+h_{k}^{2}\kappa_{k}=0$, then it follows from the similar arguments after (\ref{eigv4}) in Lemma~\ref{lemma_A} that
\begin{eqnarray}
\lambda=-\frac{h_{k}^{2}\kappa_{k}}{2}\pm\frac{1}{2}\sqrt{(h_{k}^{2}\kappa_{k})^{2}-4h_{k}^{2}\kappa_{k}}
\end{eqnarray} are the possible eigenvalues of $B_{k}^{[j]}+h_{k}^{2}A_{{\mathrm{c}}k}$. 

In summary,
\begin{eqnarray}\label{egspace} 
&&\{0\}\subseteq{\mathrm{spec}}(B_{k}^{[j]}+h_{k}^{2}A_{{\mathrm{c}}k})\subseteq\nonumber\\
&&\Big\{0,-1,-\frac{h_{k}^{2}\kappa_{k}}{2}\pm\frac{1}{2}\sqrt{(h_{k}^{2}\kappa_{k})^{2}-4h_{k}^{2}\kappa_{k}},\lambda_{1},\lambda_{2}\in\mathbb{C}:\forall\frac{\lambda_{1}^{2}+\kappa_{k} h_{k}^{2}\lambda_{1}+\kappa_{k}h_{k}^{2}}{\eta_{k}h_{k}\lambda_{1}+\mu_{k} h_{k}^{2}\lambda_{1}+\mu_{k}h_{k}^{2}}\in{\mathrm{spec}}(-L_{k})\backslash\{0\},\nonumber\\
&&\lambda_{2}^{3}+(1+h_{k}^{2}\kappa_{k})\lambda_{2}^{2}+(2h_{k}^{2}\kappa_{k}-h_{k}\kappa_{k})\lambda_{2}+h_{k}^{2}\kappa_{k}=0\Big\}. 
\end{eqnarray} Finally, the semisimplicity property of 0 can be proved by using the similar arguments as in the proof of Lemma~\ref{lemma_semisimple}. 
\end{IEEEproof}

Now we have the main result for the global convergence of the iterative process in Algorithm~\ref{MCO}.

\begin{theorem}\label{thm_HMCO}
Consider the following discrete-time switched linear model to describe the iterative process for MCO:
\begin{eqnarray}
x_{i}[k+1]&=&x_{i}[k]+h_{k}v_{i}[k+1],\quad x_{i}[0]=x_{i0},\label{DE_1}\\
v_{i}[k+1]&=&P[k]v_{i}[k]+h_{k}\eta_{k}P[k]\sum_{j\in\mathcal{N}_{k}^{i}}(v_{j}[k]-v_{i}[k])+h_{k}\mu_{k}P[k]\sum_{j\in\mathcal{N}_{k}^{i}}(x_{j}[k]-x_{i}[k])\nonumber\\
&&+h_{k}\kappa_{k}P[k](p[k]-x_{i}[k]),\quad v_{i}[0]=v_{i0},\\
p[k+1]&=&p[k]+h_{k}\kappa_{k}(x_{j}[k]-p[k]),\quad p[k]\not\in\mathcal{Z}_{p},\quad p[0]=p_{0},\\
p[k+1]&=&x_{j}[k],\quad p[k]\in\mathcal{Z}_{p},\quad k=0,1,2,\ldots,\quad i=1,\ldots,q,\label{MSO_4}
\end{eqnarray} where $x_{i}\in\mathbb{R}^{n}$, $v_{i}\in\mathbb{R}^{n}$, $p\in\mathbb{R}^{n}$, $\mu_{k},\eta_{k},\kappa_{k},h_{k}$ are randomly selected in $\Omega\subseteq[0,\infty)$, $\mathcal{Z}_{p}=\{p\in\mathbb{R}^{n}:f(x_{j})<f(p)\}$, and $x_{j}=\{x_{\min}\in\mathbb{R}^{n}:x_{\min}=\arg\min_{1\leq i\leq q}f(x_{i})\}$. Assume that for every $k\in\overline{\mathbb{Z}}_{+}$ and every $j=1,\ldots,q$:
\begin{itemize}
\item[H1)] $P[k]\in\mathbb{R}^{n\times n}$ is paracontracting and ${\mathrm{rank}}(P[k])=n$.
\item[H2)] $0<h_{k}<-\frac{\lambda+\bar{\lambda}}{|\lambda|^{2}}$ for every $\lambda\in\{-\kappa_{k},-\frac{\kappa_{k}(1+h_{k})}{2}\pm\frac{1}{2}\sqrt{\kappa_{k}^{2}(1+h_{k})^{2}-4\kappa_{k}},-\frac{\kappa_{k}h_{k}}{2}\pm\frac{1}{2}\sqrt{\kappa_{k}^{2}h_{k}^{2}-4\kappa_{k}},\lambda\in\mathbb{C}:\forall \frac{\lambda^{2}+\kappa_{k} h_{k}\lambda+\kappa_{k}}{\eta_{k}\lambda+\mu_{k} h_{k}\lambda+\mu_{k}}\in{\mathrm{spec}}(-L_{k})\backslash\{0\}\}$;
\item[H3)] $0<h_{k}<-\frac{\lambda+\bar{\lambda}}{|\lambda|^{2}}$ for every $\lambda\in\{-1,-\frac{h_{k}^{2}\kappa_{k}}{2}\pm\frac{1}{2}\sqrt{(h_{k}^{2}\kappa_{k})^{2}-4h_{k}^{2}\kappa_{k}},\lambda_{1},\lambda_{2}\in\mathbb{C}:\forall\frac{\lambda_{1}^{2}+\kappa_{k} h_{k}^{2}\lambda_{1}+\kappa_{k}h_{k}^{2}}{\eta_{k}h_{k}\lambda_{1}+\mu_{k} h_{k}^{2}\lambda_{1}+\mu_{k}h_{k}^{2}}\in{\mathrm{spec}}(-L_{k})\backslash\{0\},\lambda_{2}^{3}+(1+h_{k}^{2}\kappa_{k})\lambda_{2}^{2}+(2h_{k}^{2}\kappa_{k}-h_{k}\kappa_{k})\lambda_{2}+h_{k}^{2}\kappa_{k}=0\}$;
\item[H4)] $\|I_{2nq+n}+h_{k}A_{k}^{[j]}+h_{k}^{2}A_{{\mathrm{c}}k}\|\leq 1$ and $\|I_{2nq+n}+B_{k}^{[j]}+h_{k}^{2}A_{{\mathrm{c}}k}\|\leq1$.
\item[H5)] $\ker((h_{k}A_{k}^{[j]}+h_{k}^{2}A_{{\mathrm{c}}k})^{\mathrm{T}}(h_{k}A_{k}^{[j]}+h_{k}^{2}A_{{\mathrm{c}}k})+(h_{k}A_{k}^{[j]}+h_{k}^{2}A_{{\mathrm{c}}k})^{\mathrm{T}}+h_{k}A_{k}^{[j]}+h_{k}^{2}A_{{\mathrm{c}}k})=\ker((h_{k}A_{k}^{[j]}+h_{k}^{2}A_{{\mathrm{c}}k})^{\mathrm{T}}\\(h_{k}A_{k}^{[j]}+h_{k}^{2}A_{{\mathrm{c}}k})+(h_{k}A_{k}^{[j]}+h_{k}^{2}A_{{\mathrm{c}}k})^{2})$ and $\ker((B_{k}^{[j]}+h_{k}^{2}A_{{\mathrm{c}}k})^{\mathrm{T}}(B_{k}^{[j]}+h_{k}^{2}A_{{\mathrm{c}}k})+(B_{k}^{[j]}+h_{k}^{2}A_{{\mathrm{c}}k})^{\mathrm{T}}+B_{k}^{[j]}+h_{k}^{2}A_{{\mathrm{c}}k})=\ker((B_{k}^{[j]}+h_{k}^{2}A_{{\mathrm{c}}k})^{\mathrm{T}}(B_{k}^{[j]}+h_{k}^{2}A_{{\mathrm{c}}k})+(B_{k}^{[j]}+h_{k}^{2}A_{{\mathrm{c}}k})^{2})$.
\end{itemize}
Then the following conclusions hold: 
\begin{itemize}
\item[C1)] If $\Omega$ is a finite discrete set, then $x_{i}[k]\to p^{\dag}$, $v_{i}[k]\to \textbf{0}_{n\times 1}$, and $p[k]\to p^{\dag}$ as $k\to\infty$ for every $x_{i0}\in\mathbb{R}^{n}$, $v_{i0}\in\mathbb{R}^{n}$, $p_{0}\in\mathbb{R}^{n}$, and every $i=1,\ldots,q$, where $p^{\dag}\in\mathbb{R}^{n}$ is some constant vector.
\item[C2)] If for every positive integer $N$, there always exists $s\geq N$ such that  $h_{s}(A_{s}^{[j_{s}]}+h_{s}A_{{\mathrm{c}}s})=B_{s}^{[j_{s}]}+h_{s}^{2}A_{{\mathrm{c}}s}=h_{T}(A_{T}^{[j_{T}]}+h_{T}A_{{\mathrm{c}}T})=B_{T}^{[j_{T}]}+h_{T}^{2}A_{{\mathrm{c}}T}$ for some fixed $T\in\overline{\mathbb{Z}}_{+}$, where $j_{s},j_{T}\in\{1,\ldots,q\}$, then $x_{i}[k]\to p^{\dag}$, $v_{i}[k]\to \textbf{0}_{n\times 1}$, and $p[k]\to p^{\dag}$ as $k\to\infty$ for every $x_{i0}\in\mathbb{R}^{n}$, $v_{i0}\in\mathbb{R}^{n}$, $p_{0}\in\mathbb{R}^{n}$, and every $i=1,\ldots,q$, where $p^{\dag}\in\mathbb{R}^{n}$ is some constant vector.
\end{itemize}
\end{theorem}

\begin{IEEEproof}
Let $Z=[x_{1}^{\rm{T}},\ldots,x_{q}^{\rm{T}},v_{1}^{\rm{T}},\ldots,v_{q}^{\rm{T}},p^{\rm{T}}]^{\rm{T}}\in\mathbb{R}^{2nq+n}$. Note that (\ref{DE_1})--(\ref{MSO_4}) can be rewritten as the compact form $Z[k+1]=(I_{2nq+n}+h_{k}(A_{k}^{[j_{k}]}+h_{k}A_{{\mathrm{c}}k}))Z[k]$, $Z[k]\not\in\mathcal{S}$, and $Z[k+1]=(I_{2nq+n}+B_{k}^{[j_{k}]}+h_{k}^{2}A_{{\mathrm{c}}k})Z[k]$, $Z[k]\in\mathcal{S}$, $j_{k}\in\{1,\ldots,q\}$ is selected based on $\mathcal{Z}_{p}$. Let $h_{k}^{\dag}=\min\Big\{-\frac{\lambda+\bar{\lambda}}{|\lambda|^{2}}:\lambda\in\{-\kappa_{k},-\frac{\kappa_{k}(1+h_{k})}{2}\pm\frac{1}{2}\sqrt{\kappa_{k}^{2}(1+h_{k})^{2}-4\kappa_{k}},-\frac{\kappa_{k}h_{k}}{2}\pm\frac{1}{2}\sqrt{\kappa_{k}^{2}h_{k}^{2}-4\kappa_{k}},\lambda\in\mathbb{C}:\forall \frac{\lambda^{2}+\kappa_{k} h_{k}\lambda+\kappa_{k}}{\eta_{k}\lambda+\mu_{k} h_{k}\lambda+\mu_{k}}\in{\mathrm{spec}}(-L_{k})\backslash\{0\}\}\Big\}$. First, we show that if $h<h_{k}^{\dag}$, then $I_{2nq+n}+h_{k}(A_{k}^{[j]}+h_{k}A_{{\mathrm{c}}k})$ becomes discrete-time semistable for every $j=1,\ldots,q$ and every $k=0,1,2,\ldots$. Note that  ${\mathrm{spec}}(I_{2nq+n}+h_{k}(A_{k}^{[j]}+h_{k}A_{{\mathrm{c}}k}))=\{1+h\lambda:\forall\lambda\in{\mathrm{spec}}(A_{k}^{[j]}+h_{k}A_{{\mathrm{c}}k})\}$. Since by Lemma~\ref{lemma_A} and Assumptions H1 and H2, $A_{k}^{[j]}+h_{k}A_{{\mathrm{c}}k}$ is semistable for every $j=1,\ldots,q$ and every $k=0,1,2,\ldots$, it follows that ${\mathrm{spec}}(I_{2nq+n}+h_{k}(A_{k}^{[j]}+h_{k}A_{{\mathrm{c}}k}))=\{1\}\cup\{1+h\lambda:\forall\lambda\in{\mathrm{spec}}(A_{k}^{[j]}+h_{k}A_{{\mathrm{c}}k}),{\mathrm{Re}}\,\lambda<0\}$. Hence, $I_{2nq+n}+h_{k}(A_{k}^{[j]}+h_{k}A_{{\mathrm{c}}k})$ is discrete-time semistable for every $j=1,\ldots,q$ and every $k=0,1,2,\ldots$ if $|1+h_{k}\lambda|<1$ for every $\lambda\in{\mathrm{spec}}(A_{k}^{[j]}+h_{k}A_{{\mathrm{c}}k})$ and ${\mathrm{Re}}\,\lambda<0$. Note that $|1+h_{k}\lambda|<1$ is equivalent to $(1+h_{k}\lambda)(1+h_{k}\bar{\lambda})=|1+h_{k}\lambda|^{2}<1$, i.e., $h_{k}<-(\lambda+\bar{\lambda})/|\lambda|^{2}$. By Lemma~\ref{lemma_A}, for any $h_{k}<h_{k}^{\dag}$, $I_{2nq+n}+h_{k}(A_{k}^{[j]}+h_{k}A_{{\mathrm{c}}k})$ is discrete-time semistable for every $j=1,\ldots,q$ and every $k=0,1,2,\ldots$. Similarly, it follows from Lemma~\ref{lemma_B} and Assumptions H1 and H3 that $I_{2nq+n}+B_{k}^{[j]}+h_{k}^{2}A_{{\mathrm{c}}k}$ is discrete-time semistable for every $j=1,\ldots,q$ and every $k=0,1,2,\ldots$. And (\ref{DE_1})--(\ref{MSO_4}) can further be rewritten as an iteration $Z[k+1]=P_{k}Z[k]$, $k=0,1,2,\ldots$, where $P_{k}\in\{I_{2nq+n}+h_{k}(A_{k}^{[j]}+h_{k}A_{{\mathrm{c}}k}),I_{2nq+n}+B_{k}^{[j]}+h_{k}^{2}A_{{\mathrm{c}}k}:j=1,\ldots,q,k=0,1,2,\ldots\}=\{I_{2nq+n}+h_{k}(A_{k}^{[j]}+h_{k}A_{{\mathrm{c}}k}),I_{2nq+n}+B_{k}^{[j]}+h_{k}^{2}A_{{\mathrm{c}}k}:j=1,\ldots,q,\mu_{k},\eta_{k},\kappa_{k},h_{k}\in\Omega\}$. 

C1) By assumption, $\Omega$ is a finite discrete set. Hence, $\{I_{2nq+n}+h_{k}(A_{k}^{[j]}+h_{k}A_{{\mathrm{c}}k}),I_{2nq+n}+B_{k}^{[j]}+h_{k}^{2}A_{{\mathrm{c}}k}:j=1,\ldots,q,\mu_{k},\eta_{k},\kappa_{k},h_{k}\in\Omega\}$ is a finite discrete set. Now it follows from Assumptions H4 and H5 as well as $i$) of Lemma \ref{lemma_DTSS} that $\lim_{k\to\infty}Z[k]$ exists. The rest of the conclusion follows directly from (\ref{DE_1})--(\ref{MSO_4}).

C2) By assumption, either $h_{T}(A_{T}^{[j_{T}]}+h_{T}A_{{\mathrm{c}}T})$ or $B_{T}^{[j_{T}]}+h_{T}^{2}A_{{\mathrm{c}}T}$ appears infinitely many times in the sequence $\{P_{k}\}_{k=0}^{\infty}$. Next, it follows from Lemmas \ref{lemma_Arank} and \ref{lemma_Ah} as well as the assumption $h_{k}>0$ that $\ker(h_{k}(A_{k}^{[j_{k}]}+h_{k}A_{{\mathrm{c}}k}))=\ker(A_{k}^{[j_{k}]})=\ker(A_{s}^{[j_{s}]})=\ker(h_{s}(A_{s}^{[j_{s}]}+h_{s}A_{{\mathrm{c}}s}))$ for every $k,s\in\overline{\mathbb{Z}}_{+}$. Using the similar arguments, one can prove that $\ker(B_{k}^{[j_{k}]}+h_{k}^{2}A_{{\mathrm{c}}k})=\ker(B_{k}^{[j_{k}]})=\ker(B_{s}^{[j_{s}]})=\ker(B_{s}^{[j_{s}]}+h_{s}^{2}A_{{\mathrm{c}}s})$ for every $k,s\in\overline{\mathbb{Z}}_{+}$. Hence, it follows from Assumptions H4 and H5 as well as $ii$) of Lemma \ref{lemma_DTSS} that $\lim_{k\to\infty}Z[k]$ exists. The rest of the conclusion follows directly from (\ref{DE_1})--(\ref{MSO_4}). Note that in this case, $\Omega$ may be an infinite set.
\end{IEEEproof}





\bibliographystyle{IEEEtran}
\bibliography{Reference}

\end{document}